\documentclass[conference]{IEEEtran}
\ifCLASSINFOpdf
\else
\fi
\usepackage{booktabs}
\usepackage{tabularx}
\usepackage{threeparttable}
\usepackage{graphicx}
\usepackage{multirow}
\usepackage{amssymb}
\usepackage{url}
\usepackage{subcaption}
\usepackage{graphicx}
\usepackage{caption}
\usepackage{booktabs}
\usepackage{xcolor}
\usepackage{array}
\usepackage{colortbl}
\usepackage{threeparttable} 
\definecolor{zebra}{RGB}{242,242,242}
\definecolor{tableborder}{RGB}{242,242,242}
\begin{document}
%
\date{}

\title{Automating Function-Level TARA for Automotive Full-Lifecycle Security}

\author{\IEEEauthorblockN{Yuqiao Yang{\#}}
	\IEEEauthorblockA{UESTC\\
		yyq\_0xdq@163.com}
	\and
	\IEEEauthorblockN{Yongzhao Zhang{\#}}
	\IEEEauthorblockA{UESTC\\
		zhangyongzhao@uestc.edu.cn}
	\and
	\IEEEauthorblockN{Wenhao Liu}
	\IEEEauthorblockA{GoGoByte Technology\\
		lwh.scu@gmail.com}
        \and
	\IEEEauthorblockN{Jun Li}
	\IEEEauthorblockA{GoGoByte Technology\\
		lijun\_research@gogobyte.com}
        \and
	\IEEEauthorblockN{Pengtao Shi}
	\IEEEauthorblockA{GoGoByte Technology\\
		shipengtao@gogobyte.com     
 }
        \and
	\IEEEauthorblockN{DingYu Zhong }
	\IEEEauthorblockA{UESTC\\
		18166789795@163.com}
        \and
	\IEEEauthorblockN{Jie Yang{*}}
	\IEEEauthorblockA{UESTC\\
		jie.yang@uestc.edu.cn}
        \and
	\IEEEauthorblockN{Ting Chen{*}}
	\IEEEauthorblockA{UESTC\\
		chenting19870201@163.com}
        \and
	\IEEEauthorblockN{Sheng Cao }
	\IEEEauthorblockA{UESTC\\
		caosheng@uestc.edu.cn}
        \and
	\IEEEauthorblockN{Yuntao Ren}
	\IEEEauthorblockA{Chengdu Anheng Information\\ Technology Co., LTD\\
		atao\_uestc@163.com}
        \and
        \IEEEauthorblockN{Yongyue Wu}
	\IEEEauthorblockA{Anheng Vision(Chengdu) Information\\ Technology Co., LTD\\
		wuyongyue@isecvision.com}
        \and
        \IEEEauthorblockN{Xiaosong Zhang}
	\IEEEauthorblockA{UESTC\\
		johnsonzxs@uestc.edu.cn}    
        }

\long\def\comment#1{}
\newcommand{\zyz}[1]{ \textcolor{red}{#1}}
\newcommand{\yyq}[1]{ \textcolor{blue}{#1}}
\newcommand{\zdy}[1]{ \textcolor{brown}{#1}}
\newcommand{\projname}{DefenseWeaver}

\maketitle
\begin{abstract}
As modern vehicles evolve into intelligent and connected systems, their growing complexity introduces significant cybersecurity risks. Threat Analysis and Risk Assessment (TARA) has therefore become essential for managing these risks under mandatory regulations. However, existing TARA automation methods rely on static threat libraries, limiting their utility in the detailed, function-level analyses demanded by industry. This paper introduces \projname{}, the first system that automates function-level TARA using component-specific details and large language models (LLMs). \projname{} dynamically generates attack trees and risk evaluations from system configurations described in an extended OpenXSAM++ format, then employs a multi-agent framework to coordinate specialized LLM roles for more robust analysis. To further adapt to evolving threats and diverse standards, \projname{} incorporates Low-Rank Adaptation (LoRA) fine-tuning and Retrieval-Augmented Generation (RAG) with expert-curated TARA reports. We validated \projname{} through deployment in four automotive security projects, where it identified 11 critical attack paths, verified through penetration testing, and subsequently reported and remediated by the relevant automakers and suppliers. Additionally, DefenseWeaver demonstrated cross-domain adaptability, successfully applying to unmanned aerial vehicles (UAVs) and marine navigation systems. In comparison to human experts, DefenseWeaver outperformed manual attack tree generation across six assessment scenarios. Integrated into commercial cybersecurity platforms such as UAES and Xiaomi, DefenseWeaver has generated over 8,200 attack trees. These results highlight its ability to significantly reduce processing time, and its scalability and transformative impact on cybersecurity across industries.

\end{abstract}
\IEEEpeerreviewmaketitle
\section{Introduction}

The automotive industry is rapidly advancing toward intelligent, networked vehicles, integrating technologies like autonomous driving~\cite{autonomous_driving2}, Over-The-Air updates~\cite{OTA1},
and Advanced Driver Assistance Systems~\cite{ADAS}. 
While these innovations enhance functionality and user experience, they also increase the number of Electronic Control Units (ECUs) and the complexity of topologies and interconnectivity within the In-Vehicle Network (IVN), significantly expanding the potential attack surface of modern vehicles. By 2030, an estimated 95\% of new vehicles will be connected, creating a vast cyber threat landscape~\cite{abdelkader2021connected}. 
Another growing concern is the risk of supply chain safety, where vulnerabilities in third-party components can compromise overall security. As component interconnectivity increases, so too does the number of potential attack vectors, highlighting the need to assess both the entire vehicle and its individual components.
In response to the growing attack surface, TARA has become a cornerstone of automotive cybersecurity, systematically identifying, analyzing, and prioritizing security risks.
At its core, TARA involves generating attack trees and assessing risk levels, which together provide a structured approach for understanding and mitigating potential threats throughout the vehicle lifecycle. 
TARA is also a mandatory regulatory requirement for automotive OEMs and suppliers, in compliance with standards such as WP29 R155e~\cite{WP29_R155e} and ISO/SAE 21434~\cite{ISO/SAE_21434}. Despite its critical importance, TARA is still largely conducted manually, making it labor-intensive, time-consuming, and difficult to scale. Security analysts must repeatedly perform TARA for multiple threat scenarios, an approach that becomes increasingly impractical as vehicle systems grow in complexity and interconnectivity. This inefficiency, combined with the rise of supply chain vulnerabilities, highlights the urgent need for scalable, automated solutions.

Existing datalog-based approaches~\cite{CarVal,Saulaiman} to automate TARA for improved efficiency primarily focus on vehicle-level assessments, leaving a critical gap in addressing function-level TARA, as required by WP29 R155e and ISO/SAE 21434. Vehicle-level TARA identifies overarching threats that affect the entire vehicle but often overlooks the specific implementation details of individual components. In contrast, function-level TARA examines detailed functions or components, such as battery management systems or individual ECUs, considering their interactions, hardware configurations, software versions, communication channels, and interfaces. This granularity is also crucial in the context of supply chain risks, where vulnerabilities in third-party components can compromise vehicle security. 
Thus, TARA at the function level offers deeper insight into vulnerabilities and attack paths, making it essential for comprehensive risk management.


Moreover, existing datalog-based approaches~\cite{CarVal,Saulaiman} rely on predefined threat libraries, which pose major limitations when extending to function-level TARA. These libraries lack the granularity needed to address component-specific threats in function-level analyses and are difficult to maintain amid a rapidly evolving cybersecurity landscape. This raises a critical question: \textit{Can TARA be automated to enable detailed and adaptive function-level assessments?}

\begin{figure}[t]
    \centering
    \includegraphics[width=1\linewidth]{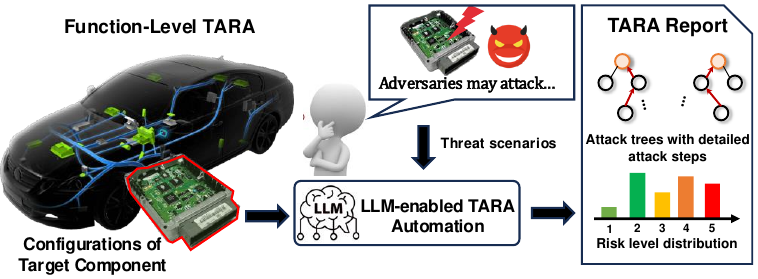}
    \caption{\projname{} is capable of automating function-level TARA by leveraging the power of LLMs for components with detailed attributes.}
    \label{fig:intro}
\end{figure}

\noindent\textbf{Our Approach:}
We introduce \textbf{\projname{}}, a novel system that automates the function-level TARA by leveraging component-specific details and the LLMs.
By incorporating detailed, component-specific information, \projname{} dynamically generates attack trees and evaluates associated risk levels. As shown in Fig.~\ref{fig:intro}, users only need to define relevant threat scenarios—\projname{} then conducts the TARA process with minimal manual input, significantly reducing the workload on security analysts. Importantly, the system produces both attack trees and risk assessments, two foundational components of TARA. Attack trees provide structured visualizations of potential threat paths, while risk assessments categorize their severity, enabling prioritized and informed mitigation. When developing \projname{}, we address the following key issues.
\textit{\textbf{Representing Complex Automotive Configurations.}} Function-level automotive configurations, created during development phases, are diagrammatic models that detail components (e.g., hardware, software), channels, interfaces, and their associated attributes within the IVN. 
A suitable representation must balance comprehensiveness, machine-readability, and conciseness to ensure efficient processing by LLMs. 
To address this, we propose the following designs: (i) OpenXSAM++ Format: a structured format to systematically represent automotive configurations, capturing detailed attributes while preserving the logical and visual topology of IVNs. (ii) Logical Path Extraction: abstraction of connectivity between units for each threat scenario, omitting specific attack techniques or procedures to reduce unnecessary complexity.
(iii) Atom Segmentation: decomposition of logical paths into atomic structures, the minimal units that preserve essential topology, enabling efficient and accurate analysis.
\textit{\textbf{Building Attack Trees and Assessing Risk Levels with LLMs.}} Generating comprehensive attack trees for function-level TARA requires detailed component-level reasoning and the ability to infer attack methods without relying on static threat libraries. 
To achieve this, we designed a multi-agent framework built on LLMs, with each agent fulfilling a specialized role:
(i) \textit{Sub-Tree Constructor}: Creates sub-trees from atomic structures, embedding detailed attack methods for granular analysis.
(ii) \textit{Attack-Tree Assembler}: Integrates sub-trees into complete attack trees, ensuring logical consistency and coherence between consecutive nodes.
(iii) \textit{Risk Assessor}: Analyzes the feasibility of attack methods and computes the overall risk level for threat scenarios, providing actionable insights into potential vulnerabilities.
Together, these agents ensure comprehensive attack trees and rational risk assessments aligned with standard TARA requirements.

\textit{\textbf{Adapting to Evolving Threat Landscapes and Diverse Standards.}} For full-lifecycle security, the TARA process must adapt to evolving threats and various evaluation standards across regions and stakeholders. To address this, we incorporate Low-Rank Adaptation (LoRA) fine-tuning and Retrieval-Augmented Generation (RAG) to dynamically provide relevant examples and tailored prompts for LLM-based agents. This ensures \projname{} accommodates diverse requirements while maintaining compliance and practicality, enhancing its adaptability and robustness.

\comment{
\zyz{
Modern vehicles are built on extremely complex electrical and electronic architectures. A typical modern car now integrates over 150 electronic control units (ECUs) and more than 100 million lines of code, reflecting unprecedented system complexity. No single manufacturer can develop all of these components in-house, so original equipment manufacturers (OEMs) rely on a vast global supply chain of specialized suppliers for hardware, software, and sub-systems. While this distributed development model accelerates innovation, it also expands the attack surface by introducing potential vulnerabilities across the supply chain.
}

\zyz{
However, this extensive reliance on suppliers has opened the door to new cyber threats, notably supply chain attacks. In a supply chain attack (sometimes called supply chain poisoning), adversaries target a supplier’s hardware, firmware, or software to inject malicious code or backdoors before it reaches the vehicle. Recent reports show that such attacks are on the rise: for example, 41\% of documented automotive cyber incidents in 2022 were attributed to compromises in the supply chain. These breaches can have far-reaching consequences, since a single compromised component may be incorporated into many vehicles, affecting multiple models or brands. The prevalence and severity of supply chain attacks underscore the urgent need to secure every link in the automotive supply chain.
}

\zyz{
To address these escalating risks, new cybersecurity regulations and standards mandate stronger protections at all levels of vehicle development. The United Nations’ UN R155 regulation (adopted in 2021) requires automakers to implement a certified Cyber Security Management System (CSMS) and to ensure that their entire supply chain (including all tier-1 and sub-tier suppliers) complies with cybersecurity requirements. This regulatory pressure effectively pushes cybersecurity accountability down to each component and supplier. In parallel, the ISO/SAE 21434:2021 standard provides a framework for managing automotive cybersecurity across the vehicle lifecycle and explicitly calls for performing Threat Analysis and Risk Assessment (TARA) at the vehicle- and component-level during development. In practice, ISO 21434 requires manufacturers and suppliers to identify potential attack paths for each critical component and to assess the associated risks early in the design process. Together, regulations like UN R155 and standards like ISO 21434 make function-level TARA a necessary part of automotive development and certification.
}

\zyz{
While both vehicle-level and component-level TARA are essential for comprehensive risk management, existing datalog-based approaches~\cite{CarVal,Saulaiman} have primarily automated only the vehicle-level assessments, leaving a critical gap in component-level TARA. Vehicle-level TARA identifies overarching threats affecting the entire vehicle but often overlooks the specific implementation details of individual components. In contrast, component-level TARA scrutinizes individual functions or components (e.g., a battery management system or a particular ECU) with fine-grained detail, accounting for hardware/software configurations, communication interfaces, and interactions among sub-components. This granular focus uncovers component-level vulnerabilities and attack paths that broader vehicle-level assessments might miss. However, extending automation to the component level remains challenging due to a key limitation of current methods: their reliance on predefined threat libraries~\cite{CarVal,Saulaiman}. Such libraries, while effective for vehicle-level analysis, lack the granularity and currency needed for component-specific assessments. Indeed, it is impractical to enumerate every potential threat for each component, and maintaining up-to-date libraries is infeasible in an evolving cybersecurity landscape. This raises a critical question: can the TARA process be automated to eliminate reliance on static threat libraries while enabling dynamic and robust component-level assessments?
}
}

\comment{

\zyz{One of the major challenges in the automotive industry is the constantly changing environment in which the product is operating. Advancements in technologies and competencies of adversarial actors (black hat), and focused research (white hat) are all playing a role in the change of the environment in which our products are operating. OEMs have decent change management processes in place and can control the rate of change of their own products, however this is not enough. The interaction and environment in which our products are operating is how risk is generated. This results in a risk assessment / TARA which is already expired before it can be released. This means that the time between your previous risk assessment and your next risk assessment is unknown and depending on your policy and assumptions, this could mean either infinite risk or no risk. The goal of course is to reduce the amount of time between risk assessments to reduce the unknown / uncertainty level to a minimum. Identifying your risk level is a great first step. Now we have to identify time based risk.}

\zyz{However, function level TARA is more challenging in the industry.}

\zyz{Why is function-level TARA is important: (i) Function-level TARA of items and components is required to be conducted. (ii) Enable full-lifecycle security assessment. One of the biggest challenges that industry is currently facing, is not the product, but the constantly changing environment in which the product is operating. Dynamic threat landscape and evolving technologies play important roles in the change of the environment in which our products are operating.}

\zyz{During the development process as well as the entire life cycle phase of all vehicles, OEMs are required to perform Threat Analysis and Risk Assessments (TARAs).}

However, performing TARA with the increasing complexity of modern automotive systems is a laborious and time consuming process. For example, a component such as a battery thermal management system might require 2-5 weeks for a thorough risk assessment~\cite{}, as security analysts must repeat the TARA process extensively, often tens or hundreds of times, to address the various threat scenarios associated with this component. Modern vehicles, moreover, contain numerous such components, further compounding the effort required.
\zyz{Complexity of TARA: (i) Repeat many times for different ECUs, functions, and threat scenarios. (ii) wide attack surfaces and various attack steps.}
}

\comment{
This is challenging due to (i) regulatory complexity, (ii) the dynamic threat landscape, and (iii) expertise dependency. First, regulatory mandates require that TARA meticulously consider individual components of a vehicle and their complex interactions over the whole lifecycle (i.e., the functional-level TARA). Modern vehicles incorporate numerous interconnected components, each requiring detailed analysis. Unlike vehicle-level TARA, which operates on a higher abstraction layer, function-level TARA delves into the details of hardware and software implementations, considering attack entries that are not accessible in vehicle-level TARA. Second, the dynamic nature of automotive technology and evolving threats necessitate continuous updates to TARA, making it difficult to maintain comprehensive and up-to-date threat databases. Third, the reliance on expert knowledge introduces variability and potential inconsistencies in risk assessments. Different analysts may produce varying results for the same threat scenario, leading to discrepancies in risk evaluations. Each function may involve tens or even hundreds of threat scenarios, requiring repeated TARA procedures to produce a comprehensive risk assessment. These challenges highlight the need for more efficient and automated approaches to TARA, which can enhance the consistency and reliability of cybersecurity assessments in the automotive industry.



As a result, automation of TARA process emerges as a critical need to enhance efficiency and reduce the dependency on manual processes. Current research attempts to achieve this by building predefined threat libraries, to build attack trees with datalog approaches~\cite{xxx,xxx,xxx}. However, these automated approaches face two significant challenges. (i) \textbf{Establishing comprehensive threat libraries is challenging.} Building a comprehensive threat library that covers all potential and updated threats of the entire vehicle type and components is a daunting and time-consuming task. Moreover, understanding and summarizing the threats of each component is also impractical due to the vendor specificity and the complexity of the systems. (ii) \textbf{Lack of generalization and customization}: Automated TARA systems based on threat libraries struggle to generalize across different industries, such as unmanned aerial vehicle or ships, and do not easily adapt to regional legislative differences. These challenges underscore the need for developing more flexible and adaptable TARA automation tools that can efficiently handle the complexities of modern automotive systems without relying solely on predefined threat libraries.

Moreover, the lack of generalization and customization in automated TARA systems based on predefined threat libraries makes it difficult to adapt to different industries and regional legislative differences. These challenges underscore the need for developing more flexible and adaptable TARA automation tools that can efficiently handle the complexities of modern automotive systems without relying solely on predefined threat libraries.
}



\comment{
\zyz{Our task: for different components and functional domains, we want to assess the risk level of each threat scenario. To reliably assess risk level, we first analyze attack paths. To analyze attack paths, we use a multi-stage attack path generation framework to generate logical attack paths to guide the subsequent analysis.}

The breakthrough of large language models (LLMs) inspires us to replace the static threat libraries of traditional datalog-based tools with the vast and dynamic knowledge base of LLMs. This insight provides three significant advantages: (i) \textbf{Enabling automated function-level TARA.} LLMs eliminate the need to establish and maintain complex, high-granularity threat libraries. It not only improves the data-log based vehicle-level TARA, but also enables the automation of function-level TARA. (ii) \textbf{Adaptation.} Unlike static libraries, LLMs can dynamically adapt to evolving threat landscapes and accommodate varying evaluation standards and preferences of automotive OEMs and suppliers. (iii) \textbf{Generalization.} By removing the constraints of predefined, automotive-related threat libraries, LLM-based TARA offers the potential for application across diverse industries, extending its utility beyond automotive cybersecurity.
}

\comment{
\textbf{Challenges:}
However, LLMs are not a panacea. Using LLMs to generate attack path trees faces three main challenges:




\comment{
\textbf{\zyz{How to process the complex visual diagrams for LLMs?}} A key issue is how to scientifically describe the increasingly complex IVN for LLMs to use. To address this challenge, we improved an existing system modeling language(SysML), named OpenXSAM++, to describe the complex IVN topology. OpenXSAM++ can systematically represent IVN structures and serve as one of the input data for LLMs. \zyz{1. Why do we only have visual inputs? 2. Why it is difficult for LLMs to process visual diagrams? 3. How to solve this problem?}
}

\textbf{\zyz{How to effectively generate attack trees and assess risks using LLMs?}} It is known that sometimes LLMs provide answers that sound reasonable but are actually incorrect, which can affect the reasonableness of attack paths. Additionally, the answers generated by LLMs have randomness, especially when the IVN structure is complex, affecting the completeness of the attack paths. Therefore, certain rules and constraints need to be established to ensure the reasonableness and completeness of the attack paths. To address this challenge, we designed a fully automated attack path tree generation framework based on constraints and LLMs—\projname{}. This framework fully leverages the advantages of LLMs and achieves attack path generation through the following four layers: (i) System Abstraction Layer: Using OpenXSAM++, different abstraction levels of path planning are performed according to different scenarios (vehicle-level TARA or functional-level TARA), extracting a hierarchical node structure. (ii) Information Splitting Layer: Each node and its related context nodes are mapped with the diverse information in OpenXSAM++, ensuring that LLMs have comprehensive and detailed background information when generating paths, thus ensuring path completeness. (iii) LLMs Vulnerability Discovery and Path Generation Layer: Using the above information, LLMs generate attack paths for each node, combining their strong knowledge reasoning capabilities to proactively discover vulnerabilities and perform accurate differential reasoning for different scenarios. (iv) Filtering and Merging Layer: The attack paths generated by LLMs are screened and filtered to ensure the reasonableness of each path. The filtered paths are finally combined into a complete attack path tree. \zyz{1. Why it is not intuitive to directly summarize attack paths using LLMs? 2. How do we solve this problem?}
}

\comment{
\textbf{\zyz{How to improve the performance and adaptation of LLM-based TARA?}} The main challenge faced during vulnerability discovery and path generation using LLMs is ensuring analysis at an appropriate level of detail. If the analysis is too detailed, it can be misleading and reduce the effectiveness and accuracy of vulnerability discovery. To address this challenge, we extract security use cases and known attack methods from real TARA projects reviewed by authoritative third parties, ensuring the details are neither too granular nor too abstract to meet the requirements of functional-level TARA. Subsequently, we use prompt engineering to facilitate LLMs' learning. This approach not only enhances the practicality of the discovered vulnerabilities but also ensures an appropriate level of detail, thereby improving the accuracy and effectiveness of the analysis. \zyz{1. Why do we need to consider the adaptation of LLM-absed TARA? 2. How to improve the adaptation and performance?}
}

\comment{
\projname{} can infer multi-stage attack paths in increasingly complex IVN and generate logical attack paths to guide subsequent analysis. To demonstrate the practicality of \projname{}, we applied it to five complete vehicles and five real components, generating realistic attack paths. Currently, \projname{} has been applied to more than ten well-known automotive OEMs and suppliers, helping them pass the ISO/SAE21434 system audit by the internationally authoritative third-party testing and certification organization, TÜV Rheinland.
}

We evaluated \projname{} across multiple dimensions to assess its effectiveness, adaptability, and real-world applicability in automating function-level TARA. Deployed across four real automotive security projects, DefenseWeaver successfully identified 11 realistic attack paths, which were validated through penetration testing and subsequently confirmed and patched by the corresponding automakers and suppliers. This demonstrated its practical value in identifying and validating critical attack paths. 
Beyond automotive applications, DefenseWeaver was tested in non-automotive sectors, including unmanned aerial vehicles (UAVs) and marine navigation systems. Its successful deployment in these safety-critical industries highlights the system's adaptability and robustness.


In comparison to human experts, DefenseWeaver consistently outperformed manual attack tree generation across six assessment scenarios, including these four automotive security projects, UAV, and marine navigation systems.
This performance was driven by its ability to avoid common human limitations, including 1) struggled to adapt to new system configurations and overlooked unconventional attacks;
2) inclusion of incorrect elements due to subjective assumptions; 3) overlooking system-specific differences.
As a result, the system excelled in \textit{novelty} (+105.00\%) and \textit{configuration alignment} (+43.68\%), offering a more comprehensive and high-quality risk assessment. Though there was a slight increase in \textit{redundancy} (-$0.90$\%), it reflects the system’s comprehensiveness in mining attack paths.


Integrated into cybersecurity management systems used by leading OEMs and suppliers like United Automotive Electronic Systems (UAES) and Xiaomi, DefenseWeaver has generated over 8,200 attack trees, showcasing its scalability and operational efficiency. Notably, the system has enabled enterprises to reduce processing time, greatly improving operational efficiency and supporting compliance with CSMS certification under WP29 R155 regulations.
In summary, this paper makes the following contributions:
\begin{itemize}
\item  We present \projname{}, the first system to automate function-level TARA using component-specific details and LLMs, significantly enhancing efficiency, accuracy, and scalability while reducing reliance on expert input.



\item We propose OpenXSAM++, a structured, machine-readable format that enables LLMs to effectively interpret detailed automotive configurations.




\item We design a multi-agent LLM framework that automates TARA, enhanced with LoRA and RAG for adaptability and robustness.




\item DefenseWeaver has been validated in four automotive projects and demonstrated its adaptability to UAVs and marine systems. It is also integrated into industry systems used by UAES and Xiaomi.



\end{itemize}

\section{Basics of Automotive Cyber Security}\label{sec:basics}


\subsection{TARA in Automotive Industries}
\begin{figure}[t]
    \centering
    \begin{subfigure}[b]{0.44\linewidth}
        \centering
        \includegraphics[height=3.7cm]{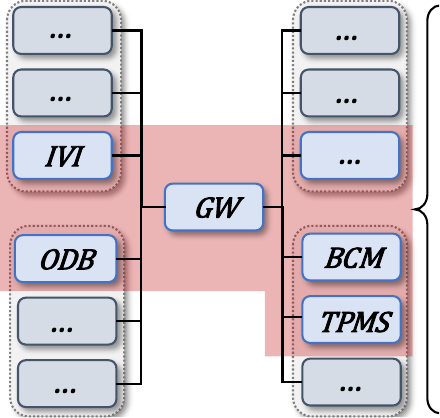}
        \caption{Vehicle-level TARA}
        \label{fig:vehicle_IVN}
    \end{subfigure}
    \hfill
    \begin{subfigure}[b]{0.50\linewidth}
        \centering
        \includegraphics[height=3.7cm]{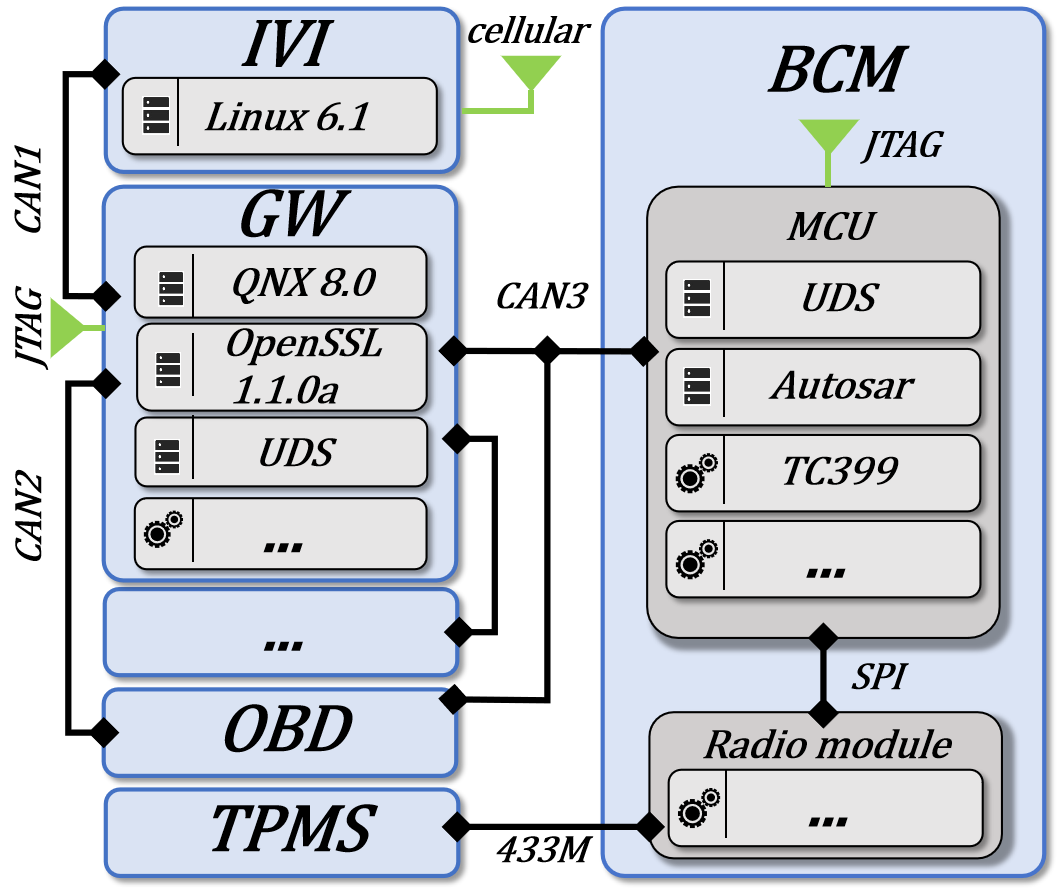}
        \caption{Function-level TARA}
        \label{fig:functional_IVN}
    \end{subfigure}
    
    \caption{Comparison of (a) vehicle-level TARA and (b) function-level TARA. Function-level TARA needs to consider extra dimensions such as hardware configurations (e.g., TC399), software versions (e.g., OpenSSL 1.1.0a), communication channels (e.g., CAN bus), interfaces (e.g., JTAG), internal connections (e.g., radio module and MCU) to comprehensively evaluate potential vulnerabilities.
    }
    \label{fig:functional_IVN vs vehicle_IVN}
\end{figure}

In the automotive industry, TARA should be conducted at different scopes, including vehicle-level TARA and function-level TARA, as shown in Figure~\ref{fig:functional_IVN vs vehicle_IVN}. Function-level TARA addresses cybersecurity threats specific to individual components (e.g., the BCM component) or groups of peripheral components (e.g., IVI, Gateway, OBD, and TPMS) that perform critical functions (Figure~\ref{fig:functional_IVN}), while vehicle-level TARA operates at a higher level and assumes attackers cannot directly access internal elements (Figure~\ref{fig:vehicle_IVN}).
This distinction leads to three main differences: (i) \textit{Function-level TARA can account for attack entries that vehicle-level TARA may overlook.} For example, a JTAG interface connecting to the MCU, as shown in Figure~\ref{fig:functional_IVN}, requires detailed information about the hardware and software of internal elements (e.g., the MCU, radio module, and SPI channel) to evaluate potential vulnerabilities effectively.
(ii) \textit{It provides more specific attack scenarios tailored to different vehicle types, even with similar IVN topologies.} For instance, while logical attack paths like IVI-GW-BCM may remain consistent across different vehicle models, the risk levels of these paths can vary significantly due to differences in OEM implementations (e.g., hardware, software, and suppliers).
(iii) \textit{It enables full-lifecycle TARA with dynamic risk assessment.} Automotive systems often undergo updates, including OTA updates, hardware replacements, and software patches. Function-level TARA offers the flexibility to analyze these changes in detail, ensuring threat analyses and risk databases are updated to reflect the latest system state. This adaptability is essential for maintaining cybersecurity across the vehicle's lifecycle, accounting for evolving vulnerabilities and system configurations.

However, today's TARA activities still rely heavily on human analysts, and this manual approach presents two critical limitations. First, cognitive biases, subjective assumptions, and incomplete attention to component-specific details can produce inconsistent or incomplete results—an issue that becomes more acute at the function level, where far finer-grained information must be considered. Second, because vehicle architectures evolve rapidly, threat analyses and risk databases require continual updates, which demand scarce specialist expertise and ongoing training. Together, these challenges underscore the need for more efficient, automated techniques that can scale function-level TARA and sustain robust cybersecurity throughout the vehicle lifecycle.

\comment{
However, implementing TARA in the automotive industry poses several challenges. First, the process is complex and resource-intensive, requiring significant investment in technology, infrastructure, and expertise to manage cybersecurity risks across the vehicle lifecycle.\yyq{Human experts designing attack trees may also introduce limitations, such as cognitive biases leading to missed unconventional attacks, subjective assumptions resulting in incorrect threat inclusions, or oversights in system-specific nuances. } Second, ensuring interoperability between diverse systems and components while maintaining strong cybersecurity is difficult, especially in heterogeneous environments with varied technologies and standards. Lastly, the dynamic nature of vehicle systems demands continuous updates to threat analyses and risk databases, necessitating specialized skills and ongoing training. These challenges underscore the need for more efficient, automated approaches to streamline TARA processes and ensure security throughout the vehicle lifecycle.
}


\comment{
In response to the increased threat of cyberattacks on connected vehicles~\cite{}, the UNECE enacted the first mandatory cybersecurity regulation, WP29 R155e~\cite{WP29_R155e}, in 2021. This regulation introduces a two-tiered certification system to ensure cybersecurity compliance: the Cyber Security Management System (CSMS) certification for Original Equipment Manufacturers (OEMs) and the Vehicle Type Approval (VTA) certification for individual vehicle types. 

\textbf{CSMS Certification.} 
The CSMS certification is granted to automotive manufacturers and focuses on their organizational processes for managing cybersecurity risks. It ensures that manufacturers have a systematic approach to identifying and mitigating cyber threats throughout the vehicle lifecycle, including development, production, and post-production phases. According to WP29 R155e, by 2024, all automotive manufacturers in UNECE member countries are required to have a certified CSMS in place~\cite{WP29_R155e}. This is necessary to ensure that all new vehicles produced and sold comply with the regulation's cybersecurity requirements.

\textbf{VTA Certification.} 
The VTA certification is granted for individual vehicle types, ensuring that each model meets specific regulatory requirements. It involves technical tests on vehicles and verifies that the vehicle type (i) has been developed following a certified CSMS and (ii) complies with UNECE's cybersecurity provisions. Notably, the approval authorities are responsible for ensuring that the vehicle type possesses capabilities to detect and respond to potential cyberattacks. By 2025, all vehicles imported into UNECE member countries must have VTA certification that demonstrates compliance with the regulation's cybersecurity requirements~\cite{WP29_R155e}.

In summary, CSMS certifies OEMs' organizational capability to manage cybersecurity risks across the vehicle lifecycle, while VTA certifies that individual vehicle models meet regulatory cybersecurity standards for market approval. These certifications impose rigorous responsibilities on OEMs and suppliers to comply with cybersecurity criteria before production and sales. Consequently, conducting comprehensive, lifecycle-spanning cybersecurity assessments is crucial for ensuring automotive security and sustaining the global automotive industry's advancement.
}

\subsection{TARA Pipeline.}
TARA, as outlined in ISO/SAE 21434, systematically identifies cybersecurity threats, evaluates associated risks, and implements countermeasures to enhance vehicle security and ensure regulatory compliance. It begins with \textbf{Item Definition}, where the system’s components and interfaces are modeled, providing a basis for identifying \textbf{Assets}—critical elements such as ECUs, communication interfaces, or sensitive data—evaluated according to confidentiality, integrity, and availability. Potential \textbf{Threat Scenarios} describe how attackers might compromise these assets, followed by \textbf{Attack Path Analysis}, often visualized with attack trees or graphs to represent all possible routes. Next, a \textbf{Feasibility Rating} estimates the effort, expertise, and resources needed for a successful attack, while an \textbf{Impact Rating} quantifies possible consequences (e.g., safety, financial, operational, privacy). These ratings combine to yield a \textbf{Risk Level}, which informs \textbf{Risk Treatment Decisions}—such as avoidance, mitigation, sharing, or acceptance. Since threats evolve over time, TARA must be continuously updated throughout the vehicle lifecycle to ensure risk assessments remain accurate and comprehensive.

\subsection{Mandatory Regulations}
In 2021, the United Nations Economic Commission for Europe (UNECE) introduced WP29 R155e, the first mandatory automotive cybersecurity regulation. It established a two-tier certification system for cybersecurity compliance: the Cyber Security Management System (CSMS) for OEMs and the Vehicle Type Approval (VTA) for individual vehicle types. The CSMS focuses on manufacturers' organizational processes for managing cybersecurity risks across the vehicle lifecycle, mandating that all OEMs in UNECE member countries hold a certified CSMS. Meanwhile, the VTA ensures each vehicle type meets specific regulatory standards through technical tests, verifying that vehicles are developed under a certified CSMS and can detect and respond to cyberattacks. 
These certifications impose rigorous responsibilities on OEMs and suppliers, requiring comprehensive, lifecycle-spanning cybersecurity assessments to uphold industry growth and regulatory standards.


\comment{
Threat Analysis and Risk Assessment (TARA) is a systematic methodology outlined in the ISO/SAE 21434 standard for identifying, analyzing, and mitigating cybersecurity risks. Its primary objective is to generate attack trees for each threat scenario, offering a detailed understanding of potential attack paths. The TARA process involves several key steps: vehicle modeling, asset identification, threat scenario identification, attack path analysis, attack feasibility rating, impact rating, risk value determination, and risk treatment.

Among these steps, attack path analysis and attack feasibility rating are the most challenging and resource-intensive. Security analysts must carefully examine complex vehicle configurations to identify potential threats and evaluate attack paths. This task requires deep expertise in system modeling and cyberattacks. Furthermore, each component or function of a vehicle type requires performing tens or even hundreds of analyses to ensure comprehensive risk coverage, further intensifying the workload for security analysts.

TARA is not a one-time activity but a continuous process that evolves throughout the vehicle's lifecycle. Updates are necessary when integrating new components, implementing software upgrades, addressing new vulnerabilities, or adapting to evolving threat landscapes. This dynamic nature significantly increases the complexity of implementing TARA. Moreover, while primarily applied to automotive systems, TARA principles are also relevant to other transportation domains, such as aircraft and ships.
}

\comment{
(i) \textit{Function-level TARA is capable of considering attack entries which is not accessible in vehicle-level TARA.} For example, there is a JTAG interface connecting to MCU as shown in Figure~\ref{fig:functional_IVN}, evaluating the potential vulnerabilities requiring detailed information about hardware and software of the internal elements (e.g, the MCU, radio module, and SPI channel).
(ii) \textit{Function-level TARA can reveal new attack paths and scenarios than vehicle-level TARA.} For instance, the Gateway component in Figure~\ref{fig:functional_IVN} also contains a JTAG interface, which may introduce additional vulnerabilities and attack vectors which are not initially launched from IVI and ODB modules.
(iii) \textit{Having the detailed information of internal hardware and software is essential to provide more specific attack steps.} For example, the version of OpenSSL 1.1.0a in the Gateway component can be affected by the CVE-2016-6309 vulnerability~\cite{}, which can be mitigated by upgrading to OpenSSL 3.3.
These detailed assessments are crucial to ensuring the robustness of each component and function. Additionally, regulations such as WP29 R155e mandate that TARA be conducted not only at the vehicle-level but also at the function-level to achieve CSMS and VTA certifications. \zyz{Attack paths (e.g., IVI-GW-BCM) without considering detailed configuration may apply a broad range of vehicles, since this attack path may share the same topology across different vehicle types. However, the risk level of this attack path may vary significantly due to the different implementation of OEMs (e.g., the hardware, software, and suppliers).} 

However, implementing TARA in the industry heavily relies on security analysts. These professionals manually analyze attack paths, assess vulnerabilities, and generate risk evaluations for each component or function. For example, each function may involve tens or even hundreds of threat scenarios, requiring repeated TARA procedures to produce a comprehensive risk assessment. This reliance on manual analysis is both labor-intensive and time-consuming, significantly increasing the cost and effort required for compliance. Moreover, the effectiveness of TARA depends on the expertise of analysts, which introduces variability into the process. Teams with different levels of experience may produce inconsistent results when assessing the same threat scenario, leading to discrepancies in risk evaluations. This reliance on human expertise is compounded by the dynamic nature of the automotive threat landscape. Vehicle components are frequently updated through Over-The-Air (OTA) updates, hardware upgrades, or software patches, requiring continuous updates to threat analyses and associated databases. In summary, current industry practices place a heavy burden on security analysts. The challenges of manual, expertise-driven processes underline the need for more efficient approaches to automating TARA, which will bring significant benefits to the automotive industry and enhance the security of connected vehicles.
}

\comment{
Challenge 1: low-level of automation.

Challenge 2: It's hard to build and maintain a comprehensive threat database.

\zyz{Tens or hundreds of threat scenarios for each component or automotive system, which means repeat for tens or hundreds of times TARA procedures for a TARA report. Moreover, the manufacturers usually need to do component-level or domain-level TARA analysis.}

\zyz{
Rapidly evolving vehicle technology and new attacks require continuous updates to domain knowledge.
}

\paragraph{Functional-Level TARA.} \zyz{define what is component-level tara and why it is difficult.}
\yyq{  According to the UN R155, vehicle original equipment manufacturers (OEMs) must obtain a CSMS compliance certificate and a VTA certificate, and demonstrate effective management of the interaction between suppliers and OEMs in cybersecurity activities. This means that not only is the whole vehicle-level TARA part of CSMS compliance, but the components are also an important part. Suppliers need to conduct TARA on the components they provide  that constitute specific functions to ensure their security. Functional-level TARA refers to the threat analysis and risk assessment conducted conducted on components within a vehicle that perform specific functions, such as a single component or a group of components.
}

\yyq{Unlike vehicle-level TARA, which operates at the level of the entire vehicle and assumes that the internal parts are not directly accessible, functional-level TARA assumes that attackers can reach the components. Therefore, it needs to consider the more detailed hardware and software implementation of the components, as well as information interactions with other related components. For example, functional-level TARA needs to consider whether the hardware interfaces of the components (such as JTAG) can be exploited by attackers to extract firmware and leak information, which vehicle-level TARA does not need to consider, as attackers will not have access to the components themselves. Because of this, the threat scenarios are more diverse than those of vehicle-level TARA. Typically,there can be dozens or even hundreds of threat scenarios for a specific function, which means that the analysis of attack paths for the same function can be repeated dozens or even hundreds of times. CSMS requires that, in addition to vehicle-level TARA, TARA reports must be provided for dozens of specific functions in a vehicle, which can be very time-consuming and labor-intensive.}

In the automotive sector, companies face two challenges when conducting TARA, including low-level of automation and the difficulty of building and maintaining comprehensive threat database.

The first is low-level automation. Although there have been some tools available to assist analysts in conducting TARA, such as OCTAVE\cite{}, EVITA\cite{EVITA} and HEAVENS\cite{HEAVENS}\cite{HEAVENS2}. In practical utilization, for instance, during attack path analysis, analysts still need to manually input information such as potential vulnerabilities in components. This heavily relies on the knowledge and experience of experts, resulting in significant time expenditure for analysts and low efficiency. Additionally, the lack of objective definitions and criteria in ISO 21434\cite{ISO/SAE_21434} result in subjective and inconsistent of TARA results. For example, the attack step of "using a CVE vulnerability in the firmware for privilege escalation" may be perceived as simple by a team that has firmware scanning tools, while a team without such tools may consider it challenging.This discrepancy implies that the two groups would adopt completely different approaches to address this attack and results in different TARA result.

 Recently some researcher have tried to automated TARA with an datalog-based approach\cite{MulVal2}\cite{MulVal3}\cite{CarVal} and improved the efficiency of TARA successfully. However, the threat libraries provided by ISO 21434 are copying existing threats from other areas, result in the lack of automotive-specific threats \zdy{such as relay attack on vehicle Passive Keyless Entry and Start (PKES) system \cite{}.Such incomplete threat libraries are insufficient to support automated TARA.} Meanwhile, as vehicle systems become increasingly complex, \zdy{TARA is not only limited to the macro level of components but also needs to consider the software and hardware details that components possess. For example, if a component uses OpenSSL, it is necessary to consider vulnerabilities such as the CVE-2016-6309 bug in the OpenSSL software.} Therefore, building a comprehensive threat library presents a significant challenge. Moreover, vehicle components are often updated throughout the vehicle's lifecycle. In the early stages of \zdy{product design phase}, vehicle components are often uncertain, and as the design matures, some components may be replaced. Even after the vehicle is sold, new software and hardware upgrades may be introduced through Over-The-Air (OTA) updates. Component updates can lead to changes in the associated threats, requiring the risk library to be continuously updated. This significantly increases the cost of maintaining the risk library.
}

\comment{
\begin{itemize}
    \item \textbf{Asset identification:} this segment identifies assets, including their cyber security properties (contains confidentiality, integrality and availability), which will lead damage scenarios if be attacked. 
    \item \textbf{Threat scenario identification:} this segment describes potential threats and vulnerabilities, including the target asset, the property threats violates and the damage scenario it will lead. For example, a user's search history on In-Vehicle Infotainment (IVI) is an asset. its confidentiality can be broken if it's under attacking, which will lead the leak of personal information. It's worth to note that a threat scenario could correspond several damage scenarios.
    \item \textbf{Impact rating:} this segment assesses the damage scenarios on four aspects:  safety, financial, operational and privacy. 
    \item \textbf{Attack path analysis:} this segment finds a potential attack path or an attack tree if there are more than one path. There are top-down approach and bottom-up approach. \zdy{This stage is time-consuming and has the potential for automation. Implementing automation could substantially improve the efficiency of TARA.}
    \item \textbf{Attack feasibility rating:} this segment assesses the feasibility of an attack path via attacking potential, CVSS or attack vector.
    \item \textbf{Risk value determination:} this segment combines impact rating and attack feasibility rating to assess the risk of a thread scenario, it values from 1 (lowest risk) to 5 (highest risk). 
    \item \textbf{Risk treatment decision:} this segment considers some methods addressing the risks (such as accept or mitigation). 
\end{itemize}
}




\comment{
Recent research~\cite{jing2023revisiting} has developed a macro-level automated TARA tool for vehicles using a datalog-based approach with an expanded threat database. \zyz{Figure?} However, this tool faces significant limitations: (i) It struggles to provide deepen component-specific analysis due to the challenges in collecting and maintaining a comprehensive micro-level threat database that adapts to the vehicle development cycle. (ii) The reliance on a threat database derived mainly from the automotive industry experience hampers its applicability to other transportation systems. Consequently, there is a critical need to develop an automated TARA tool that is independent of a threat database and adaptable across different transportation systems.
}

\comment{
\subsection{Attack Tree}

\yyq{copy from \textit{Survey: Automatic generation of attack trees and attack graphs}, Particularly in TARA, attack trees are highly suitable for analyzing attack paths for different threat scenarios, then describe the attack tree}

\zdy{Attack Tree, one of the most predominant graphical model, represents potential security threats to a system through a tree structure. Originally introduced by [x], this representation allows developers to systematically identify system vulnerabilities and supports the development of appropriate countermeasures.}

\zdy{With the comprehensible and intuitive nature, Attack Tree is considerable well-suited for the attack path analysis in TARA. They represent threat scenarios in a hierarchical structure, where each labeled node corresponds to a sub-goal of the attacker, with the root node representing the primary threat scenario. The remaining labeled nodes may function as either children of a node, representing a refinement of the parent goal into subsidiary objectives, or as leaf nodes, which signify basic actions that cannot be further decomposed. An Attack Tree is a 3-tuple$ \left ( N,\rightarrow, n_{0}   \right )$, where $N$ is a finite set of nodes, $\rightarrow$ is a finite acyclic relation of type $\rightarrow \subseteq N \times M(N)$, where $M(N)$ is the multi-set of $N$, and $n_{0}$ is the root node, such that every node in $N$ is reachable from $n_{0}$.}


\zdy{The fundamental formal model of Attack Trees incorporates two primary types of refinements: $OR$ and $AND$. $OR$ nodes indicate a disjunction (choice), where the parent node's goal is realized if at least one of the child nodes' sub-goals is achieved. Conversely, $AND$ nodes indicate conjunction (aggregation), necessitating that all child nodes' sub-goals be fulfilled. Various variants of Attack Trees have been introduced, such as the sequential conjunction refinement, $SAND$, which resembles the $AND$ node but imposes a requirement for the child nodes’ sub-goals to be achieved in a specific sequential order.Figure.1 illustrates an example of Attack Tree. There are two distinct pathways for an attacker to achieve his goal: either with authentication or without. In the scenario involving authentication, both of the refined options, ssh and rsa, must be accomplished as they are linked by an $AND$ operator. Conversely, if authentication is not required, the attacker must first obtain user privileges by completing the ftp and rsh actions, and subsequently achieve lobf. The tree structure is highly advantageous and effective for threat analysis, as it accommodates multiple attack vectors stemming from physical, technical, and human vulnerabilities.}

\begin{figure}
    \centering
    \includegraphics[width=1\linewidth]{figure/4.png}
    \caption{Attack Tree from}
    \label{fig:enter-label}
\end{figure}
}




%

\section{Related Work}
 
\textbf{Automating Tools for TARA:}
Traditional TARA tools, such as SAHARA~\cite{SAHARA}, EVITA~\cite{EVITA}, HEAVENS~\cite{HEAVENS}\cite{HEAVENS2}, and TVRA~\cite{TVRA}\cite{TVRA2}, provide systematic frameworks for identifying and assessing threats in automotive systems, relying on methodologies like attack trees\cite{schneier1999attack} or STRIDE. However, these tools are not automated, depending heavily on manual effort and expert knowledge—an increasing challenge in modern complex systems with numerous components and potential attack paths.
By contrast, datalog-based tools like MulVal~\cite{MulVal}\cite{MulVal2}\cite{MulVal3}\cite{MulVal4} automate parts of TARA by using the logic programming language to encode vulnerabilities, threats, and reasoning rules into a library, which then generates possible attack paths. Building on MulVal, Saulaiman et al.\cite{Saulaiman} and CarVal\cite{CarVal} tailored it for the automotive domain, with CarVal incorporating expert interviews to manually establish a more comprehensive threat database.

However, these datalog-based solutions remain unsuitable for function-level TARA, which demands detailed, component-specific assessments and faces high system variability. Maintaining granular, dynamic threat libraries is labor-intensive and difficult to scale. Moreover, TARA applies to a range of other systems—such as aircraft, ships, extended reality (XR), and Space Information Networks (SIN)—where constructing reusable, cross-domain threat databases remains a major obstacle. 

\textbf{Attack Tree Generation:} Attack trees are the core artifact produced during TARA.  By iteratively decomposing high-level threats into concrete attack steps, they provide both a systematic analysis framework and an intuitive medium for communicating risk among engineers and regulators~\cite{schneier1999attack,AT-forvehicle-noauto}.  Several studies have sought to automate their construction in the automotive domain~\cite{AT-M-Karray-forvehicle,AT-M-Kern-for-vehicle,AT-M-Chlup-forvehicle}, but similarly, they are rule-based and only work at the vehicle-level.



\comment{
\textbf{Automating Tools for TARA:}
Traditional TARA tools, such as SAHARA~\cite{SAHARA}, EVITA~\cite{EVITA}, HEAVENS~\cite{HEAVENS}\cite{HEAVENS2}, and TVRA~\cite{TVRA}\cite{TVRA2}, were developed to provide systematic frameworks for identifying and assessing threats in automotive systems. These methods primarily address challenges related to structuring the TARA process and ensuring comprehensive coverage of threats and risks. For instance, these approaches rely on methodologies like attack trees or STRIDE to analyze vulnerabilities and potential attack paths systematically. However, these tools are not automated—they still depend heavily on manual processes and the expertise of security analysts to design and analyze attack paths. This dependency becomes increasingly challenging when dealing with modern complex automotive systems, where the number of components, interactions, and potential attack paths can grow significantly. 

In contrast, datalog-based tools like MulVal~\cite{MulVal}\cite{MulVal2}\cite{MulVal3}\cite{MulVal4} aim to automate portions of the TARA process. MulVal is an open-source framework that uses the logic programming language Datalog to model and generate attack graphs and trees. Vulnerabilities, threats, attack techniques, and reasoning rules are encoded into a library, which the tool uses to automatically generate potential attack paths. Inspired by MulVal, Saulaiman et al.\cite{Saulaiman} and CarVal\cite{CarVal} adopted the framework for the automotive domain. These adaptations customized and modified MulVal to automatically generate attack paths tailored to vehicle-level systems. For example, CarVal incorporated insights from interviews with automotive cybersecurity experts, manually establishing an improved threat database to enable more comprehensive TARA.

However, datalog-based tools can only extend threat libraries for vehicle-level TARA by abstracting components into functional roles and enumerating potential threats at the system level (e.g., CarVal~\cite{CarVal}), leveraging standardized patterns and general knowledge of vehicle architectures. For function-level TARA, this approach is impractical due to the need for detailed assessments of specific hardware and software components, high variability across systems, and the increased complexity of attack surfaces. Maintaining the highly granular and dynamic threat libraries required for function-level TARA is labor-intensive and unsustainable. Additionally, TARA is not exclusive to the automotive industry but is widely applied in other domains such as extended aircraft systems, ships, reality (XR) systems, and Space Information Networks (SIN). Even though these domains share some commonalities in electronic systems, building comprehensive and reusable threat libraries across domains remains a challenge. 
}



\comment{
Many previous studies have proposed various TARA methods, SAHARA\cite{}, EVITA\cite{}, HEAVENS\cite{}, TVRA\cite{} among others\cite{}, aimed at addressing inherent challenges in earlier approaches. However, they fell short in automating the TARA process. With the requirement of manually devise the attack path, these methods are time-consuming and labor-intensive in modern complex automotive systems, at the same time, they are inefficient and heavily relying on the experience of analysts. MulVal\cite{}\cite{} is a datalog-based open-source framework for modeling and generating attack graphs and trees, which uses the logic programming language datalog as the modeling language, where vulnerabilities, threats, attack techniques and reasoning rules are manually entered into a library in the type of datalog. The capability of MulVal on automatic attack path generation inspired Saulaiman et al\cite{}. and CarVal\cite{}, for the purpose of automatic generation of attack path in automotive domain, they customized and modified MulVal to make it applicable to the automotive domain to automatically generate attack path. In addition, CarVal further conducting extensive interviews with automotive cybersecurity experts, then manually established an improved threat database with the data collected from the interviews, which allowed them to conduct a more reasonable TARA. 

Although Saulaiman et al.\cite{} and CarVal\cite{} automatic generate attack path based on threat libraries to conduct efficient TARA, however, it is time-consuming and labor-intensive that establishing comprehensive threat libraries and upgrading libraries whenever some of components are changed or upgraded. Therefore, both of them do not fundamentally address the challenge of being time-consuming and labor-intensive. We utilize LLMs to automatically generate attack path, with vast open knowledge data which covers the knowledge required for TARA, as a result, we achieve automated attack path generation without threat libraries, thereby circumventing the time-consuming and labor-intensive tasks of creating and updating such a library and demonstrating the generalized potential for threat analysis in other domains.
}

\comment{
\textbf{Generalized Threat Analysis Approaches in Other Domains:}
Since threat analysis methods can identify potential threats and assess risks, they play a crucial role in various domains by effectively enhancing system security through the implementation of mitigation strategies for the identified threats and risks. Qamar et al.\cite{} focused on extended reality (XR) systems, particularly VR, AR, MR, and the metaverse, and provided a detailed description of the various cybersecurity threats faced in this domain, such as privacy inference, vulnerabilities, and data breaches. They proposed a taxonomy that they believe aids in visualizing the diversity of XR threats in research and summarized potential defensive measures and recommendations to address these threats. Wu et al.\cite{} concentrated on  Space Information Network (SIN) domain and conducted a threat analysis based on cybersecurity attributes (i.e. confidentiality, integrity, and availability), which is more systematic compared to those threat analysis based on layers or functions. Shen et al.\cite{} analyzed vulnerabilities and security issues in Open Radio Access Network (O-RAN) interfaces, designed a threat model to address vulnerabilities, identify threats, and analyze threat descriptions, threat actors, threat assets, and affected components. The aforementioned work relies on manual processes and does not explicitly propose the possibility of automation. Attack tree and attack graph are often predominant method when conducting threat analysis. Multi-host Multi-stage Vulnerability Analysis (MulVal)\cite{}\cite{} was not designed specifically for TARA, in fact, it was initially employed for the assessment of enterprise network security. Ou et al.\cite{}\cite{}\cite{} conducted threat analysis in enterprise networks and achieve automatic generation of attack graphs. Although they can automatically generate attack graphs within MulVal, they do not consider automating the generation of attack graphs for TARA in the vehicle domain.
}

\textbf{Success of Large Language Models (LLMs):}
LLMs have profoundly advanced natural language processing and machine learning, sparking transformative changes across diverse fields. Since the introduction of the transformer architecture~\cite{Transformer} in 2017, models like BERT~\cite{BERT} (2018) and GPT-4~\cite{GPT-4} (2023) have demonstrated remarkable capabilities, owing to billions or even trillions of parameters and training on massive, varied datasets. They excel at generalizing across tasks and adapting to new challenges. In cybersecurity, LLMs have proven effective for vulnerability detection~\cite{LLM_for_vulnerability_detection}\cite{LLM_for_vulnerability_detection2}\cite{LLM_for_vulnerability_detection3}, code fuzzing~\cite{LLM_for_code_fuzz}\cite{LLM_for_code_fuzz2}, phishing detection~\cite{LLM_for_phishing_detection}\cite{LLM_for_phishing_detection2}\cite{LLM_for_phishing_detection3}\cite{LLM_for_phishing_detection4}, and content moderation~\cite{LLM_for_content_moderation}, leveraging fine-tuning or prompting to tailor solutions. Inspired by these advances, we ask: Can we replace static threat libraries in traditional TARA tools with the vast knowledge base of LLMs and thereby automate the TARA process?

\comment{
\noindent\textbf{Success of Large Language Models (LLMs):}
The development of LLMs has significantly advanced natural language processing and machine learning applications, driving transformative changes across various domains. Since the introduction of the transformer architecture~\cite{Transformer} in 2017, models such as BERT~\cite{BERT} (2018) and GPT-4~\cite{GPT-4} (2023) have demonstrated remarkable capabilities in addressing complex problems. These models, built with billions or even trillions of parameters and trained on massive, diverse datasets, are highly effective at generalizing across tasks and adapting to new challenges. In cybersecurity, LLMs have been successfully applied to critical tasks, including vulnerability detection~\cite{LLM_for_vulnerability_detection}\cite{LLM_for_vulnerability_detection2}\cite{LLM_for_vulnerability_detection3}, code fuzzing~\cite{LLM_for_code_fuzz}\cite{LLM_for_code_fuzz2}, phishing detection~\cite{LLM_for_phishing_detection}\cite{LLM_for_phishing_detection2}\cite{LLM_for_phishing_detection3}\cite{LLM_for_phishing_detection4}, and content moderation~\cite{LLM_for_content_moderation}. Their success lies in their ability to generalize across diverse tasks and tailor their responses to specific problems through fine-tuning or prompting. Inspired by these advancements, we pose the following question: Can we replace the threat libraries of traditional TARA tools by leveraging the vast knowledge base of LLMs and automate the TARA process?
}

\comment{
\zyz{The basics of LLMs. Three key words, "large", "language", "neural network models"} As soon as transformer architecture was presented in 2017, Large Language Models (LLMs) start the first step to change the world. From BERT (2018) to GPT-4 (2023), LLMs have achieved an amazing progress. Nowadays, LLMs are offering various new applications in science and technology domains, providing transformational impact. 

\zyz{The application of LLMs.}
In general, LLMs are based on a transformer architecture with billions or even trillions of parameters and trained on quantity datasets so that they are easy to generalize. Besides, the pretraining of LLMs is unsupervised, removing the burden of labeling large datasets. Therefore, for a pretrained LLM, one can only use prompting (extra training data is not necessary) or fine-tuning (only need a few extra training data) to complete a downstream task. Fine-tuning is prohibited in some advanced LLMs, it's appropriate for these LLMs by using prompting. As a rule, the downstream tasks are described in natural language directly in prompting. Consequently, the respond of a downstream task is influence by prompting. The quality of prompting determines the quality of respond from LLMs.  
LLMs have been taken seriously in the domain of cyber security, more and more researchers are utilizing LLMs in some important and hard problems in this domain. Such as: (i) vulnerability detection, (ii) code fuzzing, (iii) toxic content detection, (iiii) content moderation etc. These works have also inspired our work to explore the possibility in automating TARA via LLMs.

\zyz{How does our work differs from previous applications.}

}

\section{\projname{}: Approach}\label{sec:System_Design}
In this section, we present the design of \projname{}, an LLM-based tool for function-level TARA automation.
\subsection{System Overview}
\begin{figure*}
    \centering
    \includegraphics[width=0.9\linewidth]{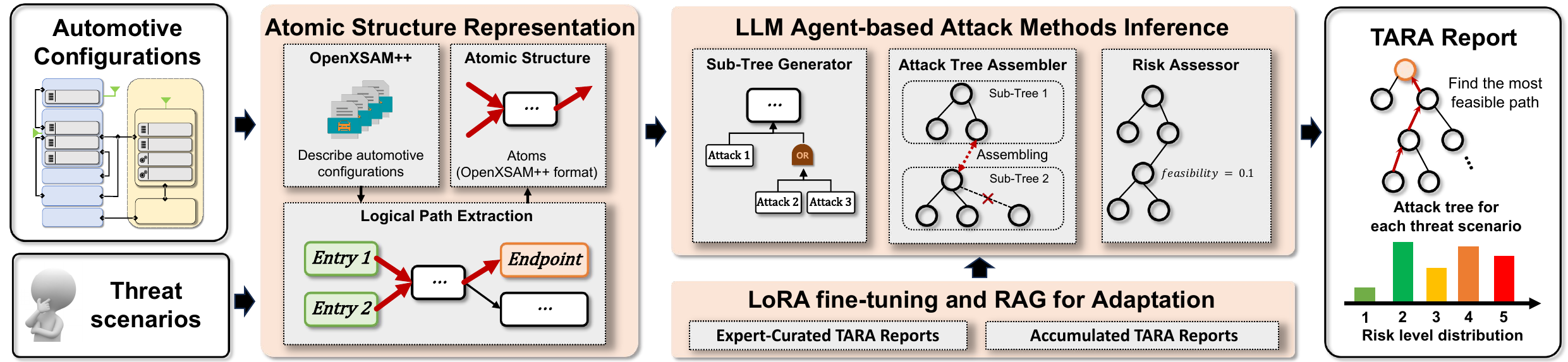}
    \caption{Framework of \projname{}. 
    Given a vehicle configuration and threat scenarios, \projname{} will automatically convert the visual diagrams into proper representations (atomic structures), generate specific attack methods for each node (sub-trees) before assembling them into attack trees, and evaluate the risk level (from 1 to 5) according to the most feasible attack path for each threat scenario.
    }
    \label{fig:overview}
\end{figure*}



\projname{} automates function-level TARA by leveraging component-specific details and the capabilities of LLMs. Unlike approaches relying on static threat libraries, \projname{} dynamically infers attack methods and evaluates risk levels using detailed component-specific information. This scalable and adaptive system overcomes the limitations of manual processes and static libraries, enhancing efficiency, accuracy, and responsiveness to evolving cybersecurity challenges. Its architecture consists of five key components, each contributing to its overall functionality and adaptability.


\noindent\textbf{Automotive Configurations and Threat Scenarios (Input):} Automotive configurations detail component attributes, including hardware setups, software versions, communication channels, interfaces, and sub-component interactions. Threat scenarios specify attack objectives, the endpoint, and entry points, providing essential context for precise function-level TARA and ensuring comprehensive vehicle configuration analysis.


\noindent\textbf{Atomic Structure Representation:} This component decomposes complex configurations into manageable units, thereby improving TARA efficiency and accuracy. Based on the structured OpenXSAM++ format, it constructs a directed graph to identify logical paths within the IVN and segments these paths into atomic structures. Each atomic structure retains essential topological and functional information, facilitating subsequent LLM-based analysis.


\noindent\textbf{LLM Agent-Based Attack Methods Inference:} \projname{} employs a multi-agent framework to dynamically infer attack methods, assigning specialized roles to LLMs such as Sub-Tree Constructor, Attack Tree Assembler, and Risk Assessor. For example, the Assembler links sub-trees and may request the Constructor to regenerate methods if inconsistencies arise.


\noindent\textbf{Fine-tuning and RAG for Adaptation:} To adapt to dynamic threats and diverse standards, \projname{} integrates  Low-Rank Adaptation (LoRA) fine-tuning and Retrieval-Augmented Generation (RAG), which  learns from expert-curated TARA reports and relevant contextual information to refine analysis and tailor it to specific requirements. This integration ensures adaptability across regions and organizational standards by leveraging real-time and domain-specific knowledge.


\noindent\textbf{TARA Report (Output):} The output is a function-level TARA report that consolidates identified attack methods, risk levels, and analysis results into an actionable document. It provides detailed insights into vulnerabilities, attack paths, and recommended mitigations, serving as a critical tool for automotive OEMs and suppliers to ensure regulatory compliance and maintain robust cybersecurity throughout the lifecycle.

Compared to datalog-based approaches, \projname{} is able to: (i) automatically identify various attack paths/methods even with identical logical paths by considering component-specific details, (ii) discover new attack surfaces, thereby adapting to evolving threat landscapes and enabling full-lifecycle assessment, and (iii) easily deploy the pipeline to other electronic systems (e.g., UAVs and ships, etc.) for cross-domain applicability and peripheral devices of vehicles (e.g., cloud services and smartphones, etc.).



\comment{
LLMs offer distinct advantages for automating function-level TARA, eliminating the need for static threat libraries by dynamically inferring potential attack methods through pre-trained knowledge. These capabilities address the limitations of traditional datalog-based approaches by significantly reducing the manual effort required to build and maintain threat libraries. However, LLMs often have difficulty interpreting and summarizing complex visual diagrams, such as those used for automotive system models, because these models can contain numerous intricate connections and hidden information (for example, potential attack methods for certain components).

As illustrated in Fig.~\ref{fig:overview}, once a vehicle configuration and threat scenarios are defined (including potential threats and their the respective attack endpoint and several entry points), \projname{} automatically generates TARA reports that include both attack trees and risk levels (ranging from 1 to 5) for each scenario. For scenarios with higher risk levels, detailed attack trees—especially the most feasible attack path—are crucial for automotive manufacturers and component suppliers. These results guide penetration testing and help prioritize security measures by allocating resources more effectively. To accomplish this, \projname{} encompasses the following modules (see Fig.~\ref{fig:overview}): (i) Atomic structure representation for vehicle configurations in Sec.~\ref{sec:xsam} (for \textit{Challenge 1}). We first introduce the OpenXSam++ structured format to fully capture vehicle configurations. Next, we construct a directed graph to identify the logical paths—representing attacks from one component to the next without detailing specific methods—from various entry points to the designated attack endpoint. We then create atomic structures for each node, store them in the OpenXSAM++ format, and use these atoms as input for further LLM-based analysis. (ii) LLM-based attack methods inference in Sec.~\ref{sec:attack_path} (for \textit{Challenge 2}). To improve both the efficiency and performance of LLM-based inference, we adopt a multi-agent technique that assigns different specialized roles to the LLMs, including a sub-tree \emph{Constructor}, an attack tree \emph{Assembler}, and a risk \emph{Assessor}. These agents collaborate to enhance the accuracy and consistency of the generated attack methods. For instance, the \emph{Assembler} agent is responsible for linking two sub-trees and may request the constructor agent to regenerate certain methods if they appear logically inconsistent. (iii) Retrieve-Augmented Generation (RAG) for system adaptation in Sec.~\ref{sec:prompt} (for \textit{Challenge 3}). Finally, to adapt to various evaluation standards and user preferences for TARA generation, we can integrate two datasets—one containing expert-reviewed TARA reports, and another containing accumulated TARA reports during daily use—both stored in the OpenXSAM++ format. These datasets provide extra context to guide the multi-agent LLMs via the RAG approach.
}

\comment{

The framework consists of three main components: (1) OpenXSAM++ data format for system modeling, (2) attack path extraction and segmentation, and (3) LLM-based attack tree generation. The OpenXSAM++ data format is used to describe the system modeling diagram, including components, channels, and interfaces. The attack path extraction and segmentation component extracts logical attack paths from the system modeling diagram and segments them into smaller, more manageable pieces. Finally, the LLM-based attack tree generation component uses the extracted attack paths to generate comprehensive attack trees for specific threat scenarios. The framework is designed to be flexible and adaptable, allowing users to easily update the system modeling diagram and re-generate attack trees as needed.

By harnessing the generalization and adaptability of LLMs, we propose a novel framework for TARA that not only automates the process but also demonstrates broader applicability to other domains requiring systematic threat analysis.

The rapid development of the IVN has brought increasingly complex information security challenges. Traditional automated attack path analysis methods based on threat libraries, due to their limitations (\yyq{Sec 2.3}), have found it difficult to cope with the growing complexity, especially in the functional-level TARA . These methods often fail to deeply analyze the attack paths of components, thus failing to meet the needs of full Life Cycle Assessment(LCA). \projname{} aims to rely not on threat libraries, but on the vast cross-domain knowledge and analytical capabilities of LLMs, to achieve a comprehensive analysis and assessment of attack paths at the functional-level of IVN systems. The core advantage of \projname{} lies in its cross-domain analytical capabilities and coverage throughout the entire LCA.

When using \projname{}, users only need to draw the system modeling diagram of the current project, select potential attack entry and exit points, and define a specific threat scenario. Subsequently, \projname{} will automatically generate an attack tree composed of multiple possible attack paths and assess the feasibility of each path. As the system continues to evolve throughout its lifecycle, components may undergo updates and iterations, such as updates to software versions or changes in hardware selection. \projname{} can flexibly respond to these changes. Users only need to make corresponding modifications to the system modeling diagram, and \projname{} can automatically re-analyze the attack paths according to these modifications, achieving LCA.

Figure \ref{fig:overview} shows the workflow of \projname{}. In order for LLMs to understand the system modeling diagram, we designed the OpenXSAM++ data format to describe system modeling, and through attack path extraction and information segmentation, we extract highly relevant information as the basis for the input of LLMs (Sec.~\ref{sec:xsam}). To reduce LLMs from generating unreasonable resultss and efficiently generate attack paths, we designed three steps for constraints, including LLMs generating attack steps, constraint and merge, and LLMs attack feasibility rating (Sec.~\ref{sec:attack_path}). 
\yyq{To further enhance the logical reasoning capabilities of LLMs and meet the differentiated needs of users, we utilize Retrieval-Augmented Generation (RAG) to extract external corpora from authoritative third-party reviewed TARA reports and the accumulated daily use by users, providing references for LLMs. (Sec.~\ref{sec:prompt}).
We detail each system component hereafter.}
}

\begin{figure}
    \centering
    \includegraphics[width=0.7\linewidth]{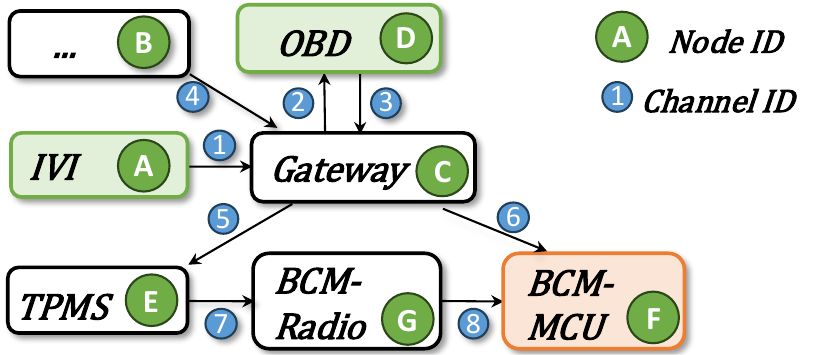}
    \caption{A simplified IVN topology with one unique attack endpoint (BCM-MCU) and two entrypoints (IVI and OBD) according to given threat scenarios (e.g., disrupt the availability of BCM-MCU). The nodes are connected with channels.
    }
    \label{fig:IVN}
\end{figure}

\subsection{Atomic Structure Representation}\label{sec:xsam}
In this section, we discuss how to efficiently describe vehicle configurations using atomic representations.

\subsubsection{Comprehensive Description of Configurations}\label{sec:OpenXSAM++}
Vehicle configurations, as illustrated in Fig.\ref{fig:IVN}, are often represented by visual diagrams that depict various components and the connections among them. These diagrams capture key design details but are difficult to interpret automatically, which is a significant challenge for LLM-based parsing (see Sec.\ref{sec:attack_path}) and dataset construction (see Sec.~\ref{sec:prompt}).

To overcome this issue, we convert these visual diagrams into a structured format called OpenXSAM++, an extension of OpenXSAM~\cite{openxsam} (\textbf{\textit{Open}} \textbf{\textit{X}}ml \textbf{\textit{S}}ecure \textbf{\textit{A}}nalysis \textbf{\textit{M}}odel). OpenXSAM is an XML-based framework designed for information exchange in automotive cybersecurity and risk management. It uses standardized, machine-readable documentation to describe assets, threats, risks, and mitigation measures. However, its original specifications do not include several essential elements and attributes needed to cover automotive configurations thoroughly.

To fulfill this gap, we add a \texttt{Software} attribute that specifies each component's operating system, software bill of materials, or active network services. We also introduce a \texttt{Hardware} attribute to outline hardware modules, chips and debugging capabilities. Furthermore, we incorporate additional elements, such as \texttt{Channel} and \texttt{Interface}, to capture the breadth of automotive components and their interconnections. While these enhancements focus on vehicle systems, they are also applicable to other electronic or electrical systems. 
\textbf{Therefore, \projname{} relies on OpenXSAM++ format for configuration representation and database construction.}

\begin{figure}[t]
    \centering
    \begin{subfigure}[b]{0.17\textwidth}
        \centering
        \includegraphics[height=3.5cm]{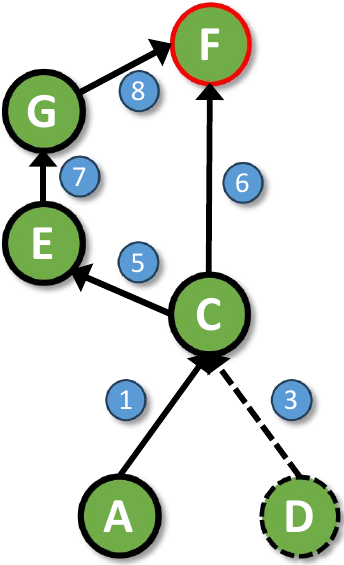}
        \caption{Logical path}
        \label{fig:Logical Attack Path}
    \end{subfigure}
    \hfill
    \begin{subfigure}[b]{0.3\textwidth}
        \centering
        \includegraphics[height=3.5cm]{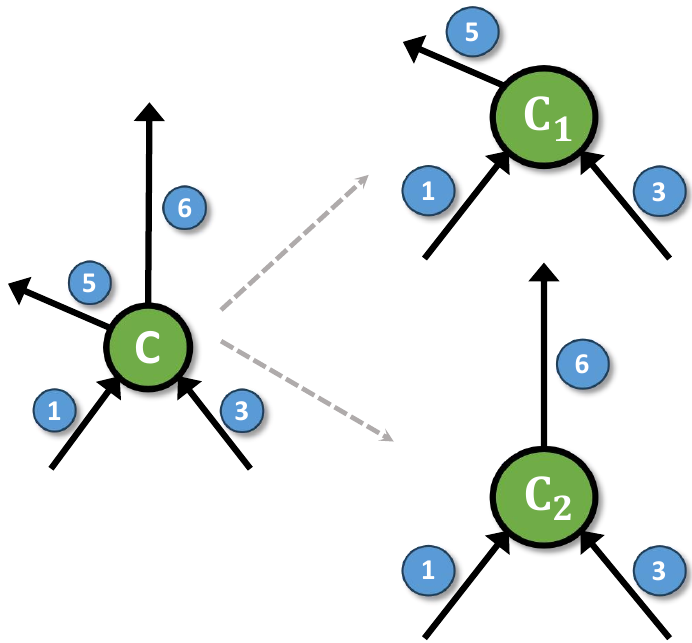}
        \caption{Atom construction ($C_1$ and $C_1$)}
        \label{fig:Segmentation}
    \end{subfigure}
    \caption{(a) Logical paths without detailed attack methods in each node, where irrelevant and redundant components (e.g., $B$) are removed. (b) The segmented node ($C$) and constructed atomic structures ($C_1$ and $C_2$) are derived based on the exit channels (e.g, channel $5$ and $6$) for sub-tree generation.}    
    \label{fig:Logical path}
\end{figure}

\subsubsection{Logical Path Extraction and Atom Construction}\label{sec:logical_path}
Vehicle configurations, with their many interconnected components, often lead to very long OpenXSAM++ descriptions that capture comprehensive details. These extensive descriptions may overwhelm LLMs when generating attack trees, as they introduce substantial irrelevant or distracting information~\cite{wu2024easilyirrelevantinputsskew}. Therefore, and as illustrated in Fig.~\ref{fig:Logical path}, \textbf{we refine the OpenXSAM++ description before fed into LLMs based on the threat scenarios in two main steps}: (i) logical path extraction to remove redundant or irrelevant components, and (ii) atom construction to split the description into several minimal units (atoms) while preserving topology.


\textbf{Logical Path Extraction.} 
Because each threat scenario defines a unique attack endpoint (e.g., BCM-MCU ($F$)) as well as several possible entry points (e.g., IVI ($A$) and OBD ($D$)), our goal is to identify every path from each entrypoint to the attack endpoint (for example, $A \rightarrow C \rightarrow F$ in Fig.~\ref{fig:Logical Attack Path}). By doing so, we discard any irrelevant or redundant elements (e.g., component $B$ and channels $2$ and $4$) and thus focus on components and channels that genuinely matter to the threat scenarios. Note that these logical paths do not include detailed attack methods per node; such methods are the core of TARA analysis and are inferred by LLMs in later steps.


To systematically generate these paths, we build a directed graph from the OpenXSAM++ description and apply a depth-first search (DFS) to find all acyclic routes connecting each entrypoint to the attack endpoint. We remove cyclic paths (for example, $D$$\rightarrow$$C$$\rightarrow$$D$), as revisiting a previously compromised component provides no further insights for practical TARA. If there are multiple possible entry points, we simply find each path independently and merge any shared segments.


\textbf{Atom Construction.}
Directly supplying all nodes and channels from the extracted logical paths to LLMs may still degrade analysis quality by overloading the model with too much information and complex interactions. To mitigate this, we break each logical path into smaller, more manageable structures (i.e., atoms), while preserving the relationships among nodes. Specifically, an atom is defined as a single node plus all its directly connected edges (i.e., channels). The channel attributes store references to any other connected nodes. For instance, in Fig.~\ref{fig:Segmentation}, node $C$ has two incoming channels and two outgoing channels. Parsing node $C$ with these channels allows the LLM to infer the previous and next nodes from the channel attributes. Since $C$ links to node $F$ through channel $6$ and to node $E$ through channel $5$, we split it into two atoms, $C_1$ and $C_2$. This ensures each atom has exactly one local attack objective (e.g., propagating the attack to $F$ via channel $6$, or to $E$ via channel $5$), thereby simplifying the subsequent inference.

\comment{
Despite using the method of extracting the logical attack path to separate high-quality information, inputting all the node and edge element information from the entire logical attack path at once often exceeds the token limit of LLMs. \zyz{and result in reduced performance} Due to this limitation, we analyze each node in the path individually, so we need to divide the highly relevant information corresponding to each node in the path.

We found that the edges (i.e., channels) of a component in the logical attack path record the external logical relationships of that component. Therefore, when splitting the information, each segment needs to include both the information of the current component and the channel information directly connected to it, ensuring the relevance between the segments. Specifically, the logical attack path in Figure 3 can be divided into four segments: [F node ID + Channel5ID], [C node ID + Channel1,3,5ID], [A node ID + Channel1ID], and [D node ID + Channel3ID]. For each segment, we can search the corresponding description information in OpenXSAM++ based on its component or channel ID and directly concatenate it.

\zyz{Define what is the minimum element.}
We found that element information that is far from a specific node tends to be of lower quality, which may cause LLMs to be less focused during analysis, thus overlooking highly relevant information. However, if we only have the information of the specific node itself, it cannot reflect the relationship between nodes. Therefore, our approach is to take the information of the specific node itself and the information of the edges directly connected to it as its highly relevant information. Based on this understanding, we can follow these steps to find the highly relevant information for each node in the tree:


We associate the corresponding highly relevant information with each node in the label tree according to the above steps, to form the basis for the prompt words, allowing LLMs to analyze each node.
}

\subsection{LLM-based Attack Methods Inference}\label{sec:attack_path} 

Inferring potential attack methods for each node is central to TARA analysis, accounting for the majority of the required time and effort. In this section, we explain how to leverage LLMs to automate the inference of attack methods.

\subsubsection{Multi-Agent Roles for Automated TARA}\label{sec:multi-agents}
Component-specific details—such as software versions, hardware configurations, communication channels, and interfaces—significantly increase the complexity of attack method inference because the number of potential attacks can grow exponentially. Additionally, generating attack trees must factor in the interactions between components and channels while assessing the feasibility of each attack method. Consequently, it is not feasible to generate complete attack trees using simple Q\&A approaches.

Recent advances in multi-agent systems have greatly enhanced the capabilities of LLMs in handling complex tasks, including software development \cite{Multi-Agent_software-development}\cite{Multi-Agent_software-development2}\cite{Multi-Agent_code_generation}\cite{Multi-Agent_code_generation2}, game simulation \cite{Multi-Agent_game-simulation}, scene simulation \cite{Multi-Agent_scene-simulation}, and multi-robot systems \cite{Multi-Agent_multi-robot-system}. The key idea of multi-agent systems is to \textbf{mimic human teamwork by dividing a complex task into several simpler subtasks, each handled by an agent with specialized skills}. Through collaboration, these specialized agents reduce the likelihood of errors, especially in tasks with high complexity. Drawing on this approach, we divide TARA analysis into three subtasks, each handled by a separate LLM-based agent. Their roles are as follows, while detailed processes are described in Sec.\ref{sec:attack tree generation} and Sec.\ref{sec:attack tree evaluation}:

\noindent\textbf{$\bullet$ Sub-Tree Constructor:} 
Generate sub-trees for each atom. Each sub-tree focuses on a single local attack objective for a specific node and includes comprehensive attack methods with corresponding logical relationships (e.g., AND/OR) among the summarized attack methods to achieve the objective.

\noindent\textbf{$\bullet$ Attack-Tree Assembler:} Merges the sub-trees produced by the \emph{Constructor} into a complete attack tree for each threat scenario. The \emph{Assembler} also collaborates with the \emph{Constructor} iteratively to improve the coherence of connected sub-trees (which are generated independently from different atoms) and to remove leaf nodes that violate user-defined constraints.

\noindent\textbf{$\bullet$ Risk Assessor:} After constructing complete attack trees for each threat scenario, the \emph{Assessor} first evaluates the feasibility of each attack method (i.e., step feasibility). It then assesses the cumulative feasibility of the entire tree (i.e., focusing on the most feasible attack path) and the potential impact of the threat scenario. Based on these assessments, the \emph{Assessor} determines the risk level for each scenario according to the ISO/SAE 21434 standard.

In practical functional-level TARA processes, there may be tens or even hundreds of threat scenarios, necessitating repeated application of the above steps for each scenario. Ultimately, the TARA reports provide a comprehensive risk distribution that summarizes the risk levels of all potential threat scenarios for the analyzed target.

\subsubsection{Attack Tree Generation}\label{sec:attack tree generation}

\begin{figure}[t]
    \centering
    \includegraphics[width=0.8\linewidth]{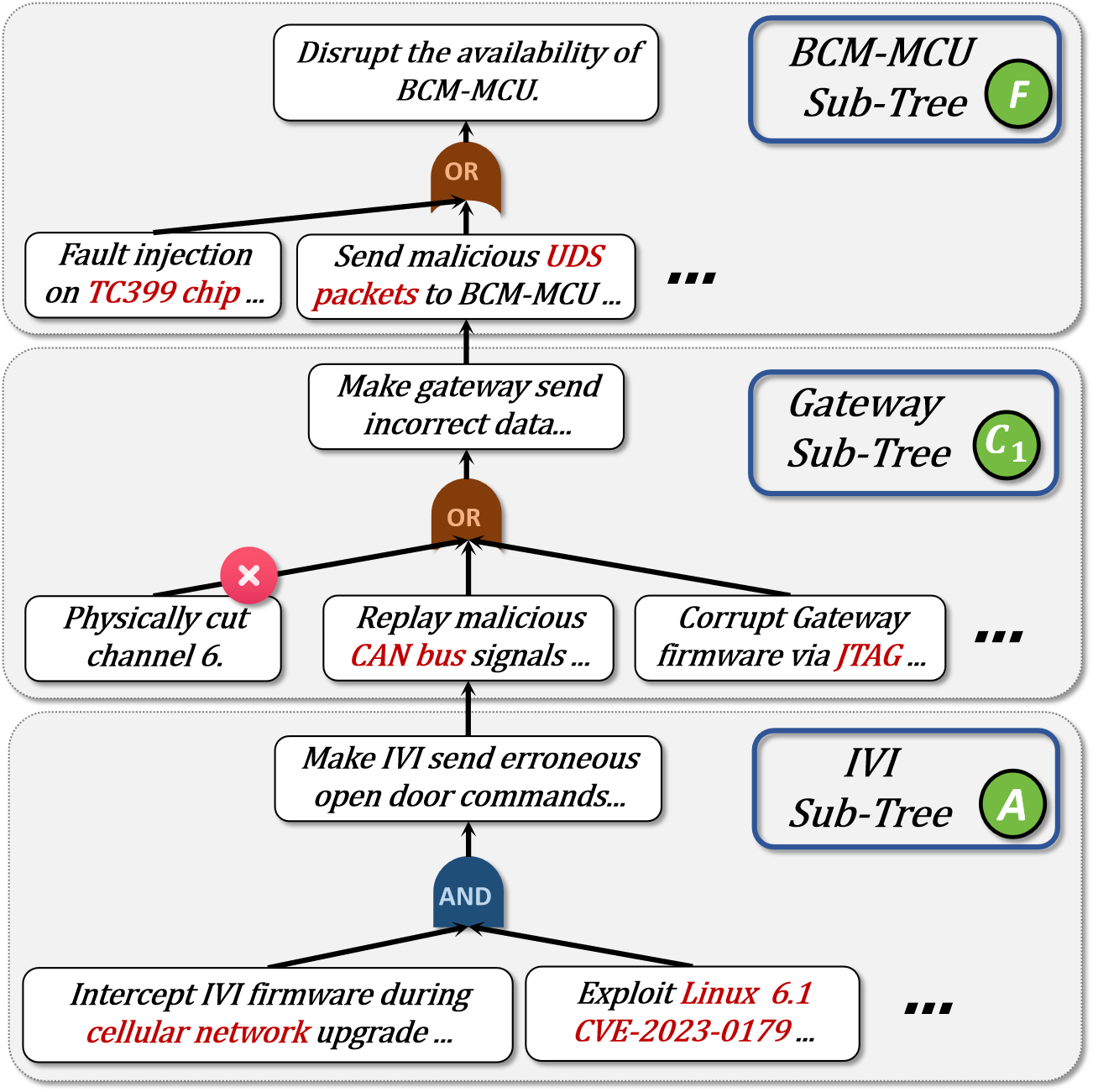}
    \caption{Simplified attack tree assembled from sub-trees for IVI ($A$), Gateway ($C_1$), and BCM-MCU ($F$). The realizations of attack methods for each node are derived based on component-specific details, which hardly be included by predefined static libraries.
    }
    \label{fig:attack_tree}
\end{figure}

Attack tree generation aims to construct comprehensive and coherent attack trees for given threat scenarios. Note that, we need to generate an attack sub-tree for each atom before assembling them into a complete attack tree. Fig.~\ref{fig:attack_tree} provides an example of an attack tree that consists of sub-trees for nodes $A$, $C_1$, and $F$, under the threat scenario ``Disrupt the availability of BCM-MCU''. The attack methods for each node are derived based on component-specific details, which are hardly included in predefined static libraries.

\paragraph{Sub-Tree Construction} 
To generate sub-trees, the \emph{Constructor} needs to summarize specific attack methods for the input atoms, relating them to the threat scenario and node attributes. However, some attributes of the node may only have simple descriptions; for example, the software attribute of the IVI might be as simple as “Linux 6.1”, which may cause LLMs to overlook some attributes, resulting in incomplete analysis. Additionally, understanding the logical connections among the generated attack methods for each node is crucial for improving the readability and quality of the entire attack tree, but it is often challenging for LLMs to consider so many details simultaneously.

To make the attack methods comprehensive and tightly connected to the threat scenario, \textbf{we adopt the concept of Chain-of-Thought (CoT) to guide the construction of sub-trees} through the following steps: (1) attack surface inference, (2) threat scenario analysis and local attack objective understanding, and (3) attack sub-tree generation. First, inferring the potential attack surfaces of the given node ensures comprehensive identification of the vulnerabilities of each component. Second, the \emph{Constructor} combines the threat scenario (e.g., ``Disrupt the availability of BCM-MCU'') and attack surfaces to infer the local attack objective (e.g., ``Make gateway send incorrect data...'' for the gateway node and ``Make IVI send erroneous lighting commands...'' for the IVI node). After that, the \emph{Constructor} formulates a series of specific attack methods closely related to the local objective, combining key information from the node, the attack surface, and its extensive cross-domain knowledge. Note that each attack method contains only one operation to better support feasibility rating in Sec.~\ref{sec:attack tree evaluation}. In other words, there might be several attack methods required to launch an attack against one attack surface.

As a result, the logical relationships among attack methods are also important to demonstrate the practical attack path. If several attack methods rely on the ``Linux 6.1'' attribute and must be executed in sequence to launch an attack, the process might involve obtaining the Linux system firmware from the IVI, reverse-engineering the firmware to identify vulnerabilities, and exploiting a known Linux vulnerability (e.g., CVE-2023-0179) to gain control of the IVI.
These sequential attack methods should be connected with a logical AND. Conversely, if completing any one of the attack methods is sufficient to achieve the attack objective, they can be connected with a logical OR. Therefore, based on the attack objectives, the \emph{Constructor} further analyzes the logical relationships between the attack methods and connects them together with logical nodes (e.g., AND and OR) to form the final attack sub-tree. For example, as shown in Fig.~\ref{fig:attack_tree}, if the attack objective is ``Make the gateway send incorrect data to the BCM-MCU,'' and accomplishing either one is sufficient to achieve the objective (i.e., Accessing the gateway via JTAG to corrupt the BCM-MCU firmware or Replaying malicious CAN bus signals on channel 6 to the BCM-MCU), they can be connected with an ``OR'' node.

\paragraph{Attack Tree Assembling} 
The \emph{Assembler} is responsible for assembling the attack sub-trees generated by the \emph{Constructor} into a complete attack tree. Therefore, the \emph{Assembler} needs to determine how to connect two sub-trees with multiple attack methods. The key observation is that \textbf{only attack methods related to the channels can propagate the attack outcomes of one component to another}, which can be easily distinguished from the attack surfaces. For example, an attacker can exploit the IVI to replay malicious CAN bus signals over Channel 1, conducting a Denial-of-Service (DoS) attack on the gateway. Consequently, the \emph{Assembler} connects the sub-tree generated for the IVI to the corresponding attack method related to the communication channel in the sub-tree of the gateway (e.g., ``Replay malicious CAN bus signals on channel 6 to BCM-MCU.'').

In addition to simply connecting sub-trees, \textbf{another important task of the \emph{Assembler} is to validate the quality of the \emph{Constructor}'s generation}. For example, since these sub-trees are generated independently, the attack methods related to the channels (shared by two nodes) may lack coherence due to the lack of a global perspective of both nodes. In such cases, the \emph{Assembler} can request the \emph{Constructor} to improve the coherence of these attack methods by providing essential information about the two sub-trees (e.g., the local attack objective of the previous node and the attack method of the input channel of the next node). Furthermore, the \emph{Assembler} can be easily customized to adapt to users' different requirements by providing explicitly defined constraints for attack methods, such as excluding social engineering attacks (e.g., stealing key fobs or passwords), physical destruction of components (e.g., damaging signal transceivers, cutting hardware, or chip replacement), and physical attacks (e.g., side-channel analysis, fault injection, or chip decapping), etc. Therefore, the attack method ``Physically cut channel 6'' will be removed.

The main difference between the \emph{Constructor} and the \emph{Assembler} is that the \emph{Constructor} focuses on generation and inference, while the \emph{Assembler} focuses on validation. Therefore, the \emph{Assembler} can double-check the generated sub-trees and request the \emph{Constructor} to regenerate some of them if necessary.

\subsubsection{Risk Assessment of Threat Scenarios}\label{sec:attack tree evaluation}
In an automotive system—or even within a single component—there can be tens or hundreds of threat scenarios. Therefore, it is required to build attack trees independently for each threat scenario in a TARA report.
The next challenge is \textbf{how to assess the risk level of each threat scenario to provide an overall analysis result to security analysts and prioritize countermeasures}. Since it is usually impossible to address all threats simultaneously and manual analysis of all attack trees requires a huge amount of human effort (as currently done in industry TARA analysis), efficient risk assessment is crucial.

\begin{figure}
    \centering
    \includegraphics[width=0.7\linewidth]{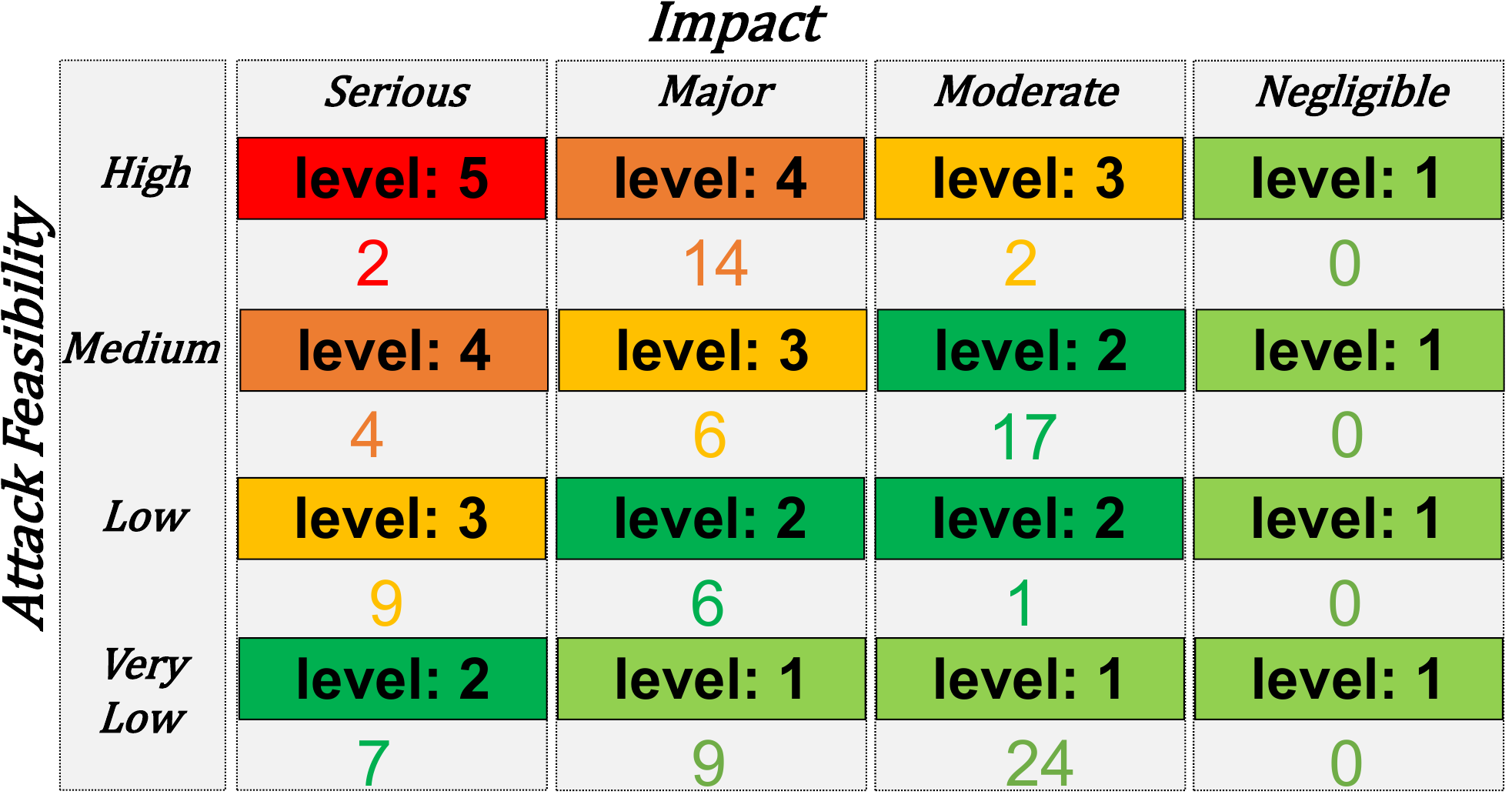}
    \caption{Risk distribution for more than $100$ threat scenarios of the IVN configuration shown in Fig.~\ref{fig:IVN}. For each threat scenario, the attack tree and risk evaluation will be generated independently.}
    \label{fig:risk_distribution}
\end{figure}

As suggested by the ISO/SAE 21434 standard, the risk level of threat scenarios can be divided into two parts: overall attack feasibility and potential impact of the threat. After determining the scores of these two factors (i.e., high, medium, low, and very low for attack feasibility, and severe, major, moderate, and negligible for potential impact), the risk levels (from 1 to 5) can be summarized in a risk distribution table. An example is shown in Fig.~\ref{fig:risk_distribution}. For example, according to the standard, a threat with \textit{High} Feasibility and \textit{Serious} Impact is assigned the highest risk level 5 and decrement the risk level as feasibility and impact decrease. Notably, threats with a risk level greater than risk level 3 (i.e., \textit{Moderate} Impact and \textit{High} Attack Feasibility; \textit{Major} Impact and \textit{Medium} Attack Feasibility; \textit{Serious} Impact and \textit{Low} Attack Feasibility) have practical inspection value and need to be verified in subsequent penetration testing. \textbf{The \emph{Risk Assessor} determines the attack feasibility and potential impact of each threat.}

\paragraph{Attack Feasibility Assessing}
It is often unreliable to directly analyze the entire attack tree to determine the feasibility of a threat scenario. Therefore, our high-level strategy is to analyze the step feasibility of each individual attack method before evaluating the cumulative feasibility score. Furthermore, for each attack method, we score it across the following five dimensions according to the ISO/SAE 21434 standard: elapsed time (ET), specialist expertise (SE), knowledge of the item or component (KoIC), window of opportunity (WoO), and equipment (Eq). For each dimension, a lower score means higher feasibility. To compute the step feasibility for each node, the \emph{Risk Assessor} is first required to choose a score from a given range for each dimension, before providing a brief explanation to ensure reasonable scores. Note that the scoring standards may vary across different regions, companies, and even products, which can be customized by users as discussed in Sec.~\ref{sec:prompt}.

The attack feasibility of a specific threat scenario is determined by the most feasible attack path (i.e., the one with the lowest overall cumulative score), although there are usually multiple attack paths in an attack tree from leaf nodes to the root node. Moreover, the cumulative feasibility of one attack path is determined by the hardest attack method (i.e., the one with the highest score). Therefore, the principles for computing the cumulative feasibility (divided into five dimensions) are as follows: (1) For sequential nodes, the cumulative scores of the current node are the maximum of the cumulative scores of its child node and its own step-feasibility scores; (2) For logical AND nodes, the cumulative scores are the highest scores among their child nodes; (3) For logical OR nodes, the cumulative scores are assigned the scores of the child node with the lowest overall score.
To summarize, the cumulative feasibility is computed in a bottom-up approach, and the cumulative feasibility of each first leaf node equals its step-feasibility score. Therefore, the cumulative scores at the root node represent the scores of the most feasible attack path, and the sum of the cumulative scores is the overall attack feasibility of the threat scenario.

\paragraph{Potential Impact Assessing} Assessing attack feasibility is more challenging since it involves technical details of the system (i.e., it is product-specific). In contrast, assessing potential impact only requires a high-level understanding of threat scenarios and system usage. In practice, impact assessment is conducted by evaluating potential consequences across four dimensions, including Safety, Financial, Operational, and Privacy. Since impact assessment is not product-specific, \projname{} scores each of the four dimensions separately, before calculating the potential impact based on the ISO 21434 standard.


\subsection{Fine-Tuning and RAG for Adaptation}\label{sec:prompt}
\textbf{To cope with evolving threat landscapes and diverse standards, \projname{} adopts a differentiated adaptation strategy}: using Retrieval-Augmented Generation (RAG) for the Risk Assessor agent and Low-Rank Adaptation (LoRA) fine-tuning for the Sub-Tree Constructor agent. The Assessor (which generates step-feasibility scores and impact scores) benefits from real-time retrieval of authoritative references and up-to-date enterprise guidelines. In contrast, the Constructor (which builds attack sub-trees) is fine-tuned via LoRA to internalize expert attack logic. We found that applying RAG to the Constructor may introduce irrelevant or incompatible examples (e.g., pulling QNX-specific nodes into a Linux analysis), thereby degrading generation quality. LoRA fine-tuning avoids this by embedding the correct patterns in the model’s parameters instead of relying on potentially noisy external retrieval.

Specifically, we leverage the OpenXSAM++ format (Section~\ref{sec:OpenXSAM++}) with three critical fields added to each analyzed node: ``Sub-Tree'' (the expert-annotated attack subtree),  ``Step-Feasibility'' and ``Impact'' (the expert-assessed scores). These fields support our two adaptation processes. For the Constructor, we construct a supervised training set using each node’s system attributes (e.g., hardware configurations, software versions) as the input and the corresponding expert ``Sub-Tree'' as the output. We then apply LoRA fine-tuning to the base LLM using this dataset, teaching the model to generate attack sub-trees in line with expert logic. This fine-tuning updates only a small portion of model weights (preserving over 95\% of the original parameters), thus retaining the model’s general language capabilities while infusing domain-specific patterns. We also apply regularization (dropout 0.3) and early stopping (halting training if validation F1 stagnates for 3 epochs) to prevent overfitting. For the Assessor, we implement RAG-based prompt augmentation by retrieving similar prior attack methods and their scores (i.e., feasibility and impact) from our knowledge base. An embedding-based similarity search~\cite{bge-m3,text2vec} finds the most relevant historical cases (from both expert and enterprise data), and the top matches are inserted into the Assessor’s input prompt~\cite{liu2023pre}. This gives the LLM concrete reference points for scoring, ensuring its assessments are grounded in authoritative examples 
and can dynamically adapt to the latest enterprise context via real-time retrieval.

Our adaptation strategy is grounded in \textbf{two primary databases} with domain knowledge, which feed the LoRA training and RAG retrieval components: (i) Expert-Curated TARA Reports: A corpus of 116 vetted automotive threat scenarios (from an industry reference library) provides over 1,000 standardized attack sub-trees and impact scores, and about 5,000 step-feasibility entries. The sub-trees serve as high-quality training targets for LoRA fine-tuning, while the score ratings populate the Assessor’s RAG reference library. (ii) Enterprise-Specific TARA Reports: \projname{} also ingests feedback from enterprise-specific assessments collected via a GUI (Section ~\ref{sec:Implementation}). When users adjust an attack tree or scores in practice, the corrected sub-tree is added as incremental training data to further refine the LoRA-based Constructor, and the updated scoring data is immediately incorporated into the RAG retrieval library.

\comment{
\subsection{LoRA fine-tuning and RAG for Adaptation
}\label{sec:prompt}
Different enterprises often have distinct evaluation standards and preferences for TARA. These variations stem from differences in technological maturity, operational requirements, and system-level focuses. For example, automotive control software suppliers may conduct particularly detailed analyses at the software level, while other stakeholders may emphasize hardware-level risks. Furthermore, LLMs can sometimes produce inaccuracies, such as incorrect logical relationships (e.g., misidentifying AND/OR nodes) or inconsistencies in feasibility ratings due to insufficient contextual information or lack of enterprise-specific expertise. 

\yyq{To address these challenges, \projname{} adopts a differentiated strategy: it incorporates Retrieval-Augmented Generation (RAG) for the \textit{Assessor} agent while employing Low-Rank Adaptation (LoRA) fine-tuning for the \textit{Constructor} agent. This design is based on the following key findings:  \textit{Assessor} agent (step feasibility ratings) requires referencing authoritative standards and dynamically adapting to the latest enterprise requirements, whereas experiments in \textit{Constructor} agent (sub-tree constructor) revealed that RAG may retrieve examples that appear similar but are incompatible, potentially introducing irrelevant or incorrect data (e.g., mistakenly incorporating QNX-specific attack nodes in a Linux system analysis), thereby degrading generation quality. In contrast, LoRA fine-tuning more effectively avoids such interference by relying not on real-time retrieval of external examples but on internal model parameters to solidify correct generation patterns. }

\yyq{Specifically: Leveraging the OpenXSAM++ format described in Sec.\ref{sec:OpenXSAM++}, we extend attributes like `Sub-Tree' and `Step-Feasibility' to each well-analyzed node in the reference corpus. These extensions enable us to use the `Sub-Tree' as the training set for LoRA fine-tuning and the `Step-Feasibility' as the database for RAG：
(i) \textbf{For the \textit{Constructor} agent }, \projname{} uses system attributes (e.g., hardware configurations, software versions) from OpenXSAM++ as input $Prompt$, domain-expert-annotated standard `Sub-Tree' from OpenXSAM++ as $Respond$ to construct supervised learning training set. The LoRA fine-tuning technique enables the model to learn expert attack logic while preserving over $95$\% of the original parameters to maintain generalization capability against novel threats. To prevent overfitting, a dropout rate of $0.3$ and early stopping (training terminates if the validation F1-score does not improve for three consecutive epochs) are applied.  
(ii) \textbf{For the \textit{Assessor} agent}, \projname{} first retrieves the most similar `Attack Methods' and their `Step-Feasibility' from expert knowledge bases and enterprise historical cases using embedding models ~\cite{bge-m3,text2vec}. The retrieved scoring examples are then integrated into the prompt ~\cite{liu2023pre} to provide the LLM with authoritative references. This design ensures both the professionalism of the scoring and the dynamic adaptability of evaluation results to the latest enterprise requirements through real-time retrieval mechanisms.  }

\yyq{The reference corpus consists of two primary sources: (i) \textbf{Expert-Curated TARA Reports: }These reports include validated attack trees and  step feasibility ratings matrices for 116 real-world threat scenarios. The attack tree portion (containing 1000+ standardized attack subtrees) is used for LoRA fine-tuning, while the  step feasibility ratings portion (covering 5000+ attack methods) supports RAG retrieval.  (ii) \textbf{Enterprise-Specific TARA Reports}: Custom enterprise libraries collect user correction records via a graphical user interface (GUI) (Sec.~\ref{sec:Implementation}). Adjusted attack trees serve as incremental training data to continuously optimize the LoRA model, while modified scoring matrices are updated in real time to the RAG retrieval library.  }
}

\comment{
While conducting in-depth research on the application of LLMs in attack path analysis, we encountered two primary issues. First, LLMs sometimes make incorrect logical node selections. For instance, the model may mistakenly connect two attack methods using an "AND" relationship when they should be connected by an "OR" relationship. This can be attributed to the fact that, although LLMs have some logical reasoning capabilities, they are fundamentally language models rather than specialized reasoning tools. Consequently, errors in handling complex logical reasoning are to be expected.

Secondly, our survey of five companies across different industries revealed that, due to varying levels of technological accumulation, each company has its own distinct criteria to assessing the feasibility of attack methods.

To address these issues of incorrect logical relationships and inconsistent feasibility evaluations, we introduced the Retrieval-Augmented Generation (RAG) model.

\subsubsection{Retrieval-Augmented Generation}\label{sec:RAG}
The root cause of the logical and evaluation issues lies in the fact that LLMs lack sufficient reference information, leading to inadequate reasoning capabilities, which in turn causes errors and inappropriateness in logical node selection and evaluation results. To address this problem, we introduced the RAG method. RAG enhances the reasoning ability of LLMs by incorporating external reference corpora, improving their performance in logical decision-making and evaluations.

Specifically, we incorporated more reference corpora into the prompts used by the large models, as mentioned in Section 4.3. The core idea is to take attribute values such as hardware, software, and interfaces from the system description information as keywords. 

By employing the BM25 algorithm with these keywords, we can find similar system descriptions in an external corpus.These keywords are then used to retrieve relevant content from external corpora using the BM25 algorithm. Then, the analysis results of these similar system descriptions are inputted into the LLMs as a reference to assist in the analysis.

It is important to emphasize that the goal of RAG is not to make the large models learn new knowledge directly but to provide appropriate reference information during their reasoning process. Additionally, RAG can accommodate different enterprises' technical levels and needs to provide customized feasibility evaluations based on their requirements.

\subsubsection{Where does the reference corpus come from?}

To enable RAG to provide reference information for LLMs, our corpus comes from two sources: S1 (expert corpus) provides references for logical and scoring reasoning, while S2 (enterprise-specific corpus) addresses the diverse requirements of different enterprises.

\textbf{S1}:The first corpus is derived from 116 real-world threat scenarios, including 1,114 attack paths and 4,942 attack methods, along with corresponding feasibility rating matrices designed by human experts. The logical relationships of all steps were constructed by experts to ensure accuracy. This corpus primarily serves as a reference for LLMs during logical node reasoning and the generation of feasibility rating matrices

\textbf{S2}: is composed of corpus data accumulated by enterprise users through their daily use of \projname{}. We provide an online web application that allows enterprise users to modify automatically generated attack trees, including feasibility rating matrices for each attack method (detailed in section 4). With the users' consent, we collect manually processed feasibility matrices on a monthly basis, creating a retrievable database that reflects the enterprise's unique scoring standards, serving as a reference for LLMs.

\subsubsection{How to build a searchable reference corpus ?}

Our core idea is that similar system descriptions should result in similar logical relationships and feasibility evaluation matrices between the attack methods generated by the LLM. Therefore, these similar logical relationships and evaluation matrices can serve as reference corpora for LLM reasoning.

When constructing the reference corpus, we use system descriptions as an index, with the indexed content being their corresponding logical relationships or feasibility evaluation matrices. This allows us to find several logical relationships or evaluation matrices that can be used as references by identifying highly similar system descriptions.

Specifically, (i) for S1, the system description information serving as the index can be generated by inputting the system modeling, attack entry and exit points, and threat scenarios from each report into \projname{}. The logical relationships, which are the indexed content, are represented by attack subtrees manually extracted from attack trees. (ii) For S2, the acquisition of the index is the same as for Source 1, and the feasibility assessment matrices, which are the indexed content, are the feasibility assessment matrices modified according to the company's differentiation.

Additionally, we have preprocessed the system description information to make it more suitable for retrieval, improving both retrieval efficiency and accuracy. Specific operations include tokenizing the attribute values in the system descriptions, extracting key feature words, and building an inverted index to facilitate efficient retrieval and matching of the corresponding logical relationships and feasibility evaluation matrices, which serve as reference corpora for LLM reasoning.
}

\section{Evaluation}
In this section, we apply \projname{} to real-world scenarios and compare its attack trees with those crafted by human experts. We demonstrate that \projname{} can identify practical attack paths (validated via penetration tests) and produce higher-quality attack trees.

\comment{
In this section, we demonstrate how \projname{} is applied to the TARA creation of real vehicles, and guide the identification of actual threats during penetration testing, where we have discovered $11$ practical attack paths that have been reported and fixed with the responsible parties. \yyq{To scientifically evaluate whether the attack trees designed by \projname{} are more advantageous than those designed by human experts, we formed a seven-member review team to score and evaluate the attack trees designed by \projname{} and those designed by the experts.}
}




\subsection{Implementation}\label{sec:Implementation}
\textbf{Setup:} We developed an interactive web application for \projname{} with a user-centered design. Users can draw or import system models and specify a threat scenario (defining the attack entry point(s) and endpoint). Based on this input, \projname{} automatically generates a detailed attack tree and evaluates the feasibility of each attack path. The interface allows users to interactively refine results: for example, they can modify or add attack methods and logic nodes, adjust feasibility ratings, or reuse portions of attack trees when system components are updated. Once an attack tree is finalized, the tool serializes it in the extended OpenXSAM++ format (including fields such as ``Sub-Tree'', ``Step-Feasibilit'' and ``Impact''). These results are stored as the training set for LoRA fine-tuning and update the RAG dataset for future analyses. The base model in our implementation is ChatGPT-4~\cite{GPT-4}.

\textbf{Integration in Industry:}
\projname{} has been integrated into leading automotive manufacturers including Xiaomi Auto and United Automotive Electronic Systems (UAES) as a core component of their cybersecurity solutions. According to system operation statistics, it has consumed around 300 million tokens in 3 months and generated more than 8,200 attack trees (around 36,500 tokens per attack tree on average). 
In daily operation, the system evaluates 90+ production threat scenarios, giving engineers continuous, fine-grained risk visibility.
Participants from these enterprise deployments highlight \projname{}'s real-world impact: The UAES security manager commented, ``We rely on \projname{} for TARA analysis and attack tree generation, which has been instrumental in achieving R155 compliance. Its efficiency makes it indispensable for our operations.'' Similarly, Xiaomi TARA manager noted, ``\projname{}'s automated attack tree generation effectively solves critical issues and greatly improves our workflow efficiency.'' These large-scale deployments prove that \projname{}'s effectiveness meets the cybersecurity demands of modern automotive development.

\comment{
We adopt a user-centered design to develop an interactive web application for \projname{}. Users can draw system-modeling diagrams directly on the web page or import existing models, and then specify threat scenarios (including the attack endpoint and entry points). Based on these inputs, \projname{} automatically constructs a detailed attack tree and rates the feasibility of potential attack paths. Moreover, the analysis results in our web application are interactive, allowing users to make personalized changes. Specifically, users can: (i) Directly modify leaf nodes (including attack methods and logical nodes) based on individual preferences. (ii) Adjust the feasibility rating of attack methods. (iii) Reuse existing attack trees when certain components are replaced or updated, thereby improving efficiency and ensuring consistency across attack trees. 
Once a finalized attack tree is completed, the web application partitions it into sub-trees again and creates new attributes to each atom's description file (e.g., \texttt{Attack Objective}, \texttt{Attack Methods} and \texttt{Step Feasibility}), then stores the analysis results in OpenXSAM++ format to build the LoRA fine-tuning training set and the RAG dataset for future use.

\yyq{Specifically, when generating attack trees, \projname{} employs LLMs that have undergone LoRA fine-tuning, while the base model can adopt general-purpose LLMs. For evaluation in this paper, we utilize ChatGPT-4\cite{GPT-4} as the base model. In conducting feasibility analysis, we implement the RAG approach by searching the OpenXsam++ corpus through embedding models (e.g., \cite{bge-m3,text2vec}), subsequently incorporating retrieved content as contextual examples to enhance LLM generation capabilities.}
}

\subsection{Open Science and Ethics Considerations}
Since \projname{} has been integrated into commercial automotive cybersecurity management platforms, we cannot fully open-source the source code of \projname{} due to proprietary licensing constraints. However, in accordance with open science principles, we will release a community version of our web application and open-source the database built from expert-curated TARA reports in~\cite{DefenseWeave}, thereby enabling scientific research in this field. All vulnerabilities discovered in real vehicles have been reported to the responsible parties and have since been resolved. We have a demo video in~\cite{demo}.

    
\subsection{Experimental Analysis on Real Vehicles}\label{CAR}
\begin{figure*}[htbp]
    \centering
    \begin{subfigure}[b]{0.24\textwidth}
        \includegraphics[width=\textwidth]{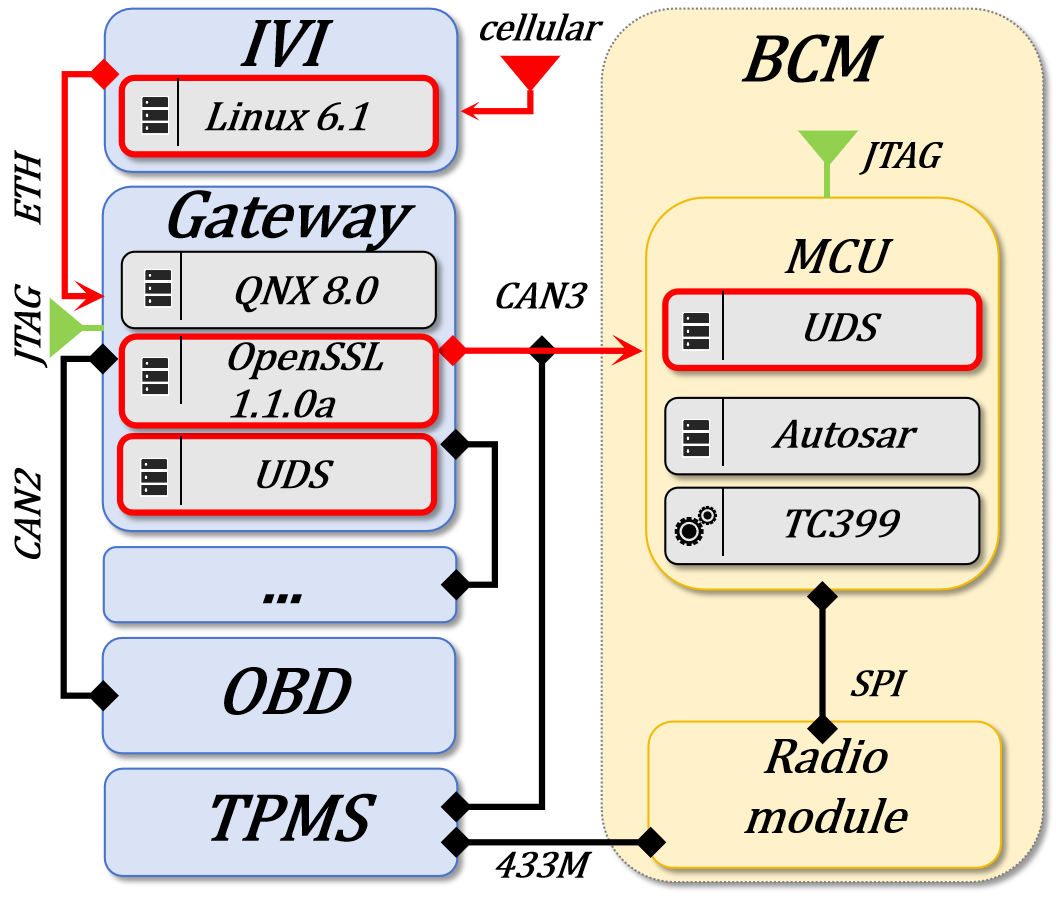}
        \caption{ Body Control
Moudle (Car $A$)}
        \label{fig:BCM_A}
    \end{subfigure}
    \hfill
    \centering
    \begin{subfigure}[b]{0.28\textwidth}
        \includegraphics[width=\textwidth]{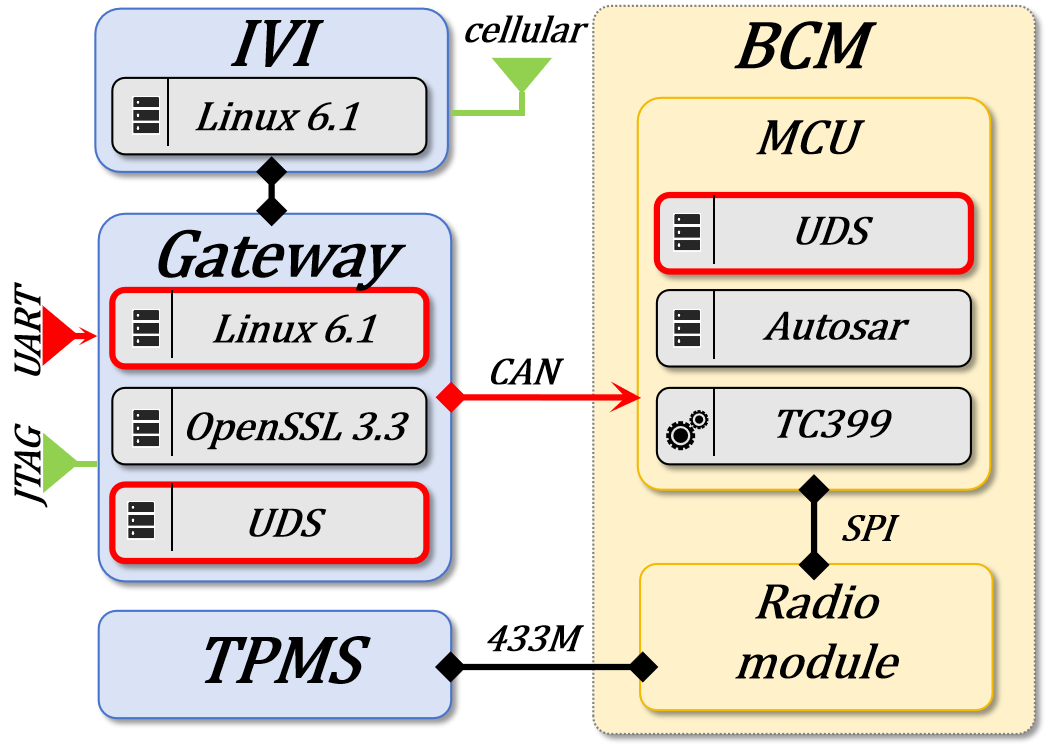}
        \caption{ Body Control
Moudle (Car $B$)}
        \label{fig:BCM_B}
    \end{subfigure}
    \hfill
    \begin{subfigure}[b]{0.37\textwidth}
        \includegraphics[width=\textwidth]{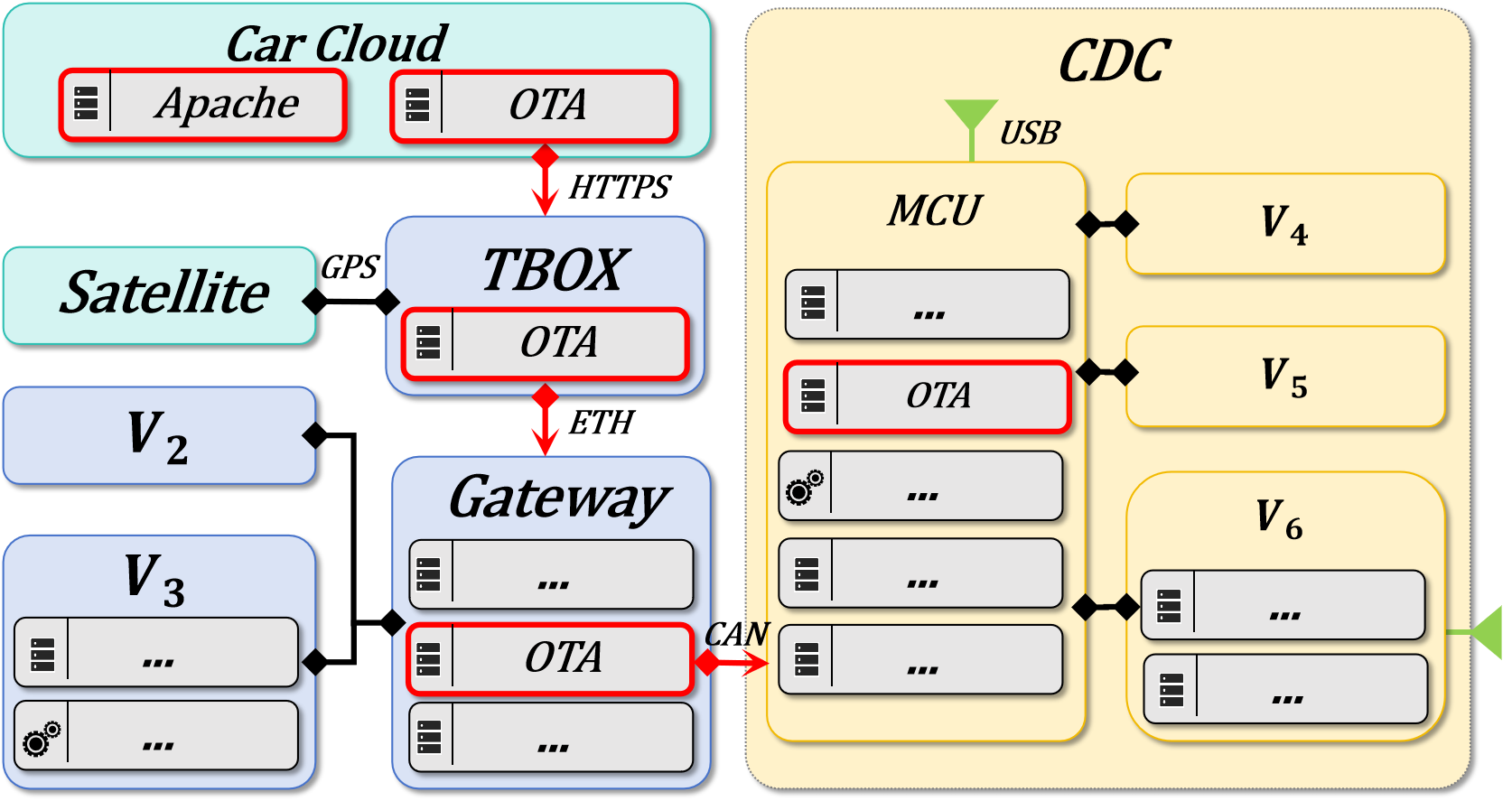}
        \caption{Cockpit Domain Controller (Car $C$)}
        \label{fig:CDC_C}
    \end{subfigure}
    \hfill
    \begin{subfigure}[b]{0.31\textwidth}
        \includegraphics[width=\textwidth]{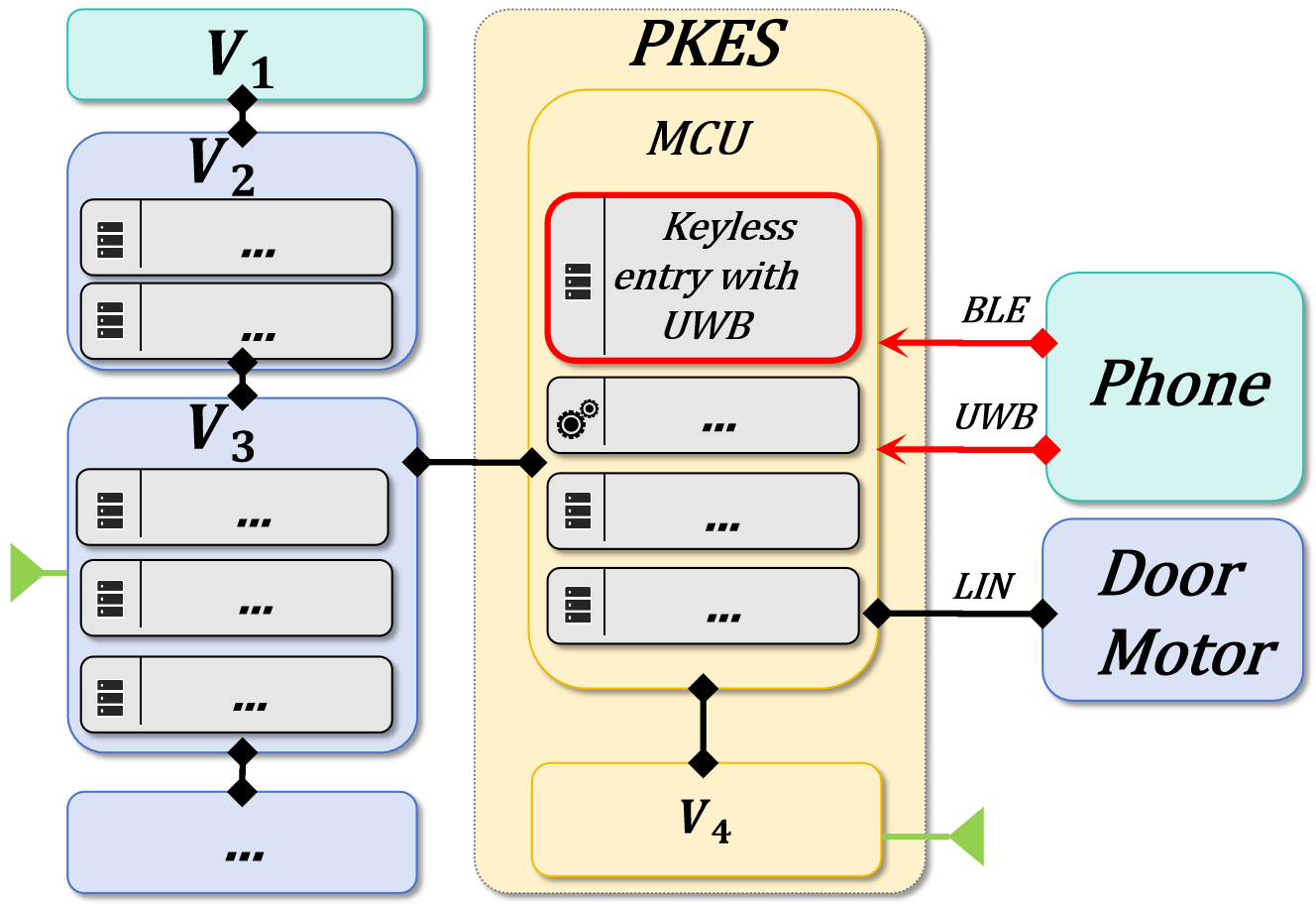}
        \caption{Passive Keyless Entry and
Start (Car $D$)}
        \label{fig:PKES_D}
    \end{subfigure}
    \hfill
    \begin{subfigure}[b]{0.345\textwidth}
        \includegraphics[width=\textwidth]{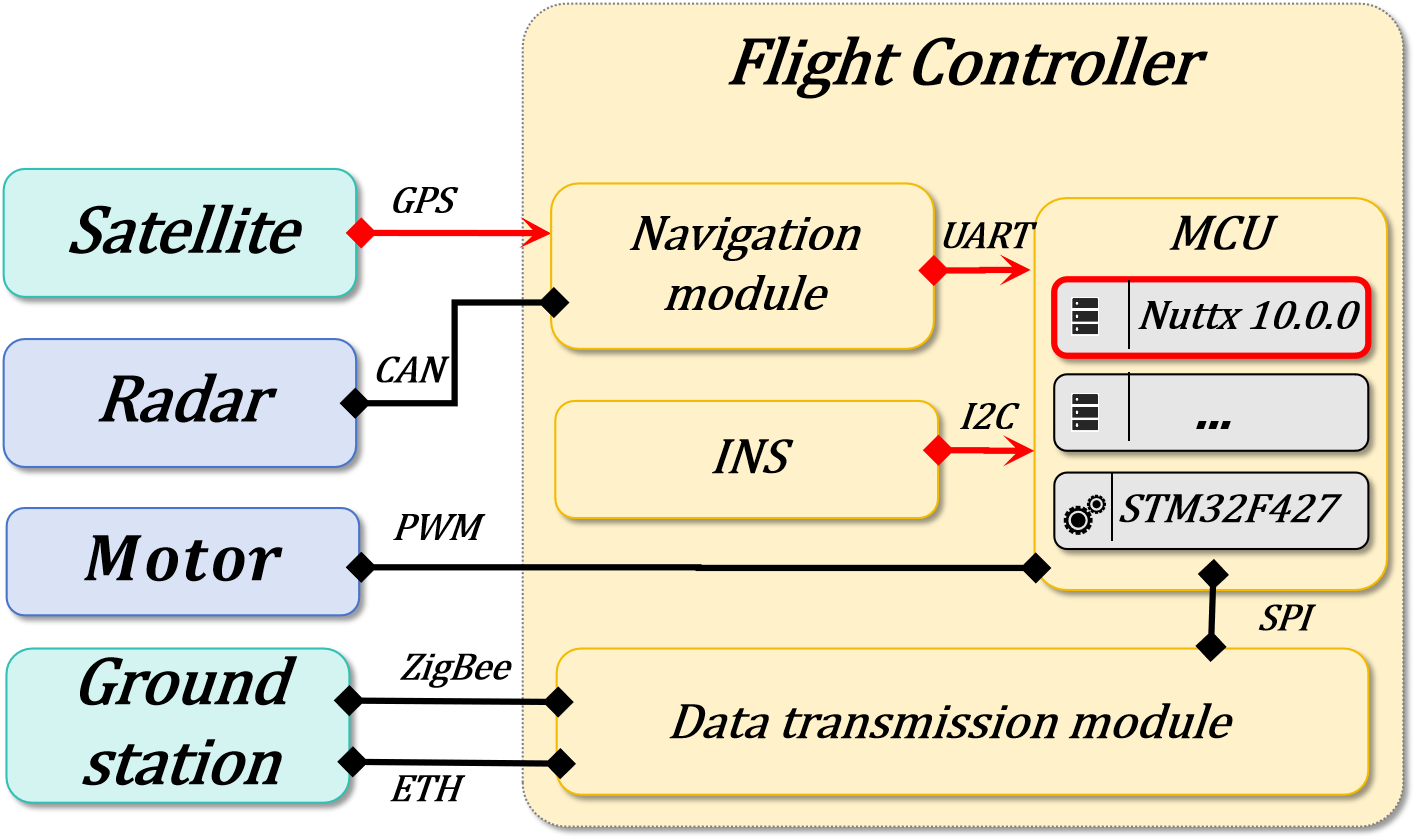}
        \caption{PX4 (UAV)}
        \label{fig:UAV_E}
    \end{subfigure}
    \hfill
    \begin{subfigure}[b]{0.305\textwidth}
        \includegraphics[width=\textwidth]{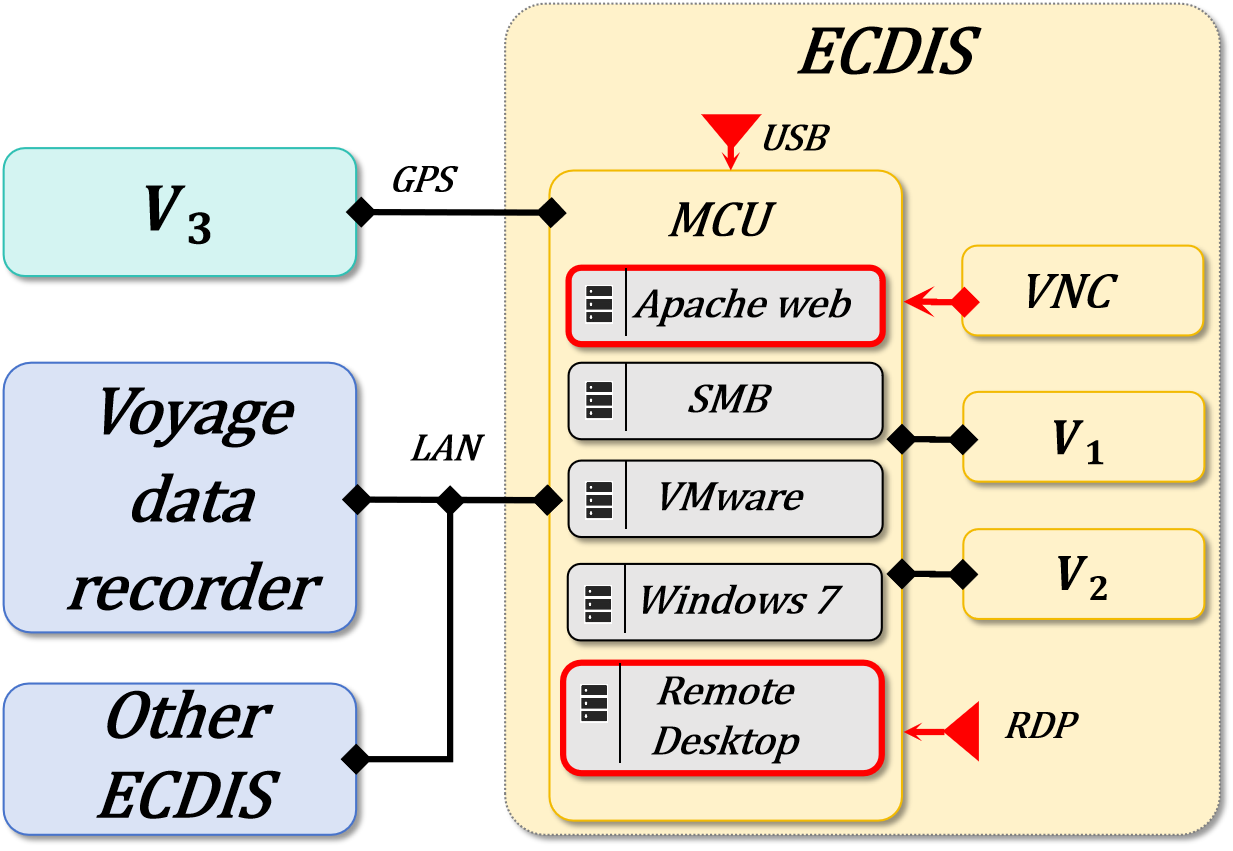}
        \caption{ECDIS (Ship)}
        \label{fig:ECDIS_F}
    \end{subfigure}
    
    \caption{Configurations of $6$ systems. (a) and (b) are BCMs of Car $A$ and Car $B$ with the same logical path, demonstrating \projname{}'s ability to generate distinct attack paths by considering the component-specific details. (c) CDC of Car $C$. \projname{} can comprehensively cover peripheral devices and cloud services in its anaylsis.
    (d) PKES of Car $D$. \projname{} successfully discovers recently emerged attack surfaces.
    (e) PX4 architecture of UAV and (f) ECDIS of Ship show that \projname{} can be applied to various electrical and electronic systems.
    }
    \label{fig:UAVandShip}
\end{figure*}

\subsubsection{Vehicles Under Examination}
To evaluate \projname{}’s effectiveness in function-level TARA, we deployed it in four real automotive security assessment projects. These involved four distinct in-vehicle components: two Body Control Modules (BCM) in different vehicles (Car A and Car B), one Cockpit Domain Controller (CDC, Car C), and one Passive Keyless Entry and Start system (PKES, Car D). The OEMs provided detailed configuration models for each component (see Fig.~\ref{fig:BCM_A}-\ref{fig:PKES_D}), which we imported into \projname{}. For each case, we defined a representative threat scenario and generated the corresponding attack tree with feasibility and risk evaluations. In accordance with industry practice (WP.29 R155 and ISO 21434), we treated any path rated with at least ``Medium'' feasibility as a candidate for penetration testing, since such paths warrant deeper security analysis. Guided by \projname{}’s output, we conducted targeted security tests on Cars A–D and ultimately confirmed 11 practical attack paths (each with risk level $\geq 3$) via proof-of-concept (POC) exploits. (Appendix A provides the complete attack trees and detailed attack steps for these scenarios.)

For the two BCM cases (Car A and Car B), we chose an identical threat scenario aiming to disable the BCM’s door control function (i.e., compromise its availability). This scenario has the same high-level attack entry point (the IVI infotainment unit) and endpoint (the BCM’s microcontroller) in both vehicles, following a logical path IVI$\rightarrow$Gateway$\rightarrow$BCM. For Car C (CDC), the threat scenario targeted the OTA update mechanism, aiming to install unauthorized firmware on the CDC. For Car D (PKES), the scenario focused on unlocking the vehicle’s doors without the owner’s authorization by exploiting the wireless key system. All four components have complex setups of hardware, software, and network interfaces, providing a rigorous testbed for \projname{}.

After \projname{} generated the initial attack trees and risk assessments for each scenario, we examined the output to identify high-feasibility attack paths for validation. All attack paths discussed below were rated at least ``Medium'' feasibility by the system, and thus merited real-world testing. We next describe the results for each scenario, including the proof-of-concept (PoC) attacks we performed and key insights derived from each case.

\comment{
To evaluate \projname{}’s effectiveness in function-level TARA, we applied it to four penetration testing projects involving real automotive components. Specifically, we examined two Body Control Modules (BCMs), a Cockpit Domain Controller (CDC), and a Passive Keyless Entry and Start (PKES) system based on the latest Car Connectivity Consortium (CCC) protocol. Each component is from a distinct vehicle type, referred to as Car $A$, Car $B$, Car $C$, and Car $D$, respectively. The OEMs for these vehicles provided detailed configurations of their respective components, as illustrated in Fig.~\ref{fig:BCM_A}, \ref{fig:BCM_B}, \ref{fig:CDC_C}, and \ref{fig:PKES_D}. Using this information, we built system models within the \projname{} web application and then defined multiple threat scenarios for each component. Note that for the BCMs in Car $A$ and Car $B$, we adopted the same threat scenario, which starts at the same attack entry point (the IVI) and ends at the same endpoint (the BCM-MCU). This allows us to compare the attack paths \projname{} generates when the underlying logical path remains identical (i.e., IVI $\rightarrow$ Gateway $\rightarrow$ BCM-MCU).

After generating attack trees and risk evaluation, both WP29 R155e and ISO 21434 require a systematic, risk-based approach to trigger deeper testing. In practice, many OEMs and suppliers adopt ``medium'' or higher feasibility as a trigger for penetration testing. Furthermore, the proportionate security measures will be implemented according the risk level.
Therefore, guided by the attack trees produced by \projname{}, we subsequently performed security analyses to exploit practical attack chains on real vehicles (Cars $A$-$D$). All the reported paths in our evaluation have a feasibility rating of at least ``medium''. The complete attack trees and detailed attack paths for the these cars are given in Appendix A.
}

\subsubsection{BCM: Identical Logical Paths but Different Attack Methods Due to Configuration Variations}
Modern vehicle E/E architectures are increasingly centralized, often placing a BCM behind a gateway that mediates inbound and outbound communications (e.g., software updates via Unified Diagnostic Services, UDS). Thus, similar threat scenarios against different BCMs may share the same high-level logical chain (IVI$\rightarrow$Gateway$\rightarrow$BCM), yet yield distinct attack paths due to differences in hardware/software configurations. To illustrate, we analyzed the BCM of two vehicles (Car A and Car B, see Fig.~\ref{fig:BCM_A} and Fig.~\ref{fig:BCM_B}) under the same scenario of disabling the door-unlock function. \projname{} produced similar logical attack trees for both, but with divergent specific exploits reflecting each vehicle’s nuances. For example, Car A’s gateway runs OpenSSL 1.1.0a, which is vulnerable to a known buffer overflow (CVE-2016-6309~\cite{cve-2016-6309}), whereas Car B’s gateway uses a newer OpenSSL 3.3 (not affected by that CVE) but exposes a UART debug interface absent in Car A’s gateway. \projname{} accordingly identified different feasible attack vectors: exploiting the OpenSSL CVE in Car A for remote code execution, versus leveraging the UART interface and a Linux kernel vulnerability in Car B.

\comment{
Modern vehicle E/E architectures are shifting toward centralized systems, consolidating multiple functions into a handful of high-performance computing units. In this setup, Body Control Modules (BCMs) typically reside behind a gateway that mediates all inbound and outbound communications—such as firmware updates and configuration changes—via Unified Diagnostic Services (UDS). This arrangement often leads to similar high-level (logical) paths for BCM-related threat scenarios (e.g., IVI$ \rightarrow$ Gateway$ \rightarrow$ BCM), but differences in hardware and software configurations can yield distinct specific attack methods.

To illustrate this, we analyzed BCMs from two distinct vehicles (i.e., Car $A$ and Car $B$), shown in Fig.\ref{fig:BCM_A} and Fig.\ref{fig:BCM_B}. In both cases, we set the BCM's microcontroller (BCM-MCU) as the attack endpoint and defined a threat scenario focused on disabling the BCM-MCU's door-opening function (i.e., compromising its availability). The logical path (IVI-GW-BCM) remains identical for both vehicles, but the detailed attack paths diverge because of variations in Gateway software and hardware. For example, Car $A$'s Gateway still uses OpenSSL 1.1.0a (which is vulnerable to CVE-2016-6309 vulnerability~\cite{cve-2016-6309}), whereas Car $B$ uses OpenSSL 3.3 (with that vulnerability fixed) but offers a UART interface that Car $A$’s Gateway lacks. Therefore, \projname{} successfully infers that attackers can exploit the CVE-2016-6309 vulnerability of Car $A$ to achieve remote control via buffer overflows, while \projname{}, however, finds UART access and privilege escalation through Linux vulnerabilities more feasible for Car $B$. The attack trees are shown in Fig.\ref{fig:attacktree_BCM_A} and \ref{fig:attacktree_BCM_B} in Appendix A.
}

\textbf{PoC attack}. We validated \projname{}’s suggested attack paths on both vehicles. \textbf{Car A}: We first obtained physical access to the BCM’s microcontroller via its JTAG debug port and dumped the firmware. From this, we reverse-engineered the BCM’s seed2key authentication algorithm for the UDS SecurityAccess service~\cite{thompson2022uds}, giving us the ability to bypass its security handshake. We then compromised the IVI unit (e.g., via an existing weakness) and sent crafted TCP packets to the gateway, exploiting the OpenSSL CVE-2016-6309 vulnerability to gain a remote shell on the gateway. Using this shell access, we injected CAN messages to the BCM, ultimately reprogramming the BCM’s firmware once the UDS authentication was bypassed. This multi-step attack was rated \textit{Medium} feasibility, with \textit{Major} potential impact, yielding an overall risk level 3. \textbf{Car B}: Here, \projname{}  also guides us to reverse-engineer the algorithm for the UDS SecurityAccess service in the BCM’s firmware. Using this knowledge, we reconstructed the BCM’s unlock authentication and then connected to the gateway via its UART interface. Through the UART, we exploited a privilege escalation vulnerability in the gateway’s Linux OS (CVE-2023-0179 in Linux 6.1) to fully compromise the gateway. From the gateway, we were able to send UDS commands to overwrite the BCM’s firmware. This Car B attack path was assessed as \textit{High} feasibility, \textit{Major} impact, and risk level 4.

\comment{
We validated the attack paths in Fig.~\ref{fig:BCM_A} and Fig.~\ref{fig:BCM_B} on Car $A$ and Car $B$, respectively,
\textbf{Car $A$:} \yyq{By dumping the BCM firmware via the MCU's JTAG port, we reverse-engineered the seed-key conversion algorithm for UDS SecurityAccess Service\cite{thompson2022uds} to reconstruct the complete authentication mechanism. After compromising the IVI, we sent malicious TCP packets to the Gateway to trigger the CVE-2016-6309 vulnerability in OpenSSL and obtain a reverse shell. Within this shell,  we send malicious CAN messages—ultimately tampering with the BCM firmware once the UDS authentication was bypassed.}  
The feasibility of this attack path is $Medium$, the potential impact of $Major$, and a final risk level of Level $3$. \textbf{Car $B$:} We also reconstructed the complete UDS authentication algorithm due to the UDS hard-coded vulnerability. Through the Gateway's UART interface, we exploited CVE-2023-0179 in Linux 6.1 to control the Gateway and tamper with the BCM firmware through the UDS protocol. This attack path is considered to have $High$ feasibility, a potential impact of $Major$, and a risk level of Level $4$.
}

\textbf{Insights.} 
Despite an identical high-level attack chain, Cars A and B demanded different exploits—highlighting two broader risk factors. (i) UDS access: Automotive security analysts often assume that UDS-based attacks require direct physical access (e.g., via OBD-II ports); however, our results show that network-facing vulnerabilities can enable remote exploitation of UDS services. (ii) Outdated software: Legacy software versions (e.g., running an obsolete OpenSSL library) can expose a vehicle to N-day vulnerabilities like CVE-2016-6309, which attackers can remotely leverage. \textbf{These findings demonstrate \projname{}’s strength in generating distinct, practical attack paths by accounting for each vehicle’s specific configuration.} In contrast, static, datalog-based TARA tools~\cite{CarVal, Saulaiman} that rely on predefined threat libraries would treat these two BCM scenarios similarly and likely miss such configuration-dependent attack vectors.

\comment{
By comparing the paths in Fig.\ref{fig:BCM_A} and Fig.\ref{fig:BCM_B}, we highlight two major risk areas: \yyq{(i) \textit{UDS Services:}  TARA experts often assume UDS exploits require physical access(e.g., via OBD-II ports or CAN bus disassembly),  neglecting how adjacent network vulnerabilities enable remote attacks.}
(ii) \textit{Legacy Software Versions:} Deploying outdated software (e.g., OpenSSL 1.1.0a) exposes systems to N-day vulnerabilities like CVE-2016-6309. Furthermore, these findings demonstrate \projname{}'s ability to generate distinct, practical attack paths—even when systems share the same high-level logical path—by leveraging the specific hardware and software configurations for each component. However, these component-specific details can not be automatically captured by datalog-based approaches~\cite{CarVal, Saulaiman}
}

\subsubsection{CDC: New Attack Vectors from Cloud Services}
Modern intelligent vehicles often include a Central Domain Controller (CDC) with Over-The-Air (OTA) capabilities, allowing manufacturers to remotely push software updates from the vehicle cloud. In addition, various external systems—such as cloud servers and mobile apps—interact with the CDC without a direct physical connection to the car, thereby introducing new attack surfaces in automotive TARA. Attackers can leverage vulnerabilities in the cloud infrastructure or in the CDC's firmware-verification logic to construct malicious update packages and bypass integrity checks, ultimately tampering with critical components. We used \projname{} to examine the OTA update process of the CDC in Car~$C$ (Fig.~\ref{fig:CDC_C}), generating an attack tree that highlights multiple threat vectors (Fig.~\ref{fig:attacktree_CDC_C} in Appendix $A$). For instance, one path involves exploiting an Apache remote-code-execution vulnerability (such as Log4j2~\cite{Log4j2}) on the server side, or obtaining SSL certificates and keys to perform a man-in-the-middle (MITM~\cite{MITM}) attack on the HTTPS channel between the vehicle cloud and the TBOX. Then we can tamper with the firmware update and push it via the TBOX and gateway to the CDC-MCU.

\textbf{PoC Attack.} We validated this attack on Car~$C$ by discovering hardcoded CA certificates\cite{CA} and private keys in the TBOX, which allowed us to bypass TLS/SSL verification and execute a MITM attack on the connection between the cloud and the TBOX. As a result, we could inject malicious code into the OTA firmware. Furthermore, reverse-engineering the firmware's verification process revealed a logical flaw: the CDC would proceed with an upgrade even if the firmware's hash check failed. Exploiting this flaw, we successfully installed our modified firmware onto the CDC-MCU. This attack path has an overall feasibility of $Medium$, a potential impact classified as $Serious$, and a final risk level of Level $4$.

\textbf{Insights.} This attack path demonstrates how OTA updates delivered through the vehicle cloud can serve as a conduit for injecting malicious code into a modern vehicle's core systems. Additional external components—such as smartphones (see Fig.~\ref{fig:PKES_D}), satellites, or ground stations—also act as potential off-vehicle entry points for functional-level TARA, expanding the overall attack surface beyond in-vehicle networks. \textbf{The results show that \projname{} can comprehensively cover peripheral devices and cloud services in its analysis, providing a more holistic view of potential attack vectors in modern vehicles.} However, it is difficult for the datalog-based approaches~\cite{CarVal, Saulaiman} to capture all of these off-vehicle attack surfaces due to their static library-based design.

\subsubsection{PKES: Discovering Unforeseen Attack Surfaces}
Passive Keyless Entry and Start (PKES) systems provide convenient vehicle access by using radio-frequency signals to unlock and start the car. Nevertheless, they are susceptible to relay attacks, wherein attackers trick the system into believing the key fob is nearby, enabling unauthorized entry or ignition. To mitigate such threats, the latest PKES implementations (Fig.~\ref{fig:PKES_D}) integrate Ultra-Wideband (UWB) technology according to the CCC protocol. UWB offers a ``secure ranging'' feature that effectively counters standard relay attacks~\cite{PKE23}. Even so, advanced PKES technology can still present unforeseen attack surfaces. Using \projname{}, we simply added a software annotation indicating that ``the latest PKES system adds UWB modules to prevent relay attacks.'' We then configured a threat scenario (attack entry at the user's phone and attack endpoint at the door motor) under the objective ``illegally open the car door.'' The resulting attack tree (Fig.~\ref{fig:attacktree_PKES_D} in Appendix $A$) reveals that bypassing PKES involves two core steps: first, intercepting UWB signals and injecting malicious ones to disrupt the normal UWB ranging process; second, using a relay attack against the system's Bluetooth Low Energy (BLE) channel. Together, these actions deceive the PKES into maintaining outdated distance data, enabling an unauthorized door unlock once the authorized user moves away.

\textbf{PoC Attack.} We validated \projname{}'s results on Car~$D$. After sniffing both the UWB and BLE signals, we identified system parameters like connection intervals and window offsets on BLE. By continually interfering with UWB ranging while the legitimate user was present, the PKES retained the user's proximity data even after the user left. We then relayed the BLE signals to complete the unlock procedure. In this scenario, the system relied on outdated UWB measurements, thus erroneously concluding that the user was still nearby. This attack path has an overall feasibility of $High$, a potential impact classified as $Serious$, and a final risk level of Level $5$.

\textbf{Insights.} This attack path illustrates how UWB—though designed to bolster PKES security—can itself become an attack surface when coupled with relay attacks on other channels. Moreover, \textbf{\projname{} demonstrates its ability to synthesize various methods (BLE, UWB manipulation, and relay attacks) into coherent, novel attack paths.} Such comprehensive analyses underscore the importance of integrating both traditional and emerging communication standards in function-level TARA. In comparison, the datalog-based approaches~\cite{CarVal, Saulaiman} can only reason the known attack surfaces based on predefined threat libraries.

\subsubsection{Responsible Disclosure and Summary}
 All penetration tests were conducted jointly with the vehicle manufacturers under ethical guidelines. In total, we identified 11 distinct vulnerabilities/attack paths across Cars A–D, each of which was promptly reported to the responsible OEM or supplier and has since been patched. Due to confidentiality agreements, we omit specific manufacturer names and certain low-level details. These four cases cover a wide range of automotive technologies (multiple ECUs, wireless interfaces, OS software, etc.), and in each case \projname{} discovered component-specific attack methods beyond the scope of existing threat libraries. Traditional approaches require analysts to manually select likely vulnerabilities from a database and write custom reasoning rules for each scenario, whereas \projname{} automates the end-to-end process of attack tree generation and feasibility evaluation. The real-world results above confirm that \projname{} can drastically reduce the manual effort while uncovering critical, non-intuitive attack paths in complex vehicle systems.

\comment{
We performed these penetration tests in close collaboration with automotive manufacturers, identifying 11 vulnerabilities that were promptly reported and subsequently resolved. Due to confidentiality agreements, we cannot disclose further details about the specific vulnerabilities or the manufacturers involved.


Collectively, these four cars and components incorporate complex IVN systems—ranging from hardware chips and wireless interfaces to diverse software components on different operating systems. Such component-specific attack methods are not covered by the static threat libraries of datalog-based approaches~\cite{CarVal, Saulaiman}. Moreover, besides the identification of threat scenarios, those methods further require users to manually select potential vulnerabilities from predefined libraries and meticulously design reasoning rules for each node. By contrast, \projname{} automates the entire attack-tree creation and evaluation process using LLMs, substantially reducing reliance on user expertise.
}



\subsection{Case Studies on Other Electronic Systems}\label{other car}


Although our primary evaluation is in the automotive domain, \projname{}’s methodology is general. We performed two case studies on non-automotive cyber-physical systems to demonstrate cross-domain applicability.


\subsubsection{Systems Under Examination}
Leveraging its broad knowledge base, \projname{} can generate attack trees for diverse embedded systems (without any code modifications). We applied it to two distinct platforms: (i) a PX4-based unmanned aerial vehicle (UAV) system with an inertial measurement unit (IMU) and GPS sensors, and (ii) an Electronic Chart Display and Information System (ECDIS) used in ship navigation (responsible for map display, route planning, etc.). In these case studies, we did not perform live penetration tests; instead, we built system models from publicly available documentation~\cite{PX4,PX42} and prior research~\cite{xu2023sok,svilicic2019raising,svilicic2019assessing}. We then compared \projname{}’s generated attack paths to known vulnerabilities reported in the literature. The complete attack trees for the UAV and ECDIS are provided in Appendix B (Fig.~\ref{fig:attacktree_UAV_E} and Fig.~\ref{fig:attacktree_ECDIS_F}, respectively). \textbf{These experiments demonstrate that \projname{}’s approach to function-level TARA can be easily extended to complex, safety-critical environments beyond automotive.}

\comment{
Leveraging the broad knowledge base of large language models, \projname{} can generate attack trees and risk evaluation for other electronic and embedded systems, such as drones and ship-borne navigation platforms. To illustrate its versatility, we applied \projname{}—without modifying any source code—to two additional systems: (i) a PX4-based UAV equipped with IMU and GPS sensors, and (ii) an ECDIS for ships, which supports precise positioning, route planning, and continuous navigation monitoring. 
Note that, in these two case studies, we did not conduct penetration testing on real systems. Instead, we constructed system models based on publicly available documentation~\cite{PX4,PX42} and research papers~\cite{xu2023sok,svilicic2019raising,svilicic2019assessing}, then compared the generated attack paths with the literature results to demonstrate the correctness and applicability of the discovered attack paths.
The complete attack trees for these systems are shown in Fig.~\ref{fig:attacktree_UAV_E} and Fig.~\ref{fig:attacktree_ECDIS_F} in Appendix B.
These case studies confirm \projname{}’s capacity to perform function-level TARA in complex, safety-critical environments beyond traditional automotive systems. 
}

\subsubsection{Unmanned Aerial Vehicle (UAV)}
Drones require thorough threat assessments, as mandated by national and international safety standards (e.g., GB 42590-2023~\cite{UAV}). These standards span the entire drone lifecycle, from data-link protection to electromagnetic compatibility. 
A recent work~\cite{xu2023sok} manually identified $2$ novel multi-round attack paths to degrade drone's sensor reliability over time. By applying \projname{} to a PX4-based UAV (Fig.~\ref{fig:UAV_E}), we identified not only these two reported $2$ multi-round attack paths in \cite{xu2023sok}, but also $1$ additional attack path against drone clusters.
(i) Electromagnetic Interference on the IMU. Broadcasting interference signals at frequencies matching the MPU6000 hardware chip can distort gyroscope outputs. The resulting unstable flight dynamics cause blurred images that lead to target misclassification.
(ii) Forged GPS Signals. Emitting counterfeit GPS data disrupts accurate positioning, again producing blurred or misplaced images and undermining flight autonomy.
(iii) Swarm-Level Manipulation for Drone Clusters. Exploiting the communication channels between drones in a swarm allows attackers to tamper with the collective positioning signals. This can delay or misroute multiple drones, potentially compromising time-critical missions like search and rescue. Though this attack path is not covered and validated in \cite{xu2023sok}, it is well discussed in~\cite{tayyab2024swarm}.
The above $3$ diverse paths highlight how \projname{} takes hardware, software, and operational context into account—ultimately revealing new drone-specific vulnerabilities (e.g., swarm communication) not covered in earlier automotive studies. 

\begin{figure*}
    \centering
    \includegraphics[width=0.9
    \linewidth]{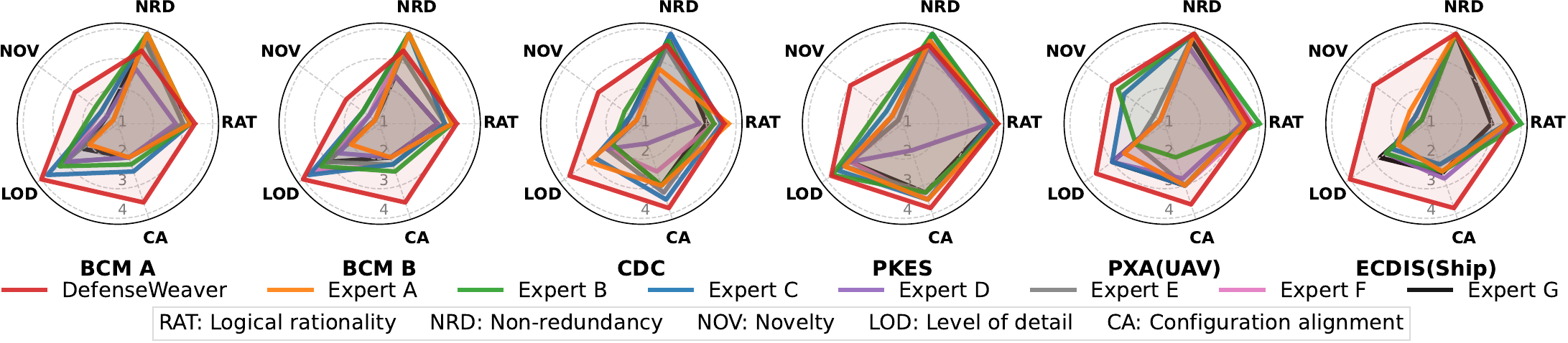}
    \caption{Comparison of \projname{} and seven human experts in different dimensions.}
    \label{fig:radar}
\end{figure*}

\subsubsection{ECDIS of Ships}
Ships also face stringent cybersecurity requirements. For instance, the UR E27 Rev.1 standard~\cite{URE27} explicitly mandates risk assessments for on-board systems and equipment. 
Existing works~\cite{svilicic2019raising,svilicic2019assessing} show that ECDIS platforms often contain high-risk vulnerabilities such as (i) outdated services (e.g., SMB or RDP) lacking authentication and (ii) third-party software (e.g., web servers) prone to N-day exploits such as Log4j2 or CVE-2021-41773.
By applying \projname{} to a typical ECDIS system (Fig.~\ref{fig:ECDIS_F}), we also identified the $2$ known threats and $1$ additional risk factor:
(i) Exploiting Remote Desktop Services. Vulnerabilities like CVE-2019-0708 enable unauthorized remote access, potentially leading to arbitrary code execution.
(ii) Leveraging Third-Party Software Flaws. Known Apache server exploits allow attackers to exfiltrate data or upload malicious code.
(iii) Malicious ENC Files and USB Interfaces. By injecting counterfeit electronic navigational chart (ENC) files through the system's USB update port, an attacker can alter critical route data. This tactic not only exploits a physical interface but also poses immediate risks to maritime navigation safety. Although this attack path is not covered and validated in ~\cite{svilicic2019raising, svilicic2019assessing}, it is involved in the ECDIS cybersecurity guidelines~\cite{ENCWG7}.



\comment{
\subsubsection{Summary of Case Studies.}
Through these two case studies—one focusing on UAV systems and another on ship navigation—\projname{} demonstrates that the same methodology used in automotive environments can be readily adapted to diverse electronic systems. By drawing on its LLM foundation, \projname{} effectively pinpoints both conventional and novel attack vectors, underscoring its potential to guide risk assessments across multiple industries.
}

\comment{
\subsection{Summary of Our Attacks}
We conclude our experimental findings by highlighting the properties of \projname{} in the following aspects:

\textbf{Function-Level TARA Automation.}
\projname{} refines its analysis down to the software and hardware levels, enabling it to detect subtle differences arising from configuration changes. Consequently, even components sharing similar logical paths can yield distinct attack trees.

\textbf{Full-Lifecycle Assessment.}
When analyzing emerging or less-studied systems, \projname{} proposes novel attack surfaces, showcasing its ability to tackle dynamic threat landscapes, thereby enabling full-lifecycle assessments.


\textbf{Automated Attack-Path Generation and Risk Evaluation.}
Unlike prior methods that require users to manually identify potential vulnerabilities for each component, \projname{} heuristically leverages LLMs to uncover unknown or overlooked security flaws in individual components, significantly reducing workload for full-lifecycle assessment.



\textbf{Cross-Domain Applicability.}
\projname{} can also handle function-level TARA for other electronic or embedded platforms (e.g., UAVs, maritime systems), demonstrating broad versatility.
}

\subsection{Comparison with Human Experts}
While the above experiments show \projname{} guiding the discovery of real attack paths, we also quantitatively evaluated the quality of its attack trees against those designed by human experts. We assembled a seven-member review team of security professionals (with backgrounds in automotive TARA, penetration testing, and cybersecurity management; see Appendix C, Table~\ref{tab:recruit}) to score attack trees generated by \projname{} versus those created by human experts. Each human expert was required to spend enough time to get familiar with system configurations for the six scenarios we considered (the four automotive components and the two additional case-study systems) before they
independently created attack trees. The review team assessed all attack trees on five key dimensions of quality – logical rationality, non-redundancy, novelty, level of detail, and configuration alignment – following ISO/SAE 21434 guidelines and industry best practices. Each dimension was rated on a 1–4 scale (higher is better). \projname{}’s trees were rated by all seven reviewers, whereas each expert-designed tree was scored only by the other six reviewers (preventing authors from rating their own work). Scores for each dimension were averaged across reviewers, and we visualized overall performance using five-dimensional radar charts (Fig.~\ref{fig:radar}). Please refer to Appendix C for further details on the expert recruitment, scoring criteria, and evaluation process.

\comment{
\zyz{
In the previous experiments, we demonstrated \projname{}’s application to real vehicle systems and validated its outputs through practical attack-path mining and penetration testing.}

\zyz{
\subsubsection{Experiment Setup} To further quantify \projname{}’s value, we next conducted a comparative user study against human experts. In this study, $7$ experienced automotive security analysts (human TARA experts A-G) were recruited to independently generate attack trees for the same six evaluation scenarios used by \projname{} (the four in-vehicle components and two cross-domain systems: BCM A, BCM B, CDC, PKES, UAV, and ECDIS). We then evaluated the quality of the resulting attack trees from \projname{} and from the human experts side-by-side. To ensure fairness and objectivity, each attack tree was scored by the other 6 reviewers using a structured rubric (no reviewer scored their own work). Please refer to Appendix C for details of the expert recruitment, experimental procedure, and scoring criteria of this evaluation.
}

\zyz{
\subsubsection{Evaluation Dimension} Consistent with automotive cybersecurity standards like ISO/SAE 21434 and industry best practices, we defined five dimensions to assess the quality of attack trees. Briefly, these are: (1) \textbf{Logical Rationality} – each attack path should be logically sound and follow security reasoning (no implausible leaps); (2) \textbf{Non-Redundancy} – the attack tree should avoid duplicate or redundant paths, focusing on unique attack vectors; (3) \textbf{System Alignment} – all attack steps should align with the actual system’s components and configuration (using only elements present in the system configurations); (4) \textbf{Level of Detail} – the attack steps should be described with sufficient technical detail and specificity (e.g., including concrete attack techniques or CVE examples); and (5) \textbf{Novelty} – the attack tree should include creative or previously unseen attack methods beyond well-known threat templates. Each dimension was scored on a four-point scale from Poor (1) to Excellent (4). The final score for each dimension is the average of all reviewers’ ratings for that criterion on a given tree.
}

\comment{
In previous experiments, we evaluated \projname{}'s application in real-world vehicle TARA through practical attack path mining, while verifying its capability to guide actual threat identification during penetration testing. To further quantify \projname{}'s practical value, this section conducts comparative experiments to analyze differences in attack tree design between \projname{} and human TARA experts, thereby measuring \projname{}'s quality improvements. 
\yyq{We formed a seven-member review team to conduct structured scoring of the attack trees generated by \projname{} and five experts (each expert designed six attack trees targeting four automotive components and two other electronic systems) across five dimensions (Rationality, Non-Redundancy, Novelty, Level of Detail, and Configuration Alignment). The evaluation process followed a structured approach: (i) all seven reviewers assessed \projname{}'s attack trees to ensure consensus, while (ii) expert-designed attack trees were evaluated only by non-authors to avoid bias. The scoring results were averaged across reviewers to ensure consistency, and the comparative outcomes across these five key dimensions were visually presented in a radar chart (Fig.~\ref{fig:radar}).}
\zyz{Please refer to Appendix C for the details of recruitment, procedure, and scoring criteria of our evaluation.}
}

}

\begin{table}[b]
\centering
\small
\caption{Overall Comparison with Human Experts (Average)}
\label{tab:results_transposed}

\begin{tabular}{
    l
    c
    c
    c
    c
    c
    c
}
\toprule
\rowcolor{gray!25}
 & \textbf{RAT\textsuperscript{1}} & \textbf{NRD\textsuperscript{2}} & \textbf{NOV\textsuperscript{3}} & \textbf{LOD\textsuperscript{4}} & \textbf{CA\textsuperscript{5}} &
 \textbf{Time} \\
\midrule

\textbf{DW\textsuperscript{6}}      & 3.62 & 3.67 & 2.73 & 3.88 & 3.71 & 0.43m \\
\rowcolor{gray!10}
\textbf{Expert}  & 3.43 & 3.70 & 1.33 & 2.74 & 2.58 & 36.12m\\
\bottomrule

\end{tabular}

\begin{tablenotes}
\footnotesize
\item \textsuperscript{1} RAT: Logical rationality
\textsuperscript{2} NRD: Non-redundancy
\textsuperscript{3} NOV: Novelty
 \textsuperscript{4} LOD: Level of detail
\textsuperscript{5} CA: Configuration alignment
\textsuperscript{6} DW: \projname{}
\end{tablenotes}
\end{table}

\textbf{Results and Analysis:} 
We identify three major limitations in the human-crafted attack trees that \projname{} was able to overcome: (i) Human experts struggled to adapt to new system configurations, often failing to consider unconventional attacks outside their experience. (ii) Some expert-designed trees introduced superfluous or incorrect elements (e.g., non-existent components) based on subjective assumptions. (3) Experts sometimes overlooked critical yet subtle system-specific differences (for instance, they treated the two BCM cases too similarly, missing Car B’s UART vector). By avoiding these issues, \projname{}’s trees exhibited significantly higher novelty and system alignment. 
On average, as shown in Table~\ref{tab:results_transposed}, the \projname{}-generated trees scored +105.00\% higher in novelty (identifying many more unconventional attack paths) and +43.68\% higher in system alignment (strictly mapping to actual system components and interfaces) than the human experts. \projname{}’s trees also contained far more detail in attack steps (+41.46\% on the detail dimension), providing granular step-by-step paths. The logical rationality of \projname{}’s attack trees was on par with experts (a slight +5.79\% gain in rationality score). Notably, \projname{}’s comprehensive approach led to only a 0.90\% increase in redundancy, indicating it did not suffer from excessive duplicate paths despite its thoroughness. 
Overall, in all six scenarios, \projname{}’s automatically generated attack trees achieved equal or higher scores than the human-crafted trees in every dimension.
Notably, the time consuming of \projname{} for each case is only 0.43min, which is significantly reduced by 98.8\% compared with that of human experts.
\textbf{These results indicate that \projname{} can produce attack trees of substantially higher quality than traditional expert-driven methods, offering more complete coverage of potential threats without sacrificing coherence or correctness.}

\comment{
Fig.~ref{fig:radar} presents radar charts comparing \projname{} and human experts across the five dimensions, and Table~\ref{tab:results_transposed} summarizes the numerical scores. In all six scenarios, \projname{}’s automatically generated attack trees achieved equal or higher scores than the human-crafted trees in every dimension, greatly reducing the averaged time computation by 98.8\% (0.43min vs. 36.12min). In terms of individual evaluation dimensions, \projname{}’s advantage is most pronounced in Novelty, Level of Detail, and Configuration Alignment, while it performed roughly on par with experts in Rationality and Non-Redundancy. As Table~\ref{tab:results_transposed} shows, the attack trees generated by \projname{} scored 105\% higher in Novelty on average (2.73 vs. 1.33 out of 4) than the human-created trees – effectively more than doubling the number of unexpected or innovative attack methods. \projname{} also achieved a much greater Level of Detail in its attack steps (3.88 vs. 2.74, a 41.46\% improvement), providing more technical depth and specifics in the attack descriptions. Similarly, its attack steps remained tightly aligned to the system configurations (3.71 vs. 2.58, a 43.68\% improvement in Configuration Alignment), whereas the human experts’ attack trees sometimes strayed from the given system configuration. By contrast, for Logical Rationality the scores were comparable (3.62 vs. 3.43, a modest 5.8\% increase in favor of \projname{}), indicating that both \projname{} and the experts generally produced logically coherent attack paths. On Non-Redundancy, the two approaches were essentially equal (\projname{} scored 3.67 vs. 3.70 for experts), meaning both tended to avoid duplicating identical attack paths in their models. In summary, \projname{} matched or exceeded human experts on every quality dimension, with dramatically better performance in introducing novel attack ideas, incorporating low-level technical details, and staying true to the system’s actual configuration.

\zyz{
\subsubsection{Insights} The comparison yielded four key insights about how \projname{}’s attack trees differ from those of human experts:}

\zyz{
\textbf{Better Novelty:} \projname{} explores attack vectors beyond experts’ imagination. We observed that \projname{} introduced many creative and unexpected attack paths that the human experts did not consider. The Novelty scores were significantly higher for \projname{} in every case (Fig.~ref{fig:radar}). For example, in the UAV scenario, the tool identified a swarm-level communication attack vector, which none of the experts independently thought to include. This indicates that human analysts, constrained by their past experiences and cognitive biases, tend to focus on known attack patterns. In contrast, the \projname{} can generate unconventional attack methods, by drawing on a broad knowledge base. \projname{}’s higher novelty translates to more exhaustive coverage of potential threats, addressing one of the critical gaps in manual TARA (where novel threats might be overlooked).
}

\zyz{
\textbf{Better Fidelity:} Human experts sometimes include irrelevant or assumed elements. The study revealed that human-crafted attack trees often incorporated components or conditions that were not actually present in the system under analysis. Experts tended to rely on personal experience and assumptions, which led to the introduction of out-of-scope elements. For instance, in the BCM case, $4$ experts assumed an Over-the-Air (OTA) update interface existed and crafted attack steps exploiting OTA-based malware injection – even though the real system had no such OTA functionality. This kind of misalignment resulted in attack paths that deviate from the actual vehicle architecture. \projname{}, by design, strictly adheres to the provided system configurations, and therefore its attack trees remained faithful to the system’s true configuration. The much higher Configuration Alignment scores for \projname{} reflect this fidelity. We note that maintaining configuration alignment is crucial for the usefulness of TARA results – an attack path that exploits a nonexistent feature (like a phantom OTA channel) cannot be acted upon and only distracts from real vulnerabilities.
}

\zyz{
\textbf{Deeper Exploitation of Details:} Experts may overlook subtle differences in similar systems. Another insight is that \projname{} makes more effective use of detailed system information, whereas human experts sometimes underutilize available details, especially when faced with systems that appear similar. In scenarios with closely related systems or components, experts often assumed the same attacks would apply and missed critical nuances. A concrete example was the pair of Body Control Modules (BCM A and BCM B): these two systems shared the same high-level architecture (IVI→Gateway→BCM communication path), and 5 out of 7 experts produced essentially the same attack tree for BCM A and BCM B, failing to account for the different access control mechanisms in each variant. This demonstrates that \projname{} can discern and leverage fine-grained configuration details to tailor the attack tree to each system variant. The human oversight here highlights a common challenge – analysts might use a one-size-fits-all mental model for similar systems, missing out on specific weaker points. \projname{}’s detailed differentiation yields more accurate and system-specific attack trees.
}

\zyz{
\textbf{Overall Quality Gains:} From a macro perspective, \projname{}’s automated approach generates higher-quality attack trees than human experts across diverse systems. It achieved high logical soundness and avoided redundancy similarly to human experts, while surpassing them in the other dimensions. 
This suggests that \projname{} can replicate the strengths of human reasoning (maintaining logic and uniqueness of paths) and simultaneously mitigate human limitations.
}
}

\comment{
\yyq{
\subsubsection{Evaluation Dimensions:} According to According to ISO/SAE 21434, our comparison is evaluated in five dimensions. (i) \textbf{Rationality} measures whether attack paths follow security logic and meet TARA’s structural rigor—high values indicate sound reasoning, while low values suggest illogical or inconsistent paths. (ii) \textbf{Non-Redundancy} evaluates the absence of duplicate paths, where high values reflect streamlined efficiency, and low values imply unnecessary complexity.  (iii) \textbf{Novelty} examines innovative threat coverage, with high values uncovering unconventional risks and low values reflecting over-reliance on predictable paths. (iv) \textbf{Level of Detail} assesses granularity, from conceptual to technical specifics—high values enable precise traceability, whereas low values result in vague or superficial analysis. (v) \textbf{Configuration Alignment} checks if attack nodes match real configurations, with high values ensuring practical relevance and low values indicating a disconnect. 
}

\subsubsection{Results and Analysis} \zyz{The results are shown in Fig.~ref{fig:radar}, which demonstrates the superior performance of \projname{} in attack tree generation compared to human experts. In summary, we have the following $4$ key insights:}

\textbf{Overall Comparison:} Based on the radar chart area comparison, \projname{} achieved a improvement of $66.15$\% on average in the quality of attack tree generation across six evaluation scenarios compared to human experts. This performance demonstrates that \projname{} can design attack trees more comprehensively and accurately in systematic attack configuration. \projname{} excels in novelty ($105.00$\% improvement), highlighting its ability to break through the inherent cognitive limitations of human thinking. The $43.68$\% improvement in configuration alignment indicates that the attack trees generated by \projname{} adhere more strictly to system configurations. The $41.46$\% improvement in step granularity reflects \projname{}'s capability to provide more detailed and complete descriptions of attack processes. In terms of attack path rationality, \projname{} performs comparably to experts ($5.79$\% improvement), confirming that most of its generated paths are considered reasonable by professionals. Although \projname{} slightly underperforms in non-redundancy ($0.90$\% decrease), this is attributed to the system's exhaustive path generation and detailed descriptions.

\textbf{Novelty:} \projname{} achieved a $105.00$\% improvement over human experts in the novelty dimension. This conclusion is supported by the fact that human experts, when faced with novel system architectures (e.g., PKES with UWB functionality), are constrained by fixed thinking patterns and struggle to design attack paths that adapt to new system features. As a result, \projname{} can discover unconventional attack paths and identify attack vectors that human experts might overlook due to cognitive biases, thereby providing a more comprehensive assessment of system vulnerabilities.

\textbf{Configuration Alignment:} \projname{} improved configuration alignment by $43.68$\%. Human experts tend to introduce hardware/software elements not defined in the system model (e.g., assuming OTA functionality in BCM where it does not exist) based on personal experience, leading to attack paths that deviate from the actual system. In contrast, the attack trees generated by \projname{} adhere more strictly to system configuration specifications, effectively avoiding subjective assumptions that human experts might introduce and ensuring that the attack trees are more closely aligned with the specific system.

\textbf{Assessing Similar Components:} Human experts are more likely to overlook critical differences in system details. For example, $3$ out of $5$ experts failed to identify the key differences in access control mechanisms between BCM A and BCM B, resulting in the same attack paths being generated for two different systems. \projname{}, however, can more systematically leverage detailed system configuration information to provide more accurate and detailed attack tree design.

Overall, \projname{} not only strictly follows system configuration specifications to avoid subjective biases but also breaks through the boundaries of human cognition to discover unconventional attack paths. Although there are minor redundancy issues, this precisely reflects the system's comprehensiveness in attack path exploration. Further optimization of description methods can enhance the user experience.
}

\comment{\zyz{Conclusion and consequences.} From a macro-evaluation perspective, based on radar chart area comparisons, \projname{} demonstrates improvements in attack tree generation quality over experts across six evaluation scenarios: achieving enhancements of $70.76$\% (BCM A), $68.61$\% (BCM B), $74.84$\% (CDC), $42.22$\% (PKES), $55.35$\% (UAV), and $85.14$\% (ECDIS), with an average improvement of $66.15$\%. This robust performance solidly proves \projname{}'s superior capability in systematic attack modeling, enabling more comprehensive and accurate identification of system vulnerabilities. \zyz{NOV, SYSAL, LOD, NRD, RAT -> overall conclusion.}

\textbf{Novelty.} \zyz{1. conclusion (higher novelty) -> 2, insights -> 3. consequences.}

\textbf{Configuration Alignment.} \zyz{1. conclusion (human experts tend to introduce according to personal experiments, for example) -> 2. consequences.}

\textbf{Assessing Similar Components.} \zyz{1. conclusion (human experts are more likely to ignore xxx) -> 2. for example, xxx -> 3. consequences.}

\zyz{summary}}

\comment{In-depth analysis reveals three key limitations of human experts in attack tree design:
(i) \textbf{When facing novel system architectures }(e.g., PKES with UWB functionality),constrained by fixed thinking patterns, experts struggle to design attack paths adapted to the new system’s characteristics.
(ii) \textbf{Experts tend to rely on personal experience and introduce hardware/software elements not defined in the system model} (e.g., assuming non-existent OTA functionality in BCM),leading to attack paths that deviate from the actual system.
(iii) \textbf{Experts underutilize system details}. For example, $3$ out of $5$ experts failed to recognize the critical differences in access control mechanisms between BCM A and B, resulting in identical attack paths for two distinct systems.
These findings explain why \projname{} excels particularly in novelty ($105.00$\% improvement) and configuration alignment ($43.68$\% improvement). \projname{} strictly adheres to modeling specifications to avoid subjective biases while surpassing human cognitive boundaries to uncover unconventional attack paths.

Specifically, the comparison between \projname{} and the average score of Expert is shown in Table~\ref{tab:results}. \projname{}'s outstanding performance in the novelty dimension (a $105.00$\% improvement over human experts) directly validates its advantage in overcoming cognitive biases inherent to human thinking. The $43.68$\% enhancement in configuration alignment demonstrates that \projname{}'s generated attack trees adhere more strictly to system modeling specifications, effectively avoiding the introduction of subjective assumptions unrelated to the actual system that human experts might make. The $41.46$\% enhancement in step granularity highlights the \projname{}'s ability to provide more detailed and complete descriptions of attack procedures, aiding security analysts in deeper comprehension. In terms of attack path rationality, \projname{} performs comparably to experts ($5.79$\% improvement), confirming that most of its generated paths are deemed reasonable by professionals. However, it slightly underperforms in non-redundancy ($0.90$\% decrease), which expert interviews attribute to the system's exhaustive path generation and highly detailed descriptions, occasionally exceeding practical analysis needs.  

Overall, \projname{} surpasses experts across key attack tree quality metrics, particularly excelling in discovering novel attack vectors and systematically leveraging system modeling information. While minor redundancy exists, this reflects the system's thoroughness in attack path exploration—a trait that can be further refined through optimized presentation to enhance usability.}

\section{Limitations and Future Work}
While \projname{} already automates the generation and assessment of attack paths, it still relies on user-provided threat scenarios, which are high-level concepts that do not account for component-specific details. Therefore, we envision integrating Microsoft's STRIDE model~\cite{STRIDE}—covering Spoofing, Tampering, Repudiation, Information Disclosure, Denial of Service, and Elevation of Privilege—to automatically derive more comprehensive sets of threat scenarios. This enhancement would further reduce manual intervention and streamline end-to-end TARA processes.

\section{Conclusion}
We introduced \projname{}, a novel system that automates function-level TARA by leveraging LLMs. Unlike existing methods bound by static threat libraries, \projname{} adapts to evolving vulnerabilities due to the adoption of LLMs, offering flexibility across diverse standards and platforms. In extensive evaluations on four real automotive security projects, \projname{} uncovered 11 practical attack paths, each validated via penetration testing and responsibly disclosed. We also deploy \projname{} in UAV and ECDIS systems to demonstrate its cross-domain applicability, revealing new attack surfaces beyond traditional automotive contexts. Integrated into commercial cybersecurity management platforms, \projname{} has produced more than 8,200 attack trees in the industry to date. 
Compared with human experts, \projname{} significantly reduced time consumption of TARA process by 98.8\%.
Overall, \projname{} provides a robust, adaptive approach to TARA, significantly advancing the state of practice in automotive cybersecurity and beyond.






%

{\footnotesize \bibliographystyle{acm}
\bibliography{sample}}

\clearpage
\section*{Appendix A: Complete Attack Trees of Real Cars}

\subsection*{Experimental Methodology}

Each attack tree comprises a \textbf{threat scenario} (root node), \textbf{logical nodes} ($AND$, $OR$), \textbf{attack objectives} ($AO-X$), and \textbf{attack methods} ($AM-X$). For each method, \projname{} performs step-by-step feasibility assessments ($Step\_F$) and calculates a cumulative feasibility score ($Cumulative\_F$). The threat scenario denotes the ultimate goal of an attack, while the attack objectives serve as intermediate goals that support it. Logical nodes connect the attack methods necessary for achieving each objective.
Using \projname{}, we generated four comprehensive attack trees for the four threat scenarios discussed in Sec~\ref{CAR} (see Fig~\ref{fig:attacktree_BCM_A}, \ref{fig:attacktree_BCM_B}, \ref{fig:attacktree_CDC_C}, \ref{fig:attacktree_PKES_D}). In these figures, we highlight in red the specific attack paths detailed in Sec~\ref{CAR}, annotated with feasibility indicators ($Step\_F$ and $Cumulative\_F$). Guided by these four trees, each showing at least a “Medium” feasibility rating and a relatively high risk level (at least 3), we identified $11$ practical attack paths  confirmed through penetration tests.
Note that the attack tree serves as a high-level guide and may not exactly mirror the real-world attack paths uncovered in full detail. The 11 verified attack paths are as follows:

\subsection*{Complete Attack Tree of BCM (Car A)}
\begin{figure*}[b!]
    \centering
    \includegraphics[width=0.9\linewidth]{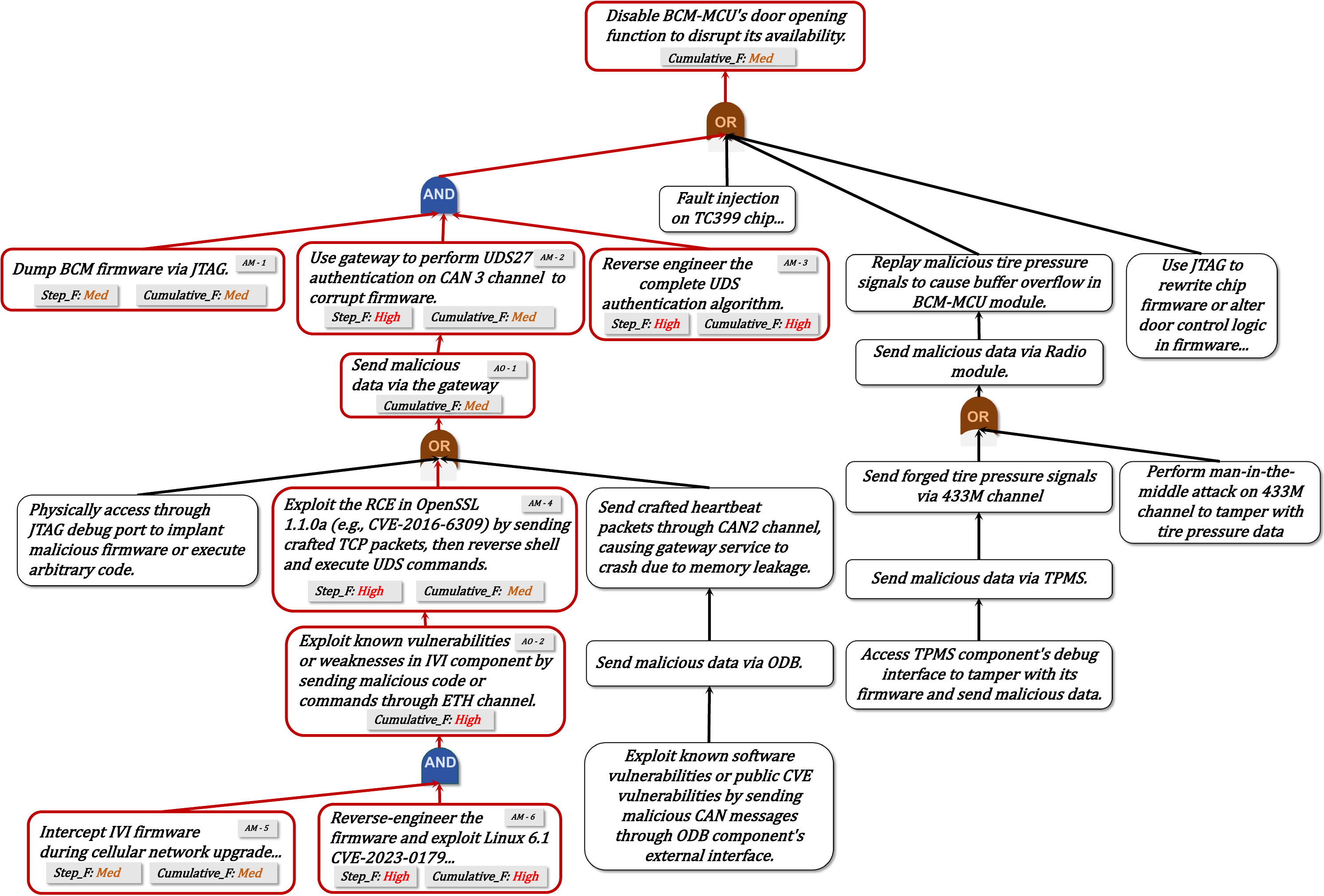}
    \caption{Attack tree BCM (Car A)}
    \label{fig:attacktree_BCM_A}
\end{figure*}

Drawing on Fig~\ref{fig:attacktree_BCM_A}, we identified three feasible paths. For this attack tree, the threat scenario's overall attack feasibility is $Medium$, the potential impact is $Major$, and the resulting risk level is $3$:

\textbf{Attack Path 1:} By dumping the BCM firmware via the MCU’s JTAG port, we reverse-engineered the seed-key conversion algorithm for UDS SecurityAccess Service to reconstruct the complete authentication mechanism. After compromising the IVI, we sent malicious TCP packets to the Gateway to trigger the CVE-2016-6309 vulnerability in OpenSSL and obtain a reverse shell. Within this shell, we send malicious CAN messages—ultimately tampering with the BCM firmware once the UDS authentication was bypassed. 

\textbf{Attack Path 2:} We demodulated TPMS data using FSK and cracked a non-standard CRC8 algorithm to forge TPMS signals. Replaying these signals triggered a buffer overflow in the BCM-MCU component.

\textbf{Attack Path 3:} We successfully detected the JTAG interface through hardware reverse engineering, including TDI, TDO, TMS, and TCK, and established a connection. However, when attempting to read the firmware further, we discovered that it has built-in read and write protection.s.

\subsection*{Complete Attack Tree of BCM (Car B)}

\begin{figure*}
    \centering
    \includegraphics[width=0.9 \linewidth]{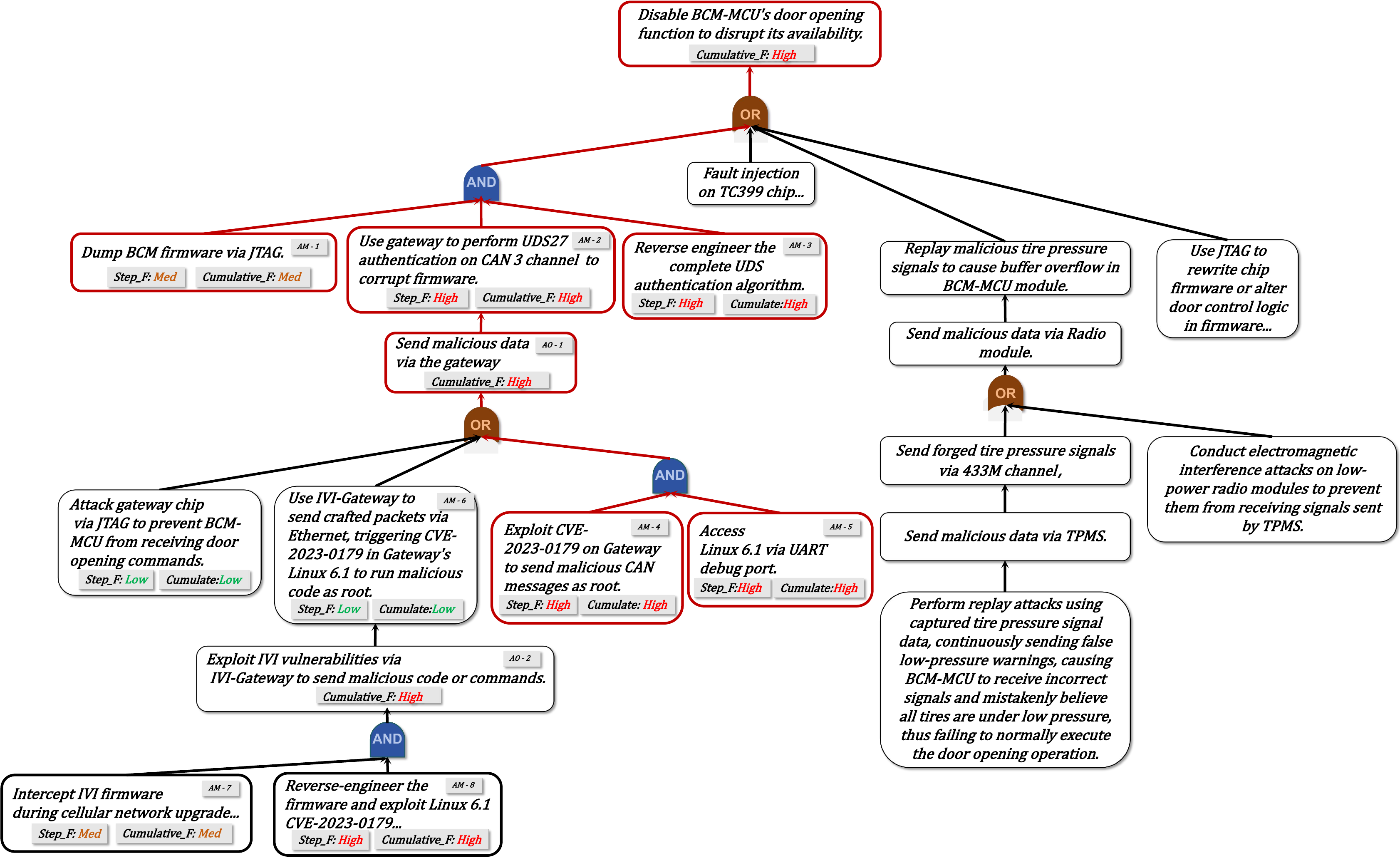}
    \caption{Attack tree BCM (Car B)}
    \label{fig:attacktree_BCM_B}
\end{figure*}

Similarly, based on Fig~\ref{fig:attacktree_BCM_B}, we discovered three paths. The threat scenario's overall attack feasibility is $High$, the potential impact is $Major$, and the final risk level is $4$:

\textbf{Attack Path 4:} As in Car A, we identified hard-coded UDS authentication in the BCM firmware. Through the Gateway's UART interface, we exploited CVE-2023-0179 in Linux 6.1 to compromise the Gateway and then tampered with the BCM firmware via the UDS protocol.

\textbf{Attack Path 5:} We launched a DoS attack by using high-power signals to attack the radio receiver module, thereby damaging its gain module.

\textbf{Attack Path 6:} Like Path 2, we can successfully connect to the JTAG interface, but due to the read protection, we are unable to read the firmware.

\subsection*{Complete Attack Tree of CDC (Car C)}
\begin{figure*}[t!]
    \centering
    \includegraphics[width=1 \linewidth]{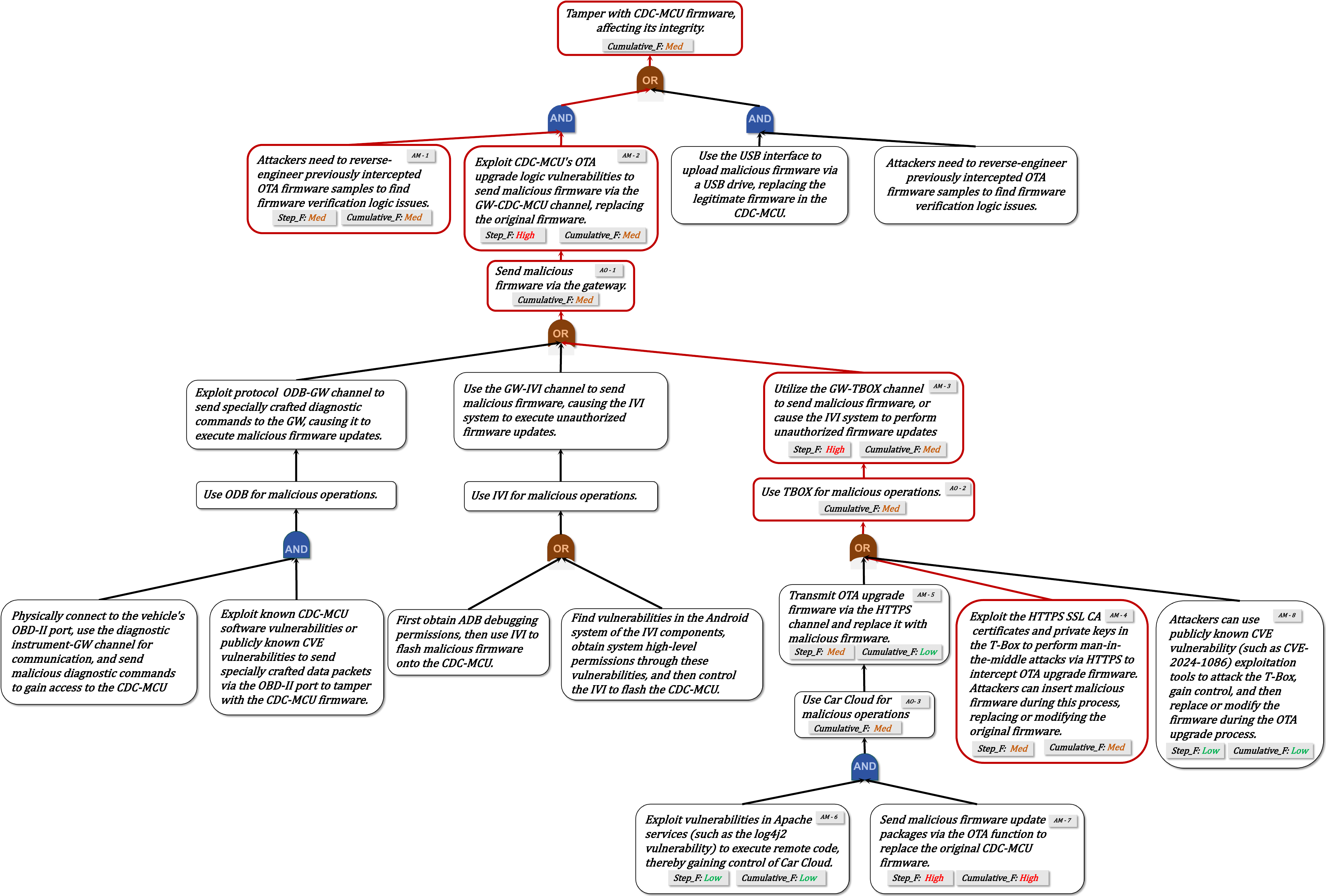}
    \caption{Attack tree CDC (Car C)}
    \label{fig:attacktree_CDC_C}
\end{figure*}

Based on Fig~\ref{fig:attacktree_CDC_C}, we identified three paths. The threat scenario's overall attack feasibility is $Medium$, the potential impact is $Serious$, and the final risk level is $4$:

\textbf{Attack Path 7:} In the T-BOX, we discovered hard-coded CA certificates and private keys, which allowed us to bypass the TLS/SSL verification and conduct a man-in-the-middle attack between the Cloud and the T-BOX. This enabled us to inject malicious code into the firmware upgrade package. Additionally, by reverse-engineering the OTA firmware package verification process in the firmware, we found a logical flaw where the CDC would proceed with the upgrade regardless of whether the firmware package hash was correct. This allowed us to successfully run the modified firmware on the CDC-MCU.

\textbf{Attack Path 8: }The CDC-MCU can be upgraded via a USB drive. By reverse-engineering the firmware, we identified a logical flaw where the CDC would proceed with the upgrade regardless of whether the firmware package hash was correct. We created a malicious firmware package on a USB drive and successfully ran the modified firmware on the CDC-MCU through the USB port.

\textbf{Attack Path 9:} We used a timing side-channel attack to brute-force the ADB access password for the IVI system. We then controlled the IVI to issue an illegal firmware rollback request. Due to a design flaw, the old firmware was updated to the CDC-MCU via the OTA function.

\subsection*{Complete Attack Tree of PKES (Car D)}

\begin{figure*}[t!]
    \centering
    \includegraphics[width=1 \linewidth]{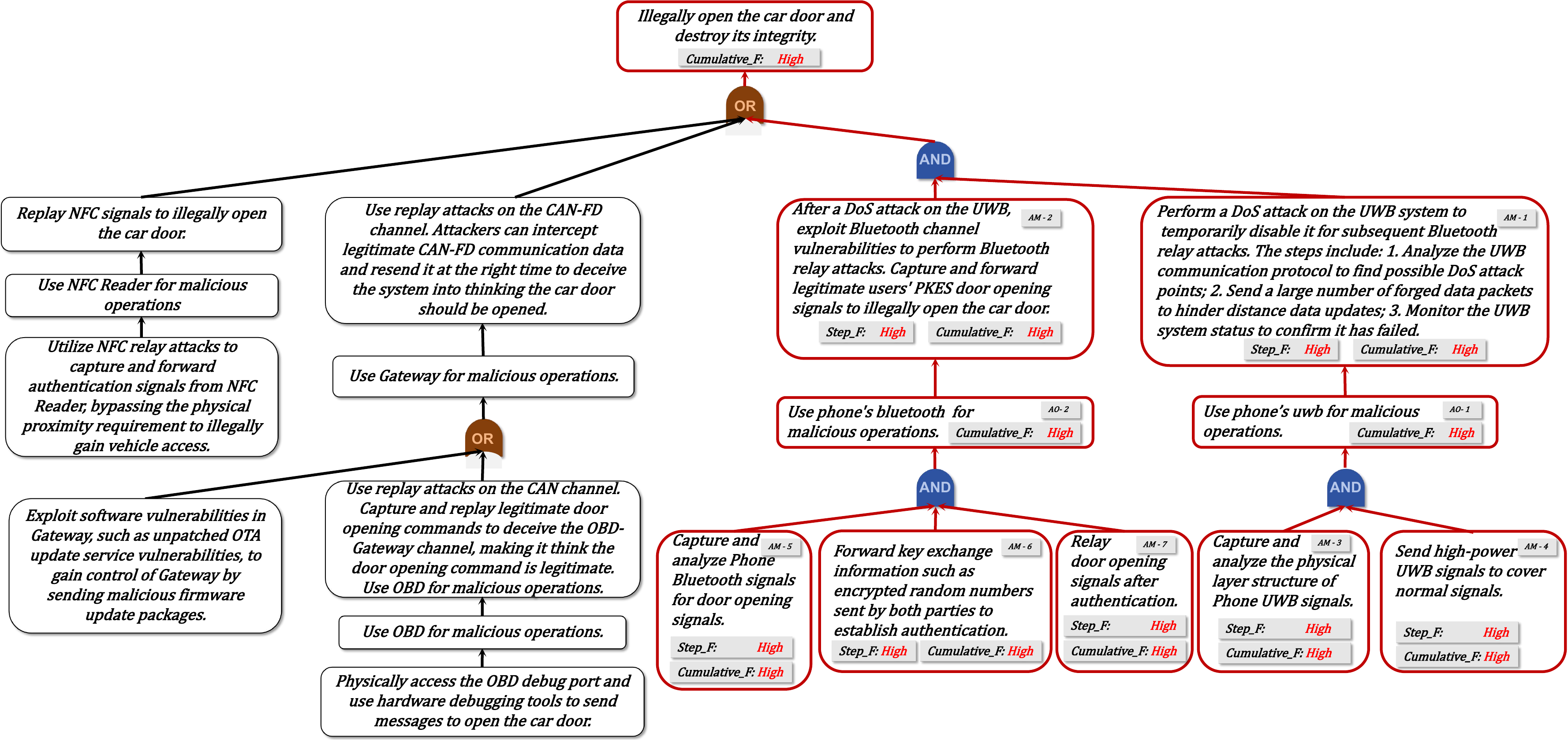}
    \caption{Attack tree PKES (Car D)}
    \label{fig:attacktree_PKES_D}
\end{figure*}

Lastly, guided by Fig~\ref{fig:attacktree_PKES_D}, we identified two feasible paths. The overall attack feasibility is $High$, the potential impact is $Serious$, and the final risk level is $5$:

\textbf{Attack Path 10:} We first sniffed the normal UWB (Ultra-Wideband) signals to analyze the physical layer structure in use, and sniffed BLE signals to confirm the exchanged information such as MAC, UUID. Subsequently, we continuously transmitted malicious UWB signals to disrupt the UWB ranging process. We discovered that when UWB ranging fails consecutively for a certain period, the car still uses the old distance data from before the interference. Therefore, as long as we interfere with the UWB signals while the car owner is still near the car, and then relay the BLE signals after the owner has moved away, the UWB ranging data will still reflect the old distance data from when the owner was near the car. The PKES still near the car and erroneously unlock the doors automatically.

\textbf{Attack Path 11:} We used a custom-made RFID relay device to receive the Select AID data transmitted by the car and relay it to another relay device positioned near the car's NFC card. The NFC card would respond with an encrypted reply, which the relay device would then forward to the vehicle's card reader, thereby unlocking the car door.

\section*{Appendix B: Complete Attack Trees of Case Studies}

\begin{figure*}[t!]
    \centering
    \includegraphics[width=1 \linewidth]{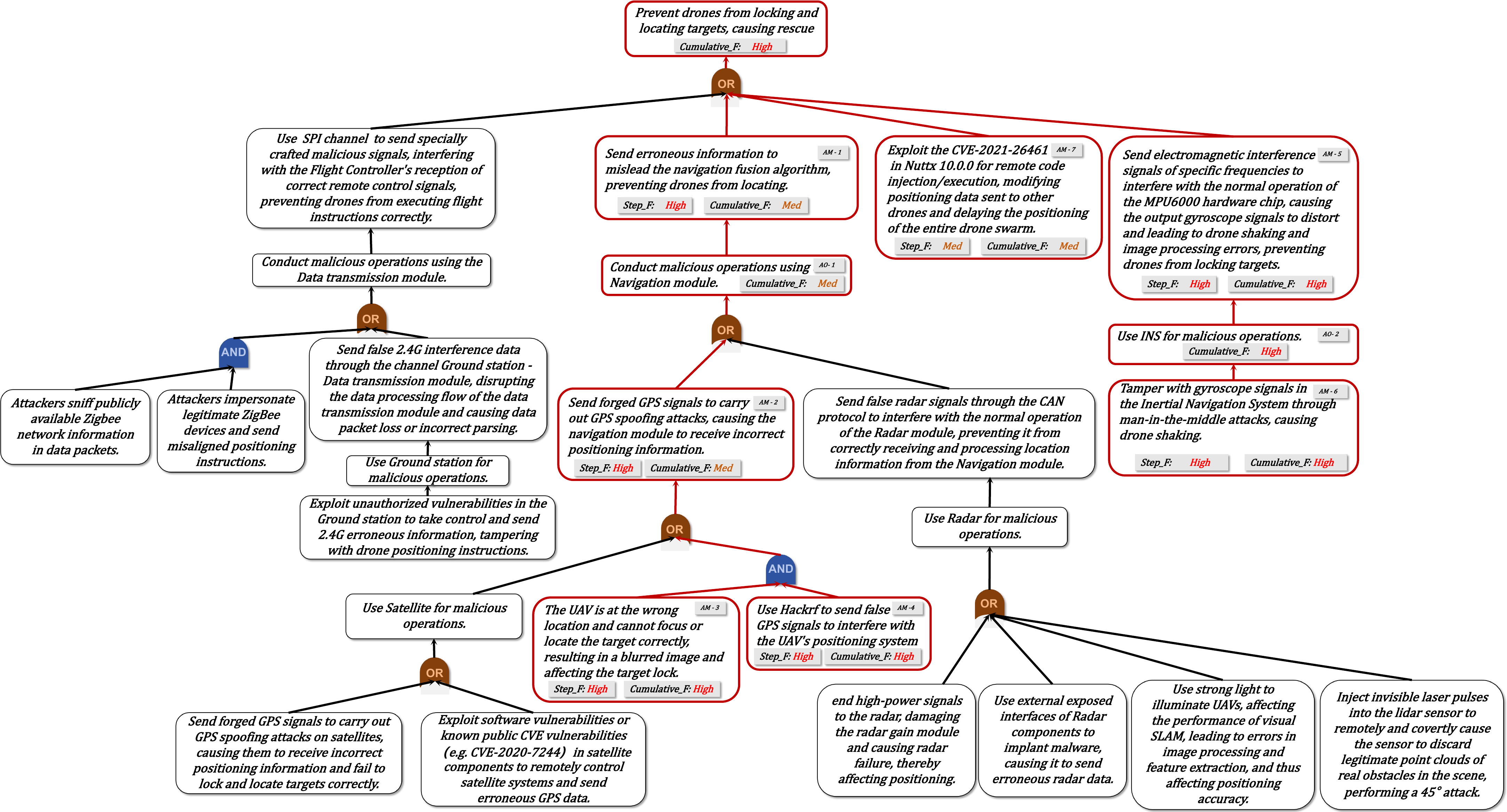}
    \caption{Attack tree UAV (Uav E)}
    \label{fig:attacktree_UAV_E}
\end{figure*}

\begin{figure*}[t!]
    \centering
    \includegraphics[width=1 \linewidth]{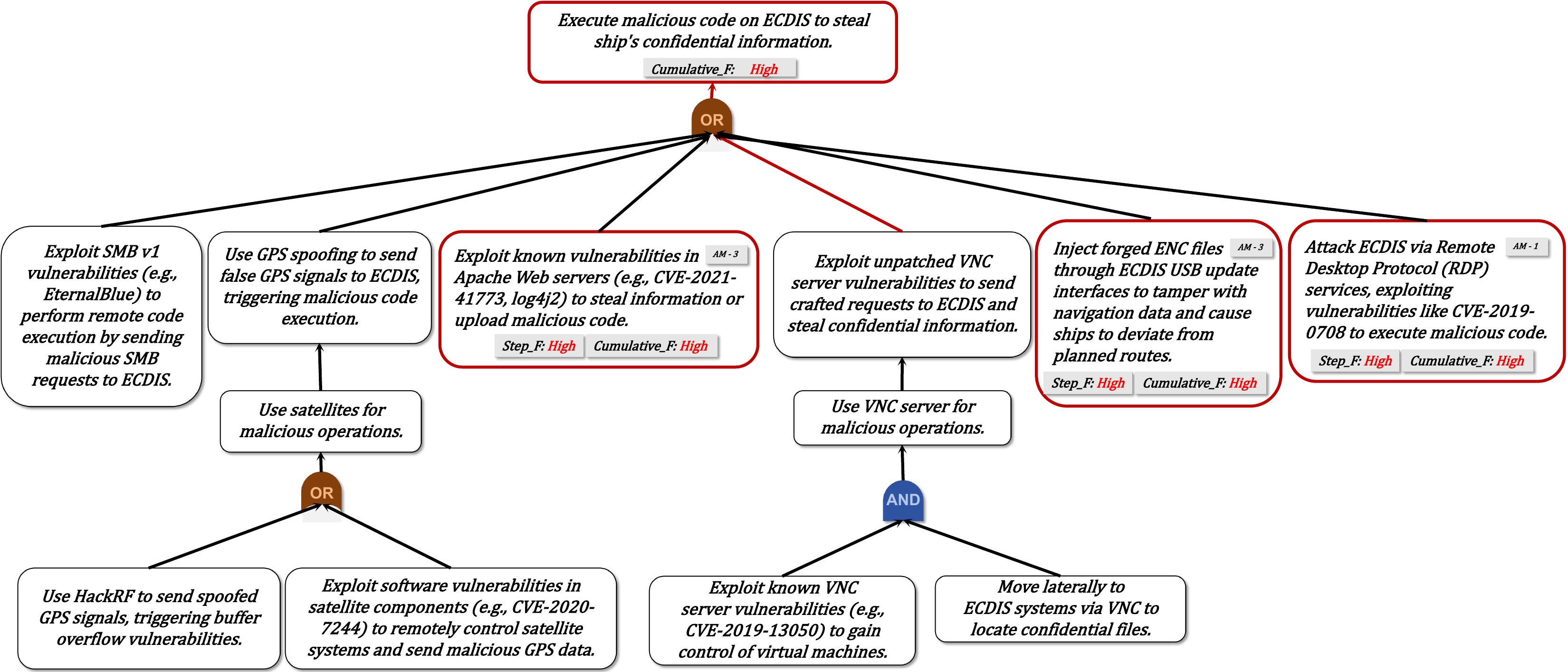}
    \caption{Attack tree ECDIS (Ship E)}
    \label{fig:attacktree_ECDIS_F}
\end{figure*}

We also generated two complete attack trees based on previous research and publicly available information (Fig~\ref{fig:attacktree_UAV_E} and Fig~\ref{fig:attacktree_ECDIS_F}), highlighting high-feasibility paths detailed in Sec~\ref{other car} in red. Figure~\ref{fig:attacktree_UAV_E} encompasses most UAV-related attack paths from~\cite{xu2023sok}, whereas Fig~\ref{fig:attacktree_ECDIS_F} includes all high-risk vulnerabilities referenced in~\cite{svilicic2019raising}.


\subsection*{Complete Attack Tree of UAV}
Fig~\ref{fig:attacktree_UAV_E} reveals three highly feasible paths, with an overall attack feasibility of $High$, a potential impact of $Major$, and a risk level of $4$:

\textbf{Attack Path 1:} By transmitting electromagnetic interference signals at specific frequencies, the normal operation of the MPU60001.0 hardware chip is disrupted, causing distortion in the gyroscope signals it outputs. This leads to drone instability and subsequent image processing errors.

\textbf{Attack Path 2:} By sending spoofed GPS signals to interfere with the drone's positioning system, the drone receives incorrect geographical information, causing its camera to fail to focus or accurately target objects, resulting in blurred images that disrupt the drone's functionality.

\textbf{Attack Path 3:} By tampering with the positioning signals transmitted between drones, the swarm algorithm of the entire drone fleet is disrupted, causing delays in reaching the target location and affecting the timeliness of drone rescue missions.

\subsection*{Complete Attack Tree of ECDIS}
Likewise, Fig~\ref{fig:attacktree_ECDIS_F} guides three highly feasible paths, with an overall attack feasibility of $High$, a potential impact of $Major$, and a final risk level of $4$:

\textbf{Attack Path 1:} Attackers can exploit vulnerabilities in the RDP service, such as CVE-2019-0708, to gain unauthorized remote desktop access and execute malicious code on the ECDIS system.

\textbf{Attack Path 2:} By exploiting known vulnerabilities in the Apache web server, such as directory traversal or remote code execution vulnerabilities, malicious code can be uploaded to the server.

\textbf{Attack Path 3:} By updating incorrect ENC files through the USB interface of the ECDIS system, the navigation data is tampered with, causing the vessel to deviate from its intended route.

\section*{Appendix C: Comparison with Human Experts}\label{Appendix: scoring}

\subsubsection{Study Setup}
To scientifically and objectively assess whether the attack trees designed by \projname{} are superior to those designed by human experts, we conducted a three-step scoring process: establishing scoring criteria, designing attack trees by experts, and scoring by experts.
\begin{table}[h]

\small 
\centering
\begin{threeparttable}
\caption{Seven-member review team}\label{tab:recruit}
\rowcolors{1}{gray!20}{white} 
\begin{tabular}{ccccc}  
\toprule

\textbf{ID} & \textbf{Exp\textsuperscript{1}} & \textbf{Co.\textsuperscript{2}} & \textbf{Position\textsuperscript{3}} & \textbf{Task\textsuperscript{4}} \\ 
\midrule
Expert A & 5 & C3 & TARA, Test & Dsgn, Scr \\
Expert B & 3 & C3 & TARA, Test & Dsgn, Scr \\
Expert C & 10 & C3 & TARA, Test & Crt, Dsgn, Scr \\
Expert D & 6 & C1 & TARA, Manag & Crt, Dsgn, Scr \\ 
Expert E & 5 & C2 & TARA, Manag & Crt, Dsgn, Scr \\ 
Expert F & 12 & C4 & TARA, Manag, Reg & Crt, Dsgn, Scr \\ 
Expert G & 5 & C4 & TARA, Manag, Reg & Crt, Dsgn, Scr \\ \bottomrule
\end{tabular}
\label{table:table_1}
\begin{tablenotes}
\small 
\item \textsuperscript{1} Years of working experience in security;
\item \textsuperscript{2} C1, C2: 1st party vehicle manufacturer,C3: 3rd party supplier; C4: TARA assessment agency.
\item \textsuperscript{3} TARA: Threat Analysis and Risk Assessment; Manag: Project manager; Reg: Regulation-related study; Test: Security testing.
\item \textsuperscript{4} Crt: establishing scoring criteria; Dsgn: designing Attack Trees; Scr: conducting scoring .
\end{tablenotes}
\end{threeparttable}
\end{table}

\textbf{Recruitment.} To ensure the professionalism and objectivity of the results, we invited experts from the automotive cybersecurity field of first-tier automotive manufacturers (2 persons), third-party suppliers (3 persons), and TARA ssessment agency (2 persons) from multiple countries (China and Germany) to form a seven-member review team, with their information presented in Table \ref{table:table_1}. On average, the team members have about 6 years of experience in the security field, with Expert c and Expert f having over 10 years of experience. Their positions include TARA, security testing, project management, and regulation study. Tasks were assigned based on their positions and interviews (5 persons establishing the scoring criteria, all experts designing the attack trees, and conducting scoring), ensuring that all participants have sufficient experience to competently perform their tasks.

\textbf{Procedure.} To ensure a thorough and unbiased evaluation of the attack trees, we assembled a seven-member review team to conduct a comprehensive scoring analysis. The team assessed the attack trees designed by DefenseWeaver and those created by seven experts (each expert contributed six attack trees targeting four automotive components and two other electronic systems). Experts selected five key scoring dimensions—rationality, non-redundancy, novelty, level of detail, and configuration alignment—based on ISO/SAE 21434 to capture different aspects of attack tree quality. Each dimension was scored on a scale of 1 to 4 points, allowing for a nuanced comparison (Table~\ref{tab:results2}). 
To maintain fairness and minimize potential bias, we implemented a structured scoring process: (i) DefenseWeaver’s attack trees were evaluated by all seven reviewers to ensure broad consensus. (ii) The experts’ attack trees were scored only by individuals not involved in their creation, preventing author-related biases. For consistency, the score for each dimension was calculated as the average based on the total points and the number of reviewers. To visually represent the overall performance differences, we plotted the five-dimensional scores into radar charts (Fig.~\ref{fig:radar}) and compared their areas. This approach allowed us to objectively quantify the comprehensive superiority of the attack trees, rather than relying on subjective judgments.
\begin{table}[h]
\centering
\caption{Overall Comparison with Human Experts}
\label{tab:results2}

\setlength{\arrayrulewidth}{0.3pt}

\begin{tabular}{
 >{\raggedright\arraybackslash}c 
    >{\raggedright\arraybackslash}c 
    c 
    c 
    >{\raggedright\arraybackslash}c 
    }
\hline
\rowcolor{gray!25}
\textbf{System} & \textbf{Dimension} & \textbf{DefenseWeaver} & \textbf{Expert} & \textbf{Improv.} \\
\hline

\multirow{5}{*}{BCM\_A} 
& Rationality        & 3.40 & 3.10 & $\uparrow$ 9.68\%    \\
\rowcolor{zebra}
& Non-redundancy     & 3.40 & 3.70 & $\downarrow$ 8.11\%  \\
& Novelty            & 2.60 & 1.30 & $\uparrow$ 100.00\%  \\
\rowcolor{zebra}
& Level of detail    & 4.00 & 3.00 & $\uparrow$ 33.33\%   \\
& System align.  & 3.60 & 2.15 & $\uparrow$ 67.44\%   \\
\hline

\multirow{5}{*}{BCM\_B} 
& Rationality        & 3.40 & 3.05 & $\uparrow$ 11.48\%   \\
\rowcolor{zebra}
& Non-redundancy     & 3.40 & 3.55 & $\downarrow$ 4.23\%   \\
& Novelty            & 2.20 & 1.25 & $\uparrow$ 76.00\%    \\
\rowcolor{zebra}
& Level of detail    & 4.00 & 2.90 & $\uparrow$ 37.93\%    \\
& System align.  & 3.60 & 2.20 & $\uparrow$ 63.64\%    \\
\hline

\multirow{5}{*}{CDC} 
& Rationality        & 3.60 & 3.30 & $\uparrow$ 9.09\%     \\
\rowcolor{zebra}
& Non-redundancy     & 3.60 & 3.30 & $\uparrow$ 9.09\%     \\
& Novelty            & 2.60 & 1.20 & $\uparrow$ 116.67\%   \\
\rowcolor{zebra}
& Level of detail    & 3.80 & 2.45 & $\uparrow$ 55.10\%    \\
& System align.  & 3.80 & 2.85 & $\uparrow$ 33.33\%    \\
\hline

\multirow{5}{*}{PKES} 
& Rationality        & 4.00 & 3.90 & $\uparrow$ 2.56\%     \\
\rowcolor{zebra}
& Non-redundancy     & 3.60 & 3.75 & $\downarrow$ 4.00\%     \\
& Novelty            & 3.00 & 1.35 & $\uparrow$ 122.22\%   \\
\rowcolor{zebra}
& Level of detail    & 3.80 & 3.35 & $\uparrow$ 13.43\%    \\
& System align.  & 3.80 & 3.05 & $\uparrow$ 24.59\%    \\
\hline

\multirow{5}{*}{UAV} 
& Rationality        & 3.67 & 3.58 & $\uparrow$ 2.51\%     \\
\rowcolor{zebra}
& Non-redundancy     & 4.00 & 3.90 & $\uparrow$ 2.56\%     \\
& Novelty            & 3.00 & 1.70 & $\uparrow$ 76.47\%    \\
\rowcolor{zebra}
& Level of detail    & 3.67 & 2.50 & $\uparrow$ 46.80\%    \\
& System align.  & 3.67 & 2.75 & $\uparrow$ 33.45\%    \\
\hline

\multirow{5}{*}{Ship} 
& Rationality        & 3.67 & 3.62 & $\uparrow$ 1.38\%     \\
\rowcolor{zebra}
& Non-redundancy     & 4.00 & 4.00 & -                    \\
& Novelty            & 3.00 & 1.20 & $\uparrow$ 150.00\%   \\
\rowcolor{zebra}
& Level of detail    & 4.00 & 2.25 & $\uparrow$ 77.78\%    \\
& System align.  & 3.80 & 2.50 & $\uparrow$ 52.00\%    \\
\hline

\multirow{5}{*}{Overall} 
& Rationality        & 3.62 & 3.43 & $\uparrow$ 5.79\%     \\
\rowcolor{zebra}
& Non-redundancy     & 3.67 & 3.70 & $\downarrow$ 0.90\%    \\
& Novelty            & 2.73 & 1.33 & $\uparrow$ 105.00\%   \\
\rowcolor{zebra}
& Level of detail    & 3.88 & 2.74 & $\uparrow$ 41.46\%    \\
& System align  & 3.71 & 2.58 & $\uparrow$ 43.68\%    \\
\hline
\end{tabular}
\end{table}

\subsubsection{Scoring Criterion}
Experts have selected five core evaluation dimensions (Table~\ref{tab:scoring})—logical rationality, non-redundancy, configuration alignment, level of detail, and novelty—based on ISO/SAE 21434 and industry white papers to comprehensively assess the quality of attack trees(as shown in Table ~\ref{tab:scoring}). The rationale for each dimension is as follows:
(i) \textbf{Logical Rationality} : ISO/SAE 21434 requires each attack path should be logically sound and follow security reasoning (no implausible leaps).  This includes ensuring that the local attack objectives are clearly defined and realistic, the attack methods are technically feasible and relevant to the context, the logical nodes accurately represent the steps and relationships in the attack process, the feasibility and impact ratings are reasonable, and the overall attack path is consistent with the defined threat scenarios;
(ii) \textbf{Non-redundancy}: According to TARA optimization principles, the attack tree should avoid duplicate or redundant paths, focusing on unique attack vectors and avoiding analytical redundancy aligns with the minimal attack tree criterion;
(iii) \textbf{Configuration Alignment}: ISO/SAE 21434 mandate that attack tree nodes strictly correspond to system modeling elements (e.g., ECUs, communication protocols), ensuring TARA's practical applicability;
(iv) \textbf{Level of Detail}: the attack steps should be described with sufficient technical detail and specificity (e.g., including concrete attack techniques or CVE examples); 
(v) \textbf{Novelty}: While ensuring compliance, the attack tree should include creative attack methods beyond well-known threat templates to enhance the comprehensiveness of threat coverage.
Each dimension was scored on a four-point scale from Poor (1) to Excellent (4). The final score for each dimension is the average of all reviewers’ ratings for that criterion on a given tree.

\begin{table}[t]
\centering
\setlength{\arrayrulewidth}{0.3pt}
\caption{Attack Tree Scoring Criteria}
\label{tab:scoring}
\begin{threeparttable}
\rowcolors{2}{gray!20}{white} 
\begin{tabularx}{\linewidth}{@{}p{2cm}p{0.8cm}p{0.8cm}>{\raggedright\arraybackslash}p{4cm}@{}}
\toprule
\textbf{Dimension} & \textbf{Score} & \textbf{Level} & \textbf{Description} \\
\midrule
Logical rationality & 1 & Poor & All paths are considered unreasonable \\
 & 2 & Limited & Only very few paths are reasonable \\
 & 3 & Good & Most paths are reasonable with some exceptions \\
 & 4 & Excellent & Nearly all paths are reasonable \\
\cmidrule(r){1-4}

Non-redundancy & 1 & Poor & All reasonable paths are completely repetitive \\
& 2 & Limited & Majority of paths are redundant \\
& 3 & Good & Few paths are repetitive \\
& 4 & Excellent & All paths are unique \\
\cmidrule(r){1-4}

Configuration alignment & 1 & Poor & Uses substantial irrelevant information \\
& 2 & Limited & Uses some relevant information mixed with irrelevant data \\
& 3 & Good & Uses 60\% of system modeling information \\
& 4 & Excellent & Uses 80\% of system modeling information \\
\cmidrule(r){1-4}

Level of detail & 1 & Poor & Incomprehensible to users \\
& 2 & Limited & Specifies target objects (e.g., attacking TBOX) \\
& 3 & Good & Specifies objects with attack techniques (e.g., MITM on TBOX's WiFi) \\
& 4 & Excellent & Includes objects, techniques, and examples (e.g., MITM on TBOX's WiFi using CVE-XXXX) \\
\cmidrule(r){1-4}

Novelty & 1 & Poor & No new attack methods provided \\
& 2 & Limited & 1-2 unexpected attack methods \\
& 3 & Good & 2-3 unexpected attack methods \\
& 4 & Excellent & More than 3 unexpected attack methods \\
\bottomrule
\end{tabularx}
\end{threeparttable}
\end{table}


\clearpage

\end{document}